\documentclass[
aps,prd,
10pt,
amsfonts,amssymb,amsmath,
preprintnumbers,
showpacs,
letterpaper,
eqsecnum,
]{revtex4-1}

\usepackage[dvipdfmx]{graphicx}
\usepackage{hyperref}

\begin{document}

\title{
New cosmological constraints on primordial black holes
}

\author{B.J.~Carr}\email[]{B.J.Carr@qmul.ac.uk}
\affiliation{
Astronomy Unit, Queen Mary University of London,
Mile End Road, London E1 4NS, United Kingdom
}
\affiliation{
Research Center for the Early Universe (RESCEU),
Graduate School of Science, The University of Tokyo,
Tokyo 113-0033, Japan
}
\affiliation{
Canadian Institute for Theoretical Astrophysics,
University of Toronto,
60 St.~George Street, Toronto, Ontario M5S 1A1, Canada
}
\author{Kazunori Kohri}\email[]{kohri@tuhep.phys.tohoku.ac.jp}
\affiliation{
Department of Physics, Tohoku University,
Sendai 980-8578, Japan
}
\affiliation{
Physics Department, Lancaster University,
Lancaster LA1 4YB, United Kingdom
}
\affiliation{
Harvard-Smithsonian Center for Astrophysics,
60 Garden Street, Cambridge, Massachusetts 02138, USA
}
\author{Yuuiti Sendouda}\email[]{sendouda@yukawa.kyoto-u.ac.jp}
\affiliation{
Yukawa Institute for Theoretical Physics,
Kyoto University,
Kyoto 606-8502, Japan
}
\affiliation{
Department of Physics,
Graduate School of Science, The University of Tokyo,
Tokyo 113-0033, Japan
}
\author{Jun'ichi Yokoyama}\email[]{yokoyama@resceu.s.u-tokyo.ac.jp}
\affiliation{
Research Center for the Early Universe (RESCEU),
Graduate School of Science, The University of Tokyo,
Tokyo 113-0033, Japan%
}
\affiliation{
Institute for the Physics and Mathematics of the Universe (IPMU),
The University of Tokyo,
Kashiwa, Chiba 277-8568, Japan
}

\date{\today}

\begin{abstract}
We update the constraints on the fraction of the Universe going into primordial black holes in the mass range $ 10^9\text{--}10^{17}\,\mathrm g $ associated with the effects of their evaporations on big bang nucleosynthesis and the extragalactic photon background.
We include for the first time all the effects of quark and gluon emission by black holes on these constraints and account for the latest observational developments.
We then discuss the other constraints in this mass range and show that these are weaker than the nucleosynthesis and photon background limits, apart from a small range $ 10^{13}\text{--}10^{14}\,\mathrm g $, where the damping of cosmic microwave background anisotropies dominates.
Finally we review the gravitational and astrophysical effects of nonevaporating primordial black holes, updating constraints over the broader mass range $ 1\text{--}10^{50}\,\mathrm g $.
\end{abstract}

\preprint{RESCEU-31/09}
\preprint{TU-852}
\preprint{YITP-09-112}

\pacs{04.70.Dy, 95.35.+d, 97.60.Lf, 98.80.Cq}

\maketitle

\section{
Introduction
\label{sec:intro}
}

Black holes with a wide range of masses could have formed in the early Universe as a result of the great compression associated with the big bang \cite{Hawking:1971ei,1966AZh....43..758Z,*1967SvA....10..602Z,Carr:1974nx}.
A comparison of the cosmological density at a time $ t $ after the big bang with the density associated with a black hole of mass $ M $ suggests that such ``primordial'' black holes (PBHs) would have a mass of order
\begin{equation}
M
\sim
  \frac{c^3\,t}{G}
\sim
  10^{15}\,\left(\frac{t}{10^{-23}\,\mathrm s}\right)\,\mathrm g.
\label{eq:Moft}
\end{equation}
This roughly corresponds to the particle horizon mass in a noninflationary model or the Hubble mass otherwise.
PBHs could thus span an enormous mass range: those formed at the Planck time ($ 10^{-43}\,\mathrm s $) would have the Planck mass ($ 10^{-5}\,\mathrm g $), whereas those formed at $ 1\,\mathrm s $ would be as large as $ 10^5\,M_\odot $\,, comparable to the mass of the holes thought to reside in galactic nuclei.
They could be even larger than this in some circumstances.
By contrast, black holes forming at the present epoch could never be smaller than about $ 1\,M_\odot $\,.

The high density of the early Universe is a necessary but not sufficient condition for PBH formation.
One possibility is that there were large primordial inhomogeneities, so that overdense regions could stop expanding and recollapse \cite{Carr:1975qj,1978AZh....55..216N,*1978SvA....22..129N,1979ApJ...232..670B}.
The quantum fluctuations arising in various inflationary scenarios are of particular interest in this context, as has been discussed in numerous papers.
In some of these scenarios the fluctuations generated by inflation are ``blue'' (i.e.\ decrease with increasing scale) and this means that the PBHs form shortly after reheating 
\cite{Carr:1993aq,Carr:1994ar,Leach:2000ea,Kohri:2007gq}.
Others involve some form of ``designer'' inflation, in which the power spectrum of the fluctuations---and hence PBH production---peaks on some scale 
\cite{Hodges:1990bf,Ivanov:1994pa,Yokoyama:1995ex,Yokoyama:1998pt,Yokoyama:1998qw,Kawasaki:1998vx,Yokoyama:1999xi,Easther:1999ws,Kanazawa:2000ea,Blais:2002gw,Blais:2002nd,Barrau:2002ru,Chongchitnan:2006wx,Nozari:2007kv,Saito:2008em}.
In other scenarios, the fluctuations have a ``running index,'' so that the amplitude increases on smaller scales but not according to a simple power law
\cite{GarciaBellido:1996qt,Randall:1995dj,Lidsey:2001nj,Easther:2004qs,Lyth:2005ze,Kohri:2007qn,Bugaev:2008bi,Alabidi:2009bk}.
Finally, PBH formation may occur due to some sort of parametric resonance effect before reheating \cite{Taruya:1998cz,Bassett:2000ha,Green:2000he,Finelli:2000gi,Kawaguchi:2007fz,Kawasaki:2007zz,Frampton:2010sw}.
In this case, the fluctuations tend to peak on a scale associated with reheating.
This is usually very small but several scenarios involve a secondary inflationary phase which boosts this scale into the macroscopic domain.

Whatever the source of the inhomogeneities, PBH formation would be enhanced if there was a sudden reduction in the pressure---for example, at the quark-hadron era \cite{Jedamzik:1996mr,Widerin:1998my,Jedamzik:1999am}---and especially likely if the early Universe went through a dustlike phase at early times as a result of either being dominated by nonrelativistic particles for a period \cite{Khlopov:1980mg,1981AZh....58..706P,*1981SvA....25..406P,1982AZh....59..639P,*1982SvA....26..391P} or undergoing slow reheating after inflation \cite{Khlopov:1985jw,Carr:1994ar}.
Another possibility is that PBHs might have formed spontaneously at some sort of phase transition even if there were no prior inhomogeneities---for example, from bubble collisions \cite{Crawford:1982yz,Hawking:1982ga,Kodama:1982sf,La:1989st,Moss:1994iq,1998PAZ...24..483K,*1998AstL...24..413K,1999YZ...62..1705K,*Konoplich:1999qq} or from the collapse of cosmic strings \cite{Hogan:1984zb,Hawking:1987bn,Polnarev:1988dh,Garriga:1993gj,Caldwell:1995fu,Cheng:1996du,MacGibbon:1997pu,Hansen:1999su,Nagasawa:2005hv} or necklaces \cite{Matsuda:2005ez,Lake:2009nq} or domain walls \cite{Berezin:1982ur,Caldwell:1996pt,Khlopov:2000js,Rubin:2000dq,Rubin:2001yw,Dokuchaev:2004kr}.

These PBH formation scenarios are reviewed in Refs.~\cite{Carr:2005zd,Khlopov:2008qy}.
Although we do not discuss them in detail here, it should be stressed that in most of them the PBH mass spectrum is narrow and centered around the mass given by Eq.~\eqref{eq:Moft} with $ t $ corresponding to the reheating epoch or the time at which the characteristic PBH scale reenters the horizon in the inflationary context or to the time of the relevant cosmological phase transition otherwise.
However, PBHs may be smaller than the horizon size at formation in some circumstances.
For example, PBH formation is an interesting application of ``critical phenomena'' \cite{Niemeyer:1997mt,Niemeyer:1999ak,Shibata:1999zs,Musco:2004ak,Musco:2008hv} and this suggests that their spectrum could be more extended and go well below the horizon mass \cite{Yokoyama:1998xd,Green:1999xm,Kribs:1999bs}.
This would also apply for PBHs formed during a dustlike phase \cite{Polnarev:1986bi}.
Note that a PBH could not be much larger than the value given by Eq.~\eqref{eq:Moft} at formation because it would then be a separate closed Universe rather part of our Universe \cite{Harada:2004pe}.
However, it could still grow subsequently as a result of accretion, so the final PBH mass could well be larger than the horizon mass at formation.

The realization that PBHs might be small prompted Hawking to study their quantum properties.
This led to his famous discovery \cite{Hawking:1974rv,Hawking:1974sw,*Hawking:1974swE} that black holes radiate thermally with a temperature
\begin{equation}
T_\mathrm{BH}
= \frac{\hbar\,c^3}{8\pi\,G\,M\,k_\mathrm B}
\sim
  10^{-7}\,\left(\frac{M}{M_\odot}\right)^{-1}\,\mathrm K,
\end{equation}
so they evaporate completely on a time scale
\begin{equation}
\tau(M)
\sim
  \frac{G^2\,M^3}{\hbar\,c^4}
\sim
  10^{64}\,\left(\frac{M}{M_\odot}\right)^3\,\mathrm{yr}\,.
\end{equation}
Only PBHs smaller than $ M_* \sim 10^{15}\,\mathrm g $ would have evaporated by the present epoch, so Eq.~\eqref{eq:Moft} implies that this effect could be important only for ones which formed before $ 10^{-23}\,\mathrm s $.
Since PBHs with a mass of around $ 10^{15}\,\mathrm g $ would be producing photons with energy of order $ 100\,\mathrm{MeV} $ at the present epoch, the observational limit on the $ \gamma $-ray background intensity at $ 100\,\mathrm{MeV} $ immediately implied that their density could not exceed about $ 10^{-8} $ times the critical density \cite{Page:1976wx}.
This suggested that there was little chance of detecting their final explosive phase at the present epoch, at least in the standard model of particle physics \cite{1979Natur.277..199P}.
It also meant that PBHs with an extended mass function could provide the dark matter only if the fraction of their mass around $ 10^{15}\,\mathrm g $ were tiny.
Nevertheless, it was soon realized that the $ \gamma $-ray background limit does not preclude PBHs having important cosmological effects \cite{Carr:1976jy}.
These are of different types, depending on the PBH mass range.

\textit{PBHs with $ M > 10^{15}\,\mathrm g $.}
These would still survive today and might be detectable by their gravitational effects.
Indeed such PBHs would be obvious dark matter candidates.
Since they formed at a time when the Universe was radiation dominated, they should be classified as nonbaryonic and so could avoid the constraints on the baryonic density associated with cosmological nucleosynthesis.
They would also be dynamically cold at the present epoch and so would be classified as cold dark matter (CDM).
In many respects, they would be like (nonbaryonic) weakly interacting massive particles (WIMPs) but they would be much more massive and so could also have the sort of dynamical, lensing, and gravitational-wave signatures associated with (baryonic) massive compact halo objects (MACHOs).
At one stage there seemed to evidence for MACHOs with $ M \sim 0.5\,M_\odot $ from microlensing observations \cite{Alcock:2000ph} and PBHs formed at the quark-hadron phase transition seemed one possible explanation for this \cite{Jedamzik:1996mr}.
The data now seems less clear but there are no constraints excluding PBHs in the sublunar range $ 10^{20}\,\mathrm g < M < 10^{26}\,\mathrm g $ \cite{Kanazawa:2000ea,Blais:2002nd,Blais:2002gw} or intermediate mass range $ 10^2\,M_\odot < M < 10^4\,M_\odot $ \cite{Mack:2006gz,Saito:2008em,Frampton:2010sw} from having an appreciable density.
Large PBHs might also influence the development of large-scale structure \cite{Meszaros:1975ef,1977A&A....56..377C,1983ApJ...275..405F,1983ApJ...268....1C,Afshordi:2003zb}, seed the supermassive black holes thought to reside in galactic nuclei \cite{1984MNRAS.206..801C,Duechting:2004dk,Khlopov:2004sc,Bean:2002kx}, generate background gravitational waves \cite{1980A&A....89....6C,Nakamura:1997sm,Ioka:1998gf,Inoue:2003di,Hayasaki:2009ug}, or produce x rays through accretion and thereby affect the thermal history of the Universe \cite{Ricotti:2007au}.

\textit{PBHs with $ M \sim 10^{15}\,\mathrm g $.}
As already noted, these would be evaporating today and, since they are dynamically cold, one would expect some of them to have clustered within the Galactic halo.
Besides contributing to the cosmological $ \gamma $-ray background, an effect we reassess in this paper, such PBHs could contribute to the Galactic $ \gamma $-ray background \cite{1996ApJ...459..487W,Lehoucq:2009ge} and the antiprotons or positrons in cosmic rays \cite{Carr:1976jy,Turner:1981ez,Kiraly:1981ci}.
They might also generate gamma-ray bursts \cite{Cline:1992ps}, radio bursts \cite{1977Natur.266..333R} and the annihilation-line radiation coming from center of the Galaxy \cite{1980A&A....81..263O,Bugaev:2008gw}.
The energy distribution of the particles emitted could also give significant information about the high-energy physics involved in the final explosive phase of black hole evaporation \cite{Halzen:1991uw}.

\textit{PBHs with $ M < 10^{15}\,\mathrm g $.}
These would have completely evaporated by now but many processes in the early Universe could have been modified by them.
For example, PBH evaporations occurring in the first second of the big bang could generate the entropy of the Universe \cite{1976PZETF..24..616Z,*1976JETPL..24..571Z}, change the details of baryogenesis \cite{Turner:1979bt,Barrow:1990he,Upadhyay:1999vk,Dolgov:2000ht,Bugaev:2001xr}, provide a source of neutrinos \cite{Bugaev:2008gw,Bugaev:2002yt} or gravitinos \cite{Khlopov:2004tn} or other hypothetical particles \cite{Green:1999yh,Lemoine:2000sq}, swallow monopoles \cite{Izawa:1984ww,Stojkovic:2004hz}, and remove domain walls by puncturing them \cite{Stojkovic:2005zh}.
If the evaporations left stable Planck-mass relics, these might also contribute to the dark matter \cite{MacGibbon:1987my,Barrow:1992hq,Carr:1994ar,Green:1997sz,Blais:2002nd,Alexeyev:2002tg,Chen:2002tu,Barrau:2003xp,Chen:2004ft,Nozari:2005ah,Alexander:2007gj}.
Most strikingly, they modify the details of cosmological nucleosynthesis, an effect we investigate in depth in this paper.
PBHs evaporating at later times could also have important astrophysical effects, such as helping to reionize the Universe \cite{He:2002vz,Mack:2008nv}.

Even if PBHs had none of these effects, it is still important to study them because each one is associated with an interesting upper limit on the fraction of the mass of the Universe which can have gone into PBHs on some mass scale $ M $\,.
This fraction is epoch dependent but its value at the formation epoch of the PBHs is of great cosmological significance, this quantity being denoted by $ \beta(M) $\,.
The current density parameter $ \Omega_\mathrm{PBH} $ (in units of the critical density) associated with unevaporated PBHs which form at a redshift $ z $ or time $ t $ is roughly related to $ \beta $ by \cite{Carr:1975qj}
\begin{equation}
\Omega_\mathrm{PBH}
\simeq
  \beta\,\Omega_\mathrm r\,(1+z)
\sim
  10^6\,\beta\,\left(\frac{t}{1\,\mathrm s}\right)^{-1/2}
\sim
  10^{18}\,\beta\,\left(\frac{M}{10^{15}\,\mathrm g}\right)^{-1/2}
\quad
(M > 10^{15}\,\mathrm g)\,,
\label{eq:roughomega}
\end{equation}
where $ \Omega_\mathrm r \sim 10^{-4} $ is the density parameter of the cosmic microwave background (CMB) and we have used Eq.~\eqref{eq:Moft}.
We have also neglected the effect of entropy production after PBH formation.
A more precise formula will be given later.
The $ (1+z) $ factor arises because the radiation density scales as $ (1+z)^4 $\,, whereas the PBH density scales as $ (1+z)^3 $\,.
Any limit on $ \Omega_\mathrm{PBH} $ therefore places a constraint on $ \beta(M) $\,.
For example, the $ \gamma $-ray limit implies $ \beta(10^{15}\,\mathrm g) \lesssim 10^{-26} $ and this is one of the strongest constraints on $ \beta $ over all mass ranges.
Another immediate constraint for PBHs with $ M > 10^{15}\,\mathrm g $ comes from requiring $ \Omega_\mathrm{PBH} $ to be less than the total matter density.
As discussed in this paper, there are also many constraints on $ \beta(M) $ for PBHs which have already evaporated, although the parameter $ \Omega_\mathrm{PBH} $ must then be reinterpreted since they no longer contribute to the cosmological density.
Note that Eq.~\eqref{eq:roughomega} assumes that the PBHs form in the radiation-dominated era, in which case $ \beta $ is necessarily small.
Indeed, $ \beta $ could only reach $ 1 $ for $ M \sim 10^{17}\,M_\odot $\,, which corresponds to the horizon mass at matter-radiation equality, and such black holes could hardly be regarded as primordial anyway.

The constraints on $ \beta(M) $ were first brought together by Novikov \textit{et al.} \cite{1979A&A....80..104N}.
An updated version of the constraints was later provided by Carr \textit{et al.} \cite{Carr:1994ar}, although this contained some errors which were corrected by Green and Liddle \cite{Green:1997sz}.
Subsequently, the $ \beta(M) $ diagram has frequently been updated and improved as the relevant effects have been studied in greater detail.
The most recent version of this diagram comes from Josan \textit{et al.} \cite{Josan:2009qn}.
The important qualitative point is that the value of $ \beta(M) $ must be tiny over almost every mass range, even if the PBH density is large today, so any cosmological model which would entail an appreciable fraction of the Universe going into PBHs is immediately excluded.
For example, this places strong constraints on the amplitude of the density inhomogeneities in the early Universe \cite{Carr:1975qj,Zaballa:2006kh,Bugaev:2008gw} and on the deviations of such inhomogeneities from Gaussianity \cite{Bullock:1996at,Ivanov:1997ia,Hidalgo:2007vk,Saito:2008em,Hidalgo:2009fp}.
One can also infer indirect limits on the spectral index of the primordial density fluctuations \cite{Carr:1994ar,Green:1997sz,Kim:1999iv,Bringmann:2001yp,Blais:2002nd,Polarski:2001jk} and constrain the reheating process which follows inflation \cite{Green:2000he,Bassett:2000ha}.
In a less conventional context, one can constrain cosmological models involving a time-varying gravitational ``constant'' \cite{Barrow:1992ay,Barrow:1996jk,Jacobson:1999vr,Carr:2000tq,Harada:2001kc} or extra dimensions \cite{Guedens:2002km,Guedens:2002sd,Clancy:2003zd,Tikhomirov:2005bt,*Tikhomirov:2006ck,Sendouda:2003dc,Sendouda:2004hz,Sendouda:2006nu}.
We do not explore such scenarios in the present paper but it should be stressed that the form of the $ \beta(M) $ constraints itself changes in this situation.

One of the main purposes of the present paper is to update the limits associated with big bang nucleosynthesis (BBN) and the extragalactic $ \gamma $-ray background (EGB) because these turn out to be the dominant ones over the mass range $ 10^9\text{--}10^{17}\,\mathrm g $.
Both of these limits have been a subject of long-standing interest but it is necessary to reassess them in the light of recent observational and theoretical developments.
On the observational front, there are new data on the light-element abundances, the neutron lifetime, and the $ \gamma $-ray background.
On the theoretical front, our view of the physics of strong interactions has changed.
It used to be assumed that hadrons are emitted from PBHs in the form of nucleons or mesons.
However, according to the modern view of quantum chromodynamics (QCD), PBHs must produce quark and gluon jets which then fragment into hadrons over the QCD distance, $ \Lambda_\mathrm{QCD}^{-1} \sim 10^{-13}\,\mathrm{cm} $ \cite{MacGibbon:1990zk}.
These hadrons then decay into the stable elementary particles of the standard model.

The consequences of quark and gluon emission for the EGB limit were first studied by MacGibbon and Carr \cite{MacGibbon:1991vc} but the analysis can now be refined because of convergence onto a standard ($ \Lambda $CDM) cosmological model and because of improved calculations of the production of photons through jet decays.
As regards the consequences of quark and gluon emission for BBN, most of the hadrons created decay almost instantaneously compared to the nucleosynthesis time scale, but long-lived ones remain in the ambient medium long enough to leave an observable signature.
Kohri and Yokoyama \cite{Kohri:1999ex} first analyzed this effect but only in the context of the relatively low-mass PBHs evaporating in the early stages of BBN.
In this paper we incorporate the effects of heavier PBHs, which evaporate after BBN, and infer constraints imposed by the dissociation and overproduction of synthesized light elements by both hadrons and high-energy photons.
Photons from unstable mesons or hadrons and bremsstrahlung emission from quarks and gluons are also appropriately incorporated.
We thus calculate for the first time the PBH constraints over the full relevant mass range.
The code employed in our calculations is similar to the one used by Kawasaki \textit{et al.} \cite{Kawasaki:2004qu} in studying the effects of decaying particles on BBN.

The plan of the paper is as follows.
Section~\ref{sec:bg} describes the background equations and defines various quantities related to PBHs.
Section~\ref{sec:evap} reviews black hole evaporation and the effects of quark-gluon emission.
Section~\ref{sec:bbn} then discusses the constraints deriving from cosmological nucleosynthesis effects, while Sec.~\ref{sec:photon} discusses the ones associated with the photon background.
Section~\ref{sec:other} combines both constraints in a single $ \beta(M) $ diagram for the mass range $ 10^9\text{--}10^{17}\,\mathrm g $ and then discusses some other (mainly less stringent) constraints in this mass range.
Section~\ref{sec:large} summarizes the most important limits in mass ranges associated with larger nonevaporating PBHs.
Section~\ref{sec:conc} collects all the constraints together into a single ``master'' $ \beta(M) $ diagram and draws some general conclusions.
It should be stressed that Secs.~\ref{sec:other} and \ref{sec:large} include quite a lot of review of previous work but it is useful to bring all the results together and we have elucidated earlier work where appropriate.
Throughout most of this paper we assume that the PBHs have a monochromatic mass function, but allowing even a small range of masses around $ 10^{15}\,\mathrm g $ would have interesting observational consequences, especially for the EGB limits.
However, this discussion is rather technical, so it is relegated to an Appendix.

\section{
Definitions and description of background cosmology
\label{sec:bg}
}

In this section, we present some relevant definitions and background equations.
We assume that the standard $ \Lambda $CDM model applies, with the age of the Universe being $ t_0 = 13.7\,\mathrm{Gyr} $, the Hubble parameter being $ h = 0.72 $ and the time of photon decoupling being $ t_\mathrm{dec} = 380\,\mathrm{kyr} $ \cite{Hinshaw:2008kr,Dunkley:2008ie}.
Throughout the paper we put $ c = \hbar = k_\mathrm B = 1 $.
The Friedmann equation in the radiation era is
\begin{equation}
H^2
= \frac{8\pi\,G}{3}\,\rho
= \frac{4\pi^3\,G}{45}\,g_*\,T^4\,,
\end{equation}
where $ g_* $ counts the number of relativistic degrees of freedom.
This can be integrated to give
\begin{equation}
t
\approx
  0.738\,
  \left(\frac{g_*}{10.75}\right)^{-1/2}\,
  \left(\frac{T}{1\,\mathrm{MeV}}\right)^{-2}\,\mathrm s,
\label{time}
\end{equation}
where $ g_* $ and $ T $ are normalized to their values at the start of the BBN epoch.
Since we are only considering PBHs which form during the radiation era (the ones generated before inflation being diluted to negligible density), the initial PBH mass $ M $ is related to the ``standard'' particle horizon mass $ M_\mathrm{PH} $ (which is not the actual particle horizon mass in the inflationary case) by
\begin{equation}
M
= \gamma\,M_\mathrm{PH}
= \frac{4\pi}{3}\,\gamma\,\rho\,H^{-3}
\approx
  2.03 \times 10^5\,
  \gamma\,
  \left(\frac{t}{1\,\mathrm s}\right)\,M_\odot\,.
\label{mass}
\end{equation}
Here $ \gamma $ is a numerical factor which depends on the details of gravitational collapse.
A simple analytical calculation suggests that it is around $ (1/\sqrt 3)^3 \approx 0.2 $ during the radiation era \cite{Carr:1975qj}, although the first hydrodynamical calculations gave a somewhat smaller value \cite{1978AZh....55..216N,*1978SvA....22..129N}.
The favored value has subsequently fluctuated as people have performed more sophisticated computations but now seems to have returned to the original one \cite{Green:2004wb}.
However, as mentioned earlier, the effect of critical phenomena could in principle reduce the value of $ \gamma $\,, possibly down to $ 10^{-4} $ \cite{Hawke:2002rf}, as could a reduction in the pressure \cite{Khlopov:1980mg,1981AZh....58..706P,*1981SvA....25..406P,1982AZh....59..639P,*1982SvA....26..391P}.
On the other hand, if the overdensity from which the PBH forms is ``noncompensated'' (i.e.\ not surrounded by a corresponding underdensity), subsequent accretion could generate an eventual PBH mass well above the formation mass, leading to an ``effective'' value of $ \gamma $ much larger than $ 1 $ \cite{1979ApJ...232..670B}.
In view of the uncertainties, we will not specify the value of $ \gamma $ in what follows.

Throughout this paper we assume that the PBHs have a monochromatic mass function, in the sense that they all have the same mass $ M $\,.
This simplifies the analysis considerably and is justified providing we only require limits on the PBH abundance at particular values of $ M $\,.
Assuming adiabatic cosmic expansion after PBH formation, the ratio of the PBH number density to the entropy density, $ n_\mathrm{PBH}/s $\,, is conserved.
Using the relation $ \rho = 3\,s\,T/4 $, the fraction of the Universe's mass in PBHs at their formation time is then related to their number density $ n_\mathrm{PBH}(t) $ during the radiation era by
\begin{equation}
\begin{aligned}
\beta(M)
& \equiv
    \frac{\rho_\mathrm{PBH}(t_\mathrm i)}{\rho(t_\mathrm i)}
  = \frac{M\,n_\mathrm{PBH}(t_\mathrm i)}{\rho(t_\mathrm i)}
  = \frac{4\,M}{3\,T_\mathrm i}\,
    \frac{n_\mathrm{PBH}(t)}{s(t)} \\
& \approx
    7.98 \times 10^{-29}\,
    \gamma^{-1/2}\,
    \left(\frac{g_{*\mathrm i}}{106.75}\right)^{1/4}\,
    \left(\frac{M}{M_\odot}\right)^{3/2}\,
    \left(\frac{n_\mathrm{PBH}(t_0)}{1\,\mathrm{Gpc}^{-3}}\right)\,,
\label{eq:beta}
\end{aligned}
\end{equation}
where the subscript ``$ \mathrm i $'' indicates values at the epoch of PBH formation and we have assumed $ s = 8.55 \times 10^{85}\,\mathrm{Gpc}^{-3} $ today.
$ g_{*\mathrm i} $ is now normalized to the value of $ g_* $ at around $ 10^{-5}\,\mathrm s $ since it does not increase much before that in the standard model and that is the period in which most PBHs are likely to form.
The current density parameter for PBHs which have not yet evaporated is given by
\begin{equation}
\Omega_\mathrm{PBH}
= \frac{M\,n_\mathrm{PBH}(t_0)}{\rho_\mathrm c}
\approx
  \left(\frac{\beta(M)}{1.15 \times 10^{-8}}\right)\,
  \left(\frac{h}{0.72}\right)^{-2}\,
  \gamma^{1/2}\,
  \left(\frac{g_{*\mathrm i}}{106.75}\right)^{-1/4}\,
  \left(\frac{M}{M_\odot}\right)^{-1/2}\,,
\label{eq:omega}
\end{equation}
which is a more precise form of Eq.~\eqref{eq:roughomega}.
The dependence on the Hubble parameter is included here but we will generally assume $ h = 0.72 $.
An immediate constraint on $ \beta(M) $ comes from the limit on the CDM density parameter, $ \Omega_\mathrm{CDM}\,h^2 = 0.110 \pm 0.006 $ with $ h = 0.72 $, so the $ 3 \sigma $ upper limit is $ \Omega_\mathrm{PBH} < \Omega_\mathrm{CDM} < 0.25 $ \cite{Hinshaw:2008kr,Dunkley:2008ie}.
This implies
\begin{equation}
\beta(M)
< 2.04 \times 10^{-18}\,
  \gamma^{-1/2}\,
  \left(\frac{\Omega_\mathrm{CDM}}{0.25}\right)\,
  \left(\frac{h}{0.72}\right)^2\,
  \left(\frac{g_{*\mathrm i}}{106.75}\right)^{1/4}\,
  \left(\frac{M}{10^{15}\,\mathrm g}\right)^{1/2}
\quad
(M \gtrsim 10^{15}\,\mathrm g)\,.
\label{eq:density}
\end{equation}
This constraint, which applies only for PBHs which have not evaporated yet, agrees with the one given by Blais \textit{et al.} \cite{Blais:2002gw} when the parameters are appropriately renormalized.
Note that the dependences on $ \gamma $ and $ g_* $ in Eq.~\eqref{eq:beta} and subsequent equations arise through the relationship between $ M $ and $ T_\mathrm i $ implied by Eqs.~\eqref{time} and \eqref{mass}.
Since $ \beta $ always appears in combination with $ \gamma^{1/2}\,g_*^{-1/4} $\,, it is convenient to define a new parameter
\begin{equation}
\beta'(M)
\equiv
  \gamma^{1/2}\,\left(\frac{g_{*\mathrm i}}{106.75}\right)^{-1/4}\,\beta(M)\,,
\label{eq:betaprime}
\end{equation}
where $ g_{*\mathrm i} $ can be specified very precisely but $ \gamma $ is rather uncertain.
Most of the constraints discussed in this paper will therefore be expressed in terms of $ \beta' $ rather than $ \beta $\,.

One can also represent the PBH constraints in terms of the ratio of the PBH density to radiation density at the evaporation epoch.
This fraction is larger than the formation fraction by a redshift factor and given by
\begin{equation}
\alpha(M)
= \beta(M)\,
  \left(\frac{1 + z_{\mathrm{form}}}{1 + z_{\mathrm{evap}}}\right)
\approx
  \beta(M)\,
  \mathrm{max}
  \biggl[
   \left(\frac{M}{M_\mathrm{Pl}}\right),
   \left(\frac{M^{3/2}}{M_\mathrm{Pl}\,M_\mathrm{eq}^{1/2}}\right)
  \biggr]\,, 
\end{equation}
where $ M_\mathrm{eq} $ is the mass of a PBH evaporating at the time of matter-radiation equality.
The first and second expressions apply for PBHs which evaporate before and after then, respectively.
A more precise expression has been derived by Green and Liddle \cite{Green:1997sz}.
They also distinguish between the fraction of the \emph{radiation} density ($ \beta $) and the fraction of the \emph{total} density $ \beta/(1+\beta) $ going into the PBHs, but the difference is negligible so long as $ \beta $ is small.
In any case, we just refer to $ \beta(M) $ in this paper.

Note that the relationship between $ \beta $ and $ \Omega_\mathrm{PBH} $ must be modified if the Universe ever deviates from the standard radiation-dominated behavior.
For example, if there is a dustlike stage for some extended early period $ t_1 < t < t_2 $\,, then one must include an extra factor $ (t_2/t_1)^{1/6} $ on the right-hand-side of Eq.~\eqref{eq:density} \cite{1982AZh....59..639P,*1982SvA....26..391P}.
This is also the period in which $ \gamma $ is likely to be small.
The expression for $ \beta(M) $ may also be modified in some mass ranges if there is a second inflationary phase \cite{Green:1997sz} or if there is a period when the gravitational constant varies \cite{Barrow:1996jk} or there are extra dimensions \cite{Sendouda:2006nu}.

In principle, the analysis can be extended to the (more realistic) case in which the PBHs have an extended mass spectrum by interpreting $ \beta(M) $ as the fractional density of PBHs at formation over a logarithmic mass interval.
More precisely, we can write
\begin{equation}
n_\mathrm{PBH}(M,t)
\equiv
  \frac{\mathrm dN_\mathrm{PBH}(M,t)}{\mathrm d\ln M}\,,
\label{eq:ndef}
\end{equation}
where $ N_\mathrm{PBH}(M,t) $ is the cumulative number density of PBHs integrated over the mass range from $ 0 $ to $ M $\,.
For a monochromatic mass function, in which all the PBHs have mass $m$, we then have $ N_\mathrm{PBH}(M,t) = n_\mathrm{PBH}(t)\,\theta(M-m) $\,, where $ \theta$ is the Heaviside function.
However, this more general analysis is only necessary if we wish to infer constraints which depend on the PBH mass spectrum itself.
Although a precisely monochromatic mass spectrum is clearly not physically realistic, one would only expect the mass function to be \emph{very} extended if the PBHs formed from \emph{exactly} scale-invariant density fluctuations \cite{Carr:1975qj};
it is not clear whether this ever applies and it is certainly not expected in an inflationary scenario.
However, it is important to stress that there are some circumstances in which the spectrum would be quite extended and this has the important implication that the constraint on one mass scale may also imply a constraint on neighboring scales.
For example, we have mentioned that the monochromatic assumption may fail badly if PBHs form through critical collapse and the way in which this modifies the form of $ \beta(M) $ has been discussed by Yokoyama \cite{Yokoyama:1998xd}.
Another important point is that if the PBHs with $ M \approx M_* $ have a spread of masses $ \Delta M \sim M_* $\,, one would expect the evaporation process itself to lead to a residual spectrum with $ n_\mathrm{PBH}(M,t_0) \propto M^3 $ for $ M \ll M_* $ \cite{Page:1976wx}; the implications of this are discussed in the Appendix.

\section{
Evaporation of primordial black holes
\label{sec:evap}
}

\subsection{
Lifetime
}

Here we briefly summarize some basic results of PBH evaporation.
As first shown by Hawking \cite{Hawking:1974rv,Hawking:1974sw,*Hawking:1974swE}, a black hole with mass $ M \equiv M_{10} \times 10^{10}\,\mathrm g $ emits thermal radiation with temperature
\begin{equation}
T_\mathrm{BH}
= \frac{1}{8\pi\,G\,M}
\approx
  1.06\,M_{10}^{-1}\,\mathrm{TeV}.
\end{equation}
This assumes that the hole has no charge or angular momentum, which is reasonable since charge and angular momentum will also be lost through quantum emission on a shorter time scale than the mass \cite{Page:1976df,Page:1976ki,Page:1977um}.
More precisely, such a black hole emits particles with energy between $ E $ and $ E + \mathrm dE $ at a rate
\begin{equation}
\mathrm d\dot N_s
= \frac{\mathrm dE}{2\pi}\,
  \frac{\Gamma_s}{e^{E/T_\mathrm{BH}}-(-1)^{2\,s}}\,,
\label{eq:bb}
\end{equation}
where $ s $ is the spin of the particle and $ \Gamma_s $ is its dimensionless absorption coefficient, whose functional shape is found in Refs.~\cite{Page:1976df,Page:1976ki,Page:1977um}.
It is related to the absorption cross section $ \sigma_s(M,E) $ by
\begin{equation}
\Gamma_s(M,E)
= \frac{E^2\,\sigma_s(M,E)}{\pi}\,.
\label{eq:grey}
\end{equation}
In the high-energy limit, $ E \gg T_\mathrm{BH} $\,, $ \sigma_s $ approaches the geometric optics limit $ \sigma_\mathrm g = 27\pi\,G^2\,M^2 $\,.
However, it is a more complicated function of  $ E $ and $ M $ at low energies and depends on the spin.
The average energies of the emitted neutrino, electron, and photon are $ E_\nu = 4.22\,T_\mathrm{BH} $\,, $ E_e = 4.18\,T_\mathrm{BH} $\,, and $ E_\gamma = 5.71\,T_\mathrm{BH} $\,, respectively.
The peak energies of the flux and power are within $ 7\,\% $ of these values \cite{MacGibbon:1990zk}.

The mass-loss rate of an evaporating black hole can be expressed as
\begin{equation}
\frac{\mathrm dM_{10}}{\mathrm dt}
= -5.34 \times 10^{-5}\,f(M)\,M_{10}^{-2}\,\mathrm s^{-1}.
\end{equation}
Here $ f(M) $ is a measure of the number of emitted particle species, normalized to unity for a black hole with $ M \gg 10^{17}\,\mathrm g $, this emitting only particles which are (effectively) massless: photons, three generations of neutrinos and antineutrinos, and gravitons.
The contribution of each relativistic degree of freedom to $ f(M) $ depends on the spin $ s $ \cite{MacGibbon:1991tj}:
\begin{equation}
\begin{aligned}
&
f_{s=0} = 0.267,
\quad
f_{s=1} = 0.060,
\quad
f_{s=3/2} = 0.020,
\quad
f_{s=2} = 0.007, \\
&
f_{s=1/2} = 0.147~\text{(neutral)}\,,
\quad
f_{s=1/2} = 0.142~\text{(charge $ \pm e $)}\,.
\end{aligned}
\label{eq:spin}
\end{equation}
Holes in the mass range $ 10^{15}\,\mathrm g < M < 10^{17}\,\mathrm g $ emit electrons but not muons, while those in the range $ 10^{14}\,\mathrm g < M < 10^{15}\,\mathrm g $ also emit muons, which subsequently decay into electrons and neutrinos.
The latter range is relevant for the PBHs which are completing their evaporation at the present epoch.

Once $ M $ falls to roughly $ 10^{14}\,\mathrm g $, a black hole can also begin to emit hadrons.
However, hadrons are composite particles made up of quarks held together by gluons.
For temperatures exceeding the QCD confinement scale, $ \Lambda_\mathrm{QCD} = 250\text{--}300\,\mathrm{MeV} $, one would expect these fundamental particles to be emitted rather than composite particles.
Only pions would be light enough to be emitted below $ \Lambda_\mathrm{QCD} $\,.
Above this temperature, it has been argued \cite{MacGibbon:1990zk} that the particles radiated can be regarded as asymptotically free, leading to the emission of quarks and gluons.
Since there are $ 12 $ quark degrees of freedom per flavor and $ 16 $ gluon degrees of freedom, one would expect the emission rate (i.e., the value of $ f $) to increase suddenly once the QCD temperature is reached.
If one includes just $ u $\,, $ d $\,, and $ s $ quarks and gluons, Eq.~\eqref{eq:spin} implies that their contribution to $ f $ is $ 3 \times 12 \times 0.14 + 16 \times 0.06 \approx 6 $, compared to the pre-QCD value of about $ 2 $.
Thus the value of $ f $ roughly quadruples, although 
there will be a further increase in $ f $ at somewhat higher temperatures due to the emission of the heavier quarks.
After their emission, quarks and gluons fragment into further quarks and gluons until they cluster into the observable hadrons when they have travelled a distance $ \Lambda_\mathrm{QCD}^{-1} \sim 10^{-13}\,\mathrm{cm} $.
This is much larger than the size of the hole, so gravitational effects can be neglected.
Note that the peak energy of the PBH emission reaches $ \Lambda_\mathrm{QCD} $ once $ M $ falls to a critical mass $ M_\mathrm q \approx 2 \times 10^{14}\,\mathrm g $, although the large increase in $ f(M) $ only occurs at a somewhat smaller mass.

If we sum up the contributions from all the particles in the standard model up to $ 1\,\mathrm{TeV} $, corresponding to $ M_{10} \sim 1 $, this gives $ f(M) = 15.35 $.
Integrating the mass-loss rate over the time \cite{MacGibbon:1991tj}, we find a lifetime
\begin{equation}
\tau
\approx
  407\,
  \left(\frac{f(M)}{15.35}\right)^{-1}\,M_{10}^3\,\mathrm s.
\label{eq:tau}
\end{equation}
This can be inverted to give the mass of a PBH evaporating at time $ \tau $ after the big bang:
\begin{equation}
M
\approx
  1.35 \times 10^9\,
  \left(\frac{f(M)}{15.35}\right)^{1/3}\,
  \left(\frac{\tau}{1 \,\mathrm s}\right)^{1/3}\,\mathrm g.
\label{eq:lifetime}
\end{equation}
A PBH evaporating before the end of BBN at $ t \sim 10^3\,\mathrm s $ therefore has $ M \lesssim 10^{10}\,\mathrm g $ and $ T_\mathrm{BH} \gtrsim 1\,\mathrm{TeV} $.
The critical mass for which $ \tau $ equals the age of the Universe is denoted by $ M_* $\,.
If one assumes the currently favored age of $ 13.7\,\mathrm{Gyr} $, one finds
\begin{equation}
M_*
\approx
  1.02 \times 10^{15}\,
  \left(\frac{f_*}{15.35}\right)^{1/3}\,\mathrm g
\approx
  5.1 \times 10^{14}\,\mathrm g,
\end{equation}
where $ f_* $ is the value of $ f $ associated with the temperature $ T_\mathrm{BH}(M_*) $\,.
The indicated value of $ M_* $ corresponds to $ f_* = 1.9 $ and $ T_\mathrm{BH} = 21\,\mathrm{MeV} $.
At this temperature muons and some pions are emitted, so the value of $ f_* $ accounts for this.
Although QCD effects are initially small for PBHs with $ M = M_* $\,, only contributing a few percent, it should be noted that they will become important once $ M $ falls to $ M_\mathrm q \approx 0.4\,M_* $ since the peak energy becomes comparable to $ \Lambda_\mathrm{QCD} $ then.
As discussed below, this means that an appreciable fraction of the time-integrated emission from the PBHs evaporating at the present epoch goes into quark and gluon jet products.

It should be stressed that the above analysis is not exactly correct because the value of $ f(M) $ in Eq.~\eqref{eq:lifetime} should really be the weighted average of $ f(M) $ over the lifetime of the black hole.
This accords with the more precise calculation of MacGibbon \cite{MacGibbon:1991tj}.
In fact, the update of the calculation in Ref.~\cite{MacGibbon:2007yq}, using the modern value $ \tau = 13.7\,\mathrm{Gyr} $, gives $ M_* = 5.00 \times 10^{14}\,\mathrm g $, which is slightly smaller than our value.
However, we will continue to use Eq.~\eqref{eq:lifetime} because we need the rough scaling with $ f(M) $ later in the paper.
The weighted average of $ f(M) $ is well approximated by $ f(M) $ unless one happens to be close to a particle mass threshold, so the error involved in our approximation is usually small.
For example, since the lifetime of a black hole of mass $ 0.4\,M_* $ is roughly $ 0.25 \times (0.4)^3 = 0.016 $ that of an $ M_* $ black hole, one expects the value of $ M_* $ to be overestimated by a few percent.
This explains the small difference between our own and MacGibbon's more precise calculation.

\subsection{
Particle spectra
\label{sec:spectra}
}

Particles injected from PBHs have two components: the \emph{primary} component, which is the direct Hawking emission, and the \emph{secondary} component, which comes from the decay of the hadrons produced by fragmentation of primary quarks and gluons, and by the decay of gauge bosons.
For example, the total photon spectrum emitted by a low-mass PBH can be written as
\begin{equation}
\frac{\mathrm d\dot N_\gamma}{\mathrm dE_\gamma}(E_\gamma,M)
= \frac{\mathrm d\dot N^\mathrm{pri}_\gamma}{\mathrm dE_\gamma}(E_\gamma,M)
  + \frac{\mathrm d\dot N^\mathrm{sec}_\gamma}{\mathrm dE_\gamma}(E_\gamma,M)\,,
\label{eq:prisec}
\end{equation}
with similar expressions for other particles.
In order to treat QCD fragmentation, we use the \textsc{PYTHIA} code, a Monte Carlo event generator \cite{Sjostrand:2006za}.
This code has been constructed to fit hadron fragmentation in the high-energy regime, in particular reproducing the latest experimental results for center-of-mass energies below $ 200\,\mathrm{GeV} $.
We use this package for even larger center-of-mass energies by assuming that standard model physics can still be applied.
It also gives the energy spectrum of created particles such as nucleons and pions, and photons from unstable mesons/hadrons and bremsstrahlung during the hadron fragmentation are now appropriately incorporated.
However, the bremsstrahlung from secondary charged leptons and long-lived charged hadrons is not considered since it is subdominant.

The spectrum of secondary photons generated by PBHs was first analyzed by MacGibbon and Webber \cite{MacGibbon:1990zk}.
It is peaked around $ E_\gamma \simeq m_{\pi^0}/2 \approx 68\,\mathrm{MeV} $, independent of the Hawking temperature, because it is dominated by the $ 2\gamma $-decay of soft neutral pions which are practically at rest.
As $ E_\gamma $ increases, it then falls as $ E_\gamma^{-1} $ before cutting off as $ E_\gamma^2 \exp(-E_\gamma/T_\mathrm{BH}) $ above $ T_\mathrm{BH} $\,.
The peak flux can be expressed as
\begin{equation}
\frac{\mathrm d\dot N_\gamma^\mathrm{sec}}{\mathrm dE_\gamma}
(E_\gamma=m_{\pi^0}/2)
\simeq
  2\,\frac{\mathrm d\dot N_{\pi^0}}{\mathrm dE_{\pi^0}}(E_{\pi^0}=m_{\pi^0})
\simeq
  2\,\sum_{i=q,g} \mathcal B_{i \to \pi^0}(\bar E,m_{\pi^0})\,
  \frac{\bar E}{m_{\pi^0}}\,
  \frac{\mathrm d\dot N_i^\mathrm{pri}}{\mathrm dE_i}
  (E_i=\bar E)\,,
\label{eq:peakfluxaprx}
\end{equation}
where $ \mathcal B_{q,g \to\pi^0}(E_\mathrm{jet},E_{\pi^0}) $ is the fraction of the jet energy $ E_\mathrm{jet} $ going into neutral pions of energy $ E_{\pi^0} $\,.
This is of order $ 0.1 $ and fairly independent of jet energy.
The second approximation in Eq.~\eqref{eq:peakfluxaprx} uses the fact that most of the primary particles have of order of the average energy $ \bar E \approx 4.4\,T_\mathrm{BH} $\,.
From Eq.~\eqref{eq:bb}, we then find that
\begin{equation}
\frac{\mathrm d\dot N_i^\mathrm{pri}}{\mathrm dE_i}(E_i=4.4\,T_\mathrm{BH})
\simeq
  \frac{1}{2\pi}\,
  \frac{27 \times 4.4^2 \times (8\pi)^{-2}}{e^{4.4}-(-1)^{2\,s}}
\approx
  1.6 \times 10^{-3}
\end{equation}
(with units of $ \hbar^{-1} $) for any value of $ s $ in the geometric optics limit.
Thus the energy dependence of Eq.~\eqref{eq:peakfluxaprx} comes entirely from the factor $ \bar E \approx 4.4\,T_\mathrm{BH} $ and is proportional to the Hawking temperature.
A similar analysis applies for the secondary production of protons and antiprotons, this also scaling as the black hole temperature.
\begin{figure}[ht]
\begin{center}
\includegraphics{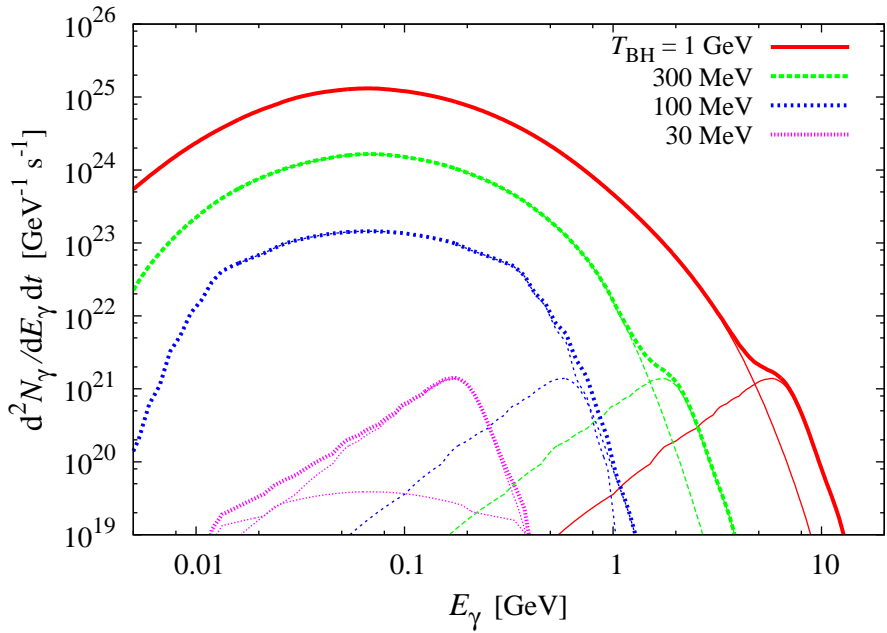}
\end{center}
\caption{
Instantaneous emission rate of photons for four typical black hole temperatures.
For each temperature, the curve with the peak to the right (left) represents the primary (secondary) component and the thick curve denotes their sum.
\label{fig:rate}
}
\end{figure}

The emission rates of primary and secondary photons for four typical temperatures are shown in Fig.~\ref{fig:rate}.
These results might be compared with those obtained by \textsc{HERWIG} \cite{Corcella:2000bw}, the code used by MacGibbon and Webber \cite{MacGibbon:1990zk}.
Although \textsc{HERWIG} and \textsc{PYTHIA} have been continually updated to match almost $ 20 $ years of high-energy accelerator runs, it should be stressed that there have been few new $ e^+ e^- $ accelerator experiments near $ \Lambda_\mathrm{QCD} $\,.
Most new $ e^+ e^- $ experiments have been at the highest accelerator energies, while recent experiments examining behavior near $ \Lambda_\mathrm{QCD} $ have involved heavy ion (i.e.\ quark-gluon plasma) rather than $ e^+ e^- $ collisions, the latter being more closely analogous to the black hole situation.
Hence there has been little recent advance in our understanding of processes near $ \Lambda_\mathrm{QCD} $ and any accelerator code differences probably come from different choices for $ u $ and $ d $ quark masses and the value of $ \Lambda_\mathrm{QCD} $\,.
These parameters can be changed arbitrarily in those codes anyway.

If one integrates the instantaneous spectrum $ \mathrm d\dot N_\gamma/\mathrm dE_\gamma $ over time for $ M \le M_* $\,, one obtains the more observationally significant quantity $ \mathrm dN_\gamma/\mathrm dE_\gamma $\,.
At most energies this is dominated by the emission associated with the \emph{initial} PBH mass $ M $\,, since the mass only changes in the relatively short period towards the end of evaporation.
However, PBHs with $ M \le M_* $ also have a high-energy tail for $ E_\gamma \gg T_\mathrm{BH}(M) $ due to their final phase of evaporation.
(For observations at the present epoch, this energy must be reduced by a factor $ 1 + z_\mathrm{evap}(M) $ where $ z_\mathrm{evap}(M) $ is the evaporation redshift.)
The precise form of this high-energy tail has been investigated by MacGibbon \cite{MacGibbon:1991tj} and it is discussed qualitatively in the Appendix, so we just summarize the conclusion here.
For $ M > M_\mathrm q $\,, one obtains
\begin{equation}
\frac{\mathrm dN_\gamma}{\mathrm dE_\gamma}
\propto  
\begin{cases}
E_\gamma^3 & (E_\gamma < M^{-1})\,, \\
E_\gamma^{-3} & (M^{-1} < E_\gamma < \Lambda_\mathrm{QCD})\,,
\end{cases}
\label{eq:energytail1}
\end{equation}
where the lower energy part comes from the initial Rayleigh--Jeans contribution.
The form of the tail changes for $ E_\gamma > \Lambda_\mathrm{QCD} $ due to secondary emission, involving an intermediate $ E_\gamma^{-1} $ regime, but it still falls off as $ E_\gamma^{-3} $ at very high energies.
For $ M < M_\mathrm q $\,, the spectrum is \emph{always} dominated by the secondary photons.
As $ M $ increases towards $ M_* $\,, the primary photons generate a \emph{low-energy} redshifted tail which scales as $ E_\gamma^{1/2} $ \cite{MacGibbon:1991tj}.

It should be noted that ratio of the time-integrated secondary flux to primary flux drops rapidly once $ M $ goes above $ M_* $\,.
This can be understood as follows.
We have already seen that a black hole with $ M = M_* $ will emit quarks efficiently once its mass gets down to $ M_\mathrm q \approx 0.4\,M_* $ and this corresponds to an appreciable fraction of its original mass.
On the other hand, a simple calculation shows that a PBH with somewhat larger initial mass, $ M = (1+\mu)\,M_* $\,, and hence a slightly longer lifetime, $ \tau \approx (1+\mu)^3\,t_0 $\,, will today have a mass
\begin{equation}
m
\equiv
  M(t_0)
\approx
  (3\,\mu)^{1/3}\,(1 + \mu + \mu^2/3)^{1/3}\,M_*\,,
\label{eq:mu}
\end{equation}
providing $ \mu $ is not too small (as explained below).
Here we have assumed $ f(M) \approx f_* $\,, which should be a good approximation for $ m > M_\mathrm q $ since the value of $ f $ only changes slowly above the QCD threshold.
For $ \mu \gg 1 $, Eq.~\eqref{eq:mu} just gives $ m \approx \mu\,M_* \approx M $\,, as expected.
For $ \mu \ll 1 $, the second term can be neglected and $ m $ exceeds $ M_\mathrm q $ providing $ (3\,\mu)^{1/3} > 0.4 $, which corresponds to $ \mu > 0.02 $.
Equation~\eqref{eq:mu} must be modified for $ \mu < 0.02 $ since the value of $ f $ increases by a factor of around $ 4 $ as $ M $ falls below $ M_\mathrm q $\,.
If we assume that it jumps discontinuously from $ f_* $ to $ \alpha\,f_* $ at $ M_\mathrm q $\,, then the mass falls to $ M_\mathrm q $ at a time
\begin{equation}
t_\mathrm q
\approx
  \frac{1 - (0.4)^3 + 3\,\mu}
       {1 - (0.4)^3\,(1 - \alpha^{-1})}\,
  t_0
\end{equation}
and this leads to a current mass
\begin{equation}
m
\approx
  (3\,\alpha\,\mu)^{1/3}\,M_*
\quad
(\mu \lesssim 0.02)\,.
\end{equation}
This differs from the expression given by Eq.~\eqref{eq:mu} because of the $\alpha$ factor but it goes to zero at $ \mu = 0 $, as expected.
The denominator in the expression for $ t_\mathrm q $ arises because the lifetime of a black hole of mass close to $ M_* $ is slightly increased by the change of $ f $\,.

\begin{figure}[ht]
\begin{center}
\includegraphics{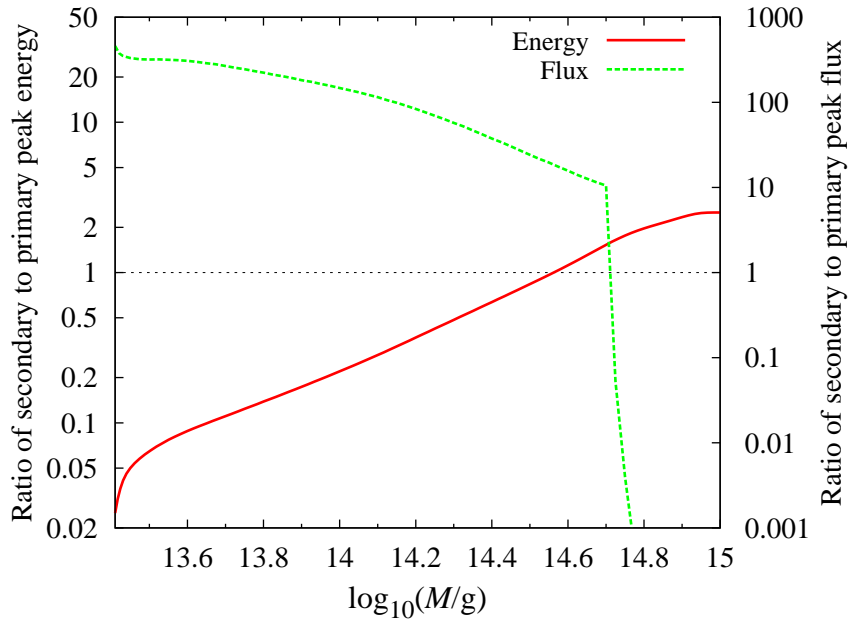}
\end{center}
\caption{
Ratios of the time-integrated secondary to primary peak energies (solid red) and fluxes (dashed green).
\label{fig:ratios}
}
\end{figure}

The fact that $ m $ falls below $ M_\mathrm q $ only when $ \mu $ is less than the tiny value $ 0.02 $ means that the fraction of the black hole mass going into secondaries falls off sharply above $ M_* $\,.
Indeed, this fraction must be around $ 0.4\,(1 - m/M_\mathrm q) $ for $ \mu \lesssim 0.02 $, since the unevaporated mass $ m $ does not contribute, and it must be exponentially reduced by the Wien blackbody factor $ \propto \exp(-\chi\,m/M_\mathrm q) $ for $ \mu \gtrsim 0.02 $, where $ \chi $ is a parameter which depends only weakly on $ \mu $\,.
A transition occurs between the $ \mu < 0.02 $ and $ \mu > 0.02 $ regimes.
The ratio of the secondary to primary peak energies and the ratio of the time-integrated fluxes are shown in Fig.~\ref{fig:ratios}; they are qualitatively well explained by the above arguments.
Note that the forms of the curves at around $ \log_{10}(M/\mathrm g) = 13.4 $ are associated with interactions of photons with the background Universe at high redshifts.
These are discussed later and affect the secondary particles more than the primary ones.

\subsection{
Photosphere effects
}

There has been some dispute in the literature about the interactions between emitted particles around an evaporating black hole.
The usual assumption \cite{1984PhLB..143...92O} that there is no interaction between emitted particles has been disputed by Heckler \cite{Heckler:1995qq}, who claims that QED interactions could produce an optically thick photosphere once the black hole temperature exceeds $ T_\mathrm{BH} = 45\,\mathrm{GeV} $.
He has proposed that a similar effect may operate at an even lower temperature, $ T_\mathrm{BH} \approx 200\,\mathrm{MeV} $, due to QCD effects \cite{Heckler:1997jv}.
Variants of these models and their astrophysical implications have been studied by Cline \textit{et al.} \cite{Cline:1998xk}, Kapusta \cite{Kapusta:2000xt} and Daghigh \textit{et al.} \cite{Daghigh:2001gy,Daghigh:2002fn,Daghigh:2006dt}.
However, MacGibbon \textit{et al.} \cite{MacGibbon:2007yq} have analyzed these claims and identified a number of physical and geometrical effects which invalidate them.
There are two key problems.
First, the particles must be causally connected in order to interact and only a negligible fraction of the emitted charged particles satisfies this constraint.
The standard cross sections are further reduced because the particles are created at a finite time and do not go back to the infinite past.
Second, a scattered particle requires a minimum distance to complete each bremsstrahlung interaction, with the consequence that there is unlikely to be more than one complete bremsstrahlung interaction per particle near the black hole \cite{Page:2007yr}.
This relates to what is termed the Landau--Pomeranchuk--Migdal effect \cite{Landau:1953um,Landau:1953gr,Migdal:1956tc}.

MacGibbon \textit{et al.} conclude that the emitted particles do not interact sufficiently to form a QED photosphere and, using analogous QCD arguments, that the conditions for QCD photosphere formation could only be temporarily satisfied (if at all) when the black hole temperature is of order the QCD confinement scale, $ \Lambda_\mathrm{QCD} = 250\text{--}300\,\mathrm{MeV} $.
Even in this case, the strong damping of the Hawking production of QCD particles and the limited available energy per particle around this threshold should suffice to suppress it.
In any case, no QCD photosphere persists once the black hole temperature climbs above $ \Lambda_\mathrm{QCD} $\,.
They also consider the suggestions of Belyanin \textit{et al.} \cite{1996MNRAS.283..626B} that plasma interactions between emitted particles could form a photosphere but conclude that this too is implausible.
A final possibility is that some form of string photosphere might arise due to ``stretched horizon'' effects, as discussed in Ref.~\cite{Bugaev:2009ad}, but this would only apply at energies too high to be of interest here.
Throughout this paper, we therefore usually adopt the standard assumption that no photosphere ever forms, although we do return to this issue in discussing PBH explosions.

It must be stressed that there may well be interactions between emitted particles in the background Universe even if there is no interaction near the black hole, so there is a sense in which one may have a \emph{cosmological} photosphere.
Although there is no cosmological QCD photosphere in the period after $ 10^{-5}\,\mathrm s $ relevant to BBN, there will be a cosmological QED photosphere before around $ 10\,\mathrm s $ since background electrons and positrons will then be as numerous as photons.
However, the situation is different in the cosmological photosphere context because the emitted particles are not interacting with each other.
Also, in the black hole photosphere context, the particle density is much higher near the hole and falls off with the square of distance until it reaches the background cosmological value.

\section{
Constraints on PBHs from big bang nucleosynthesis
\label{sec:bbn}
}

\subsection{
Historical overview
}
In this section we will be concerned with the effects of PBH evaporations on BBN, 
presenting for the first time the full analysis of the associated constraints on the PBH mass spectrum.
There is also an indirect limit on much larger mass scales associated with PBH \emph{formation} because the density inhomogeneities at around neutron-proton freeze-out would modify BBN if they were large enough to generate PBHs.
This corresponds to mass scales around $ 10^5\,M_\odot $ but the associated limit is not definitive because it assumes a particular formation mechanism for the PBHs.

Since the effects of PBH evaporation on BBN have been a subject of long-standing interest, we start with a brief review of previous work and then explain why it is necessary to reassess the constraints.
All the limits will be expressed in terms of the parameter $ \beta'(M) $ defined by Eq.~\eqref{eq:betaprime}.
Vainer and Nasel'skii \cite{1978AZh....55..231V,*1978SvA....22..138V} studied the effects of the injection of high-energy neutrinos and antineutrinos from PBHs.
This changes the epoch at which the weak interactions freeze out and thereby the neutron-to-proton ratio at the onset of nucleosynthesis.
This results in an increase of $ {}^4\mathrm{He} $ production, so by demanding that the primordial abundance satisfies $ Y_p < 0.33 $, the most conservative constraint available at the time, they concluded that the ratio of the PBH density to the baryon density at BBN should be smaller than $ 10\text{--}10^4 $ for $ M = 10^9\text{--}3 \times 10^{11}\,\mathrm g $.
Equation~\eqref{eq:omega} implies that this corresponds to a limit
\begin{equation}
\beta'(M)
< 3 \times (10^{-18}\text{--}10^{-15})\,M_{10}^{1/2}
\quad
(M = 10^9\text{--}3 \times 10^{11}\,\mathrm g)\,.
\end{equation}

These authors also studied the effects of entropy generation by PBHs but obtained only the modest constraint that their density should be less than that of the background radiation at evaporation.
More detailed numerical analysis of this effect was carried out by Miyama and Sato \cite{Miyama:1978mp}, who calculated the entropy production from the PBHs with $ M = 10^9\text{--}10^{13}\,\mathrm g $ which evaporated during or after BBN.
If these PBHs were too abundant, the baryon-to-entropy ratio at nucleosynthesis would be increased, which would result in overproduction of $ {}^4\mathrm{He} $ and underproduction of $ \mathrm D $.
They demanded that the primordial mass fractions of these elements satisfy $ Y_p < 0.29 $ and $ D_p > 1 \times 10^{-5} $.
One can approximate their constraint with a single power law:
\begin{equation}
\beta'(M)
< 10^{-15}\,M_{10}^{-5/2}
\quad
(M = 10^9\text{--}10^{13}\,\mathrm g)\,.
\end{equation}

Zel'dovich \textit{et al.} \cite{1977PAZh....3..208Z,*1977SvAL....3..110Z} studied the effect of PBH emission of high-energy nucleons and antinucleons.
They claimed that this would increase the deuterium abundance due to capture of free neutrons by protons and spallation of $ {}^4\mathrm{He} $.
They inferred the upper limits:
\begin{equation}
\beta'(M)
<
\begin{cases}
6 \times 10^{-18}\,M_{10}^{-1/2}
& (M = 10^9\text{--}10^{10}\,\mathrm g)\,, \\
6 \times 10^{-22}\,M_{10}^{-1/2}
& (M = 10^{10}\text{--}5 \times 10^{10}\,\mathrm g)\,, \\
3 \times 10^{-23}\,M_{10}^{5/2}
& (M = 5 \times 10^{10}\text{--}5 \times 10^{11}\,\mathrm g)\,, \\
3 \times 10^{-21}\,M_{10}^{-1/2}
& (M = 5 \times 10^{11}\text{--}10^{13}\,\mathrm g)\,.
\end{cases}
\end{equation}
Vainer \textit{et al.} \cite{1978PAZh....4..344V,*1978SvAL....4..185V} then performed numerical studies of neutron-antineutron injection using a nucleosynthesis network.
They found that the spallation of $ {}^4\mathrm{He} $ resulted in extra $ \mathrm D $ production.
They took the neutron lifetime to be $ 918\,(\pm 14)\,\mathrm s $ and used the observational results $ Y_p = 0.29 \pm 0.04 $ and $ D_p = 5 \times 10^{-5} $.
On the assumption that PBHs were produced from scale-invariant density fluctuations, rather than having a monochromatic spectrum, they obtained the constraint $
\beta' < 10^{-26} $ for the relevant mass range.
Later Lindley \cite{1980MNRAS.193..593L} considered the photodissociation of deuterons produced in nucleosynthesis by photons from evaporating PBHs with $ M > 10^{10}\,\mathrm g $.
He found the constraint
\begin{equation}
\beta'(M)
< 3 \times 10^{-20}\,M_{10}^{1/2}
\quad
(M > 10^{10}\,\mathrm g)
\end{equation}
and concluded that photodestruction was comparable to the extra production of deuterons discussed by Zel'dovich \textit{et al.} \cite{1977PAZh....3..208Z,*1977SvAL....3..110Z}.
All the above limits are shown in Fig.~1 of Ref.~\cite{Carr:1994ar}.

Observational data on both the light-element abundances and the neutron lifetime have changed considerably since these classic papers were written.
Much more significant, however, are developments in our understanding of the fragmentation of quark-antiquark pairs or gluons emitted from a PBH into hadrons within the QCD distance.
Most of the hadrons created decay almost instantaneously compared to the time scale of nucleosynthesis, but long-lived ones (such as pions, kaons, and nucleons) remain long enough in the ambient medium to leave an observable signature on BBN.
These effects were first discussed by Kohri and Yokoyama \cite{Kohri:1999ex} but only in the context of the relatively low-mass PBHs evaporating in the early stages of BBN.
In the rest of this section, we extend and update their calculations.

\subsection{
Particle injection
}

As mentioned above, in the Hawking temperature range of interest here, the relevant high-energy hadrons are not primary particles but decay products from hadron fragmentation.
At first these hadrons scatter off background photons and electrons because these particles are much more abundant than background nucleons.
We incorporate thermalization through the electromagnetic interactions described in Appendix~B of Ref.~\cite{Kawasaki:2004qu}.
Specifically, for hadron species $ H_i $\,, we take into account Coulomb scattering ($ H_i + e^\pm \to H_i + e^\pm $), Compton scattering ($ H_i + \gamma \to H_i + \gamma $), the Bethe--Heitler process ($ H_i + \gamma \to H_i + e^+ + e^- $) and the photo-pion process ($ H_i + \gamma \to H'_i + \pi $).
Coulomb scattering dominates in the temperature regime of interest and thermalizes relativistic charged particles on a timescale
\begin{equation}
\tau_\mathrm{ch}
\simeq
  3 \times 10^{-11}\,
  \left(\frac{E}{1\,\mathrm{GeV}}\right)\,
  \left(\frac{T}{0.02\,\mathrm{MeV}}\right)^{-2}\,
  \mathrm s,
\end{equation}
where $ E $ is the kinetic energy of the charged meson \cite{Reno:1987qw}.
This equals the hadron interaction timescale at a cosmological temperature $ T \approx 0.04\,m_e \approx 0.02\,\mathrm{MeV} $, both times being much less than the cosmic expansion time then, so thermalization occurs above this temperature.
For neutrons, Coulomb scattering occurs through their magnetic moment and the cross section is suppressed by a factor of $ T^2/m_n^2 $\,, so they are thermalized only for $ T \gtrsim 0.09\,\mathrm{MeV} $.

In the early stages of BBN, therefore, we expect emitted particles to be quickly thermalized and to have the usual kinetic equilibrium distributions before they interact with ambient nucleons.
In addition, as we show later, it is reasonable to treat the emitted hadrons as being homogeneously distributed.
In order to estimate the minimal time scale for strong interactions, we therefore use the thermally averaged cross sections $ \langle\sigma\,v\rangle^{H_i}_{N\to N'} $ for the strong interactions between hadrons and ambient nucleons $ N $ (i.e.\ protons or neutrons).
For the hadron process $ N + H_i \to N' + \cdots $\,, the strong interaction rate is
\begin{equation}
\Gamma^{H_i}_{N \to N'}
= n_N\,\langle\sigma\,v\rangle^{H_i}_{N \to N'}
\simeq
  10^8\,
  f_N\,
  \left(\frac{\eta_\mathrm i}{10^{-9}}\right)\,
  \left(\frac{\langle\sigma\,v\rangle^{H_i}_{N \to N'}}{10\,\mathrm{mb}}\right)\,
  \left(\frac{T_\nu}{2\,\mathrm{MeV}}\right)^3 \,
  \mathrm s^{-1},
\end{equation}
where $ \eta_\mathrm i $ is the initial baryon-to-photon ratio ($ n_\mathrm b/n_\gamma $ at $ T \gtrsim 10\,\mathrm{MeV} $, where $ n_\mathrm b = n_p + n_n $), $ n_N $ is the number density of the initial nucleon species, corresponding to a fraction $ f_N \equiv n_N/n_\mathrm b $\,, and $ T_\nu $ is the neutrino temperature.
Thus we only need to consider particles with lifetime longer than $ \mathcal O(10^{-8})\,\mathrm s $.
The corresponding mesons are $ \pi^+ $\,, $ \pi^- $\,, $ K^+ $\,, $ K^- $\,, and $ K_L $\,, while the baryons are $ p $\,, $ \bar p $\,, $ n $\,, and $ \bar n $\,.
In this work, we do not include effects induced by kaons because they are smaller than those induced by pions.

\subsection{
Particle reactions
}

High-energy particles emitted by PBHs would modify the standard BBN scenario in three different ways:
(1) high-energy mesons and antinucleons induce extra interconversion between background protons and neutrons even after the weak interaction has frozen out in the background Universe;
(2) high-energy hadrons dissociate light elements synthesized in BBN, thereby reducing $ {}^4\mathrm{He} $ and increasing $ \mathrm D $, $ \mathrm T $, $ {}^3\mathrm{He} $, $ {}^6\mathrm{Li} $, and $ {}^7\mathrm{Li} $;
(3) high-energy photons generated in the cascade further dissociate $ {}^4\mathrm{He} $ to increase the abundance of lighter elements even more.
We analyze (1) following Ref.~\cite{Kohri:1999ex} and (2) and (3) following Ref.~\cite{Kawasaki:2004qu}, which studied the effects of the hadronic decay of long-lived massive particles on BBN in the context of decaying dark matter scenarios.
Since the presentation in those references are very detailed, we do not repeat them in this paper but only focus on the points specific to PBHs.

The Boltzmann equations corresponding to processes (1)--(3) are schematically written as
\begin{equation}
\begin{aligned}
\frac{\mathrm dn_N}{\mathrm dt} + 3\,H\,n_N
& = \left[\frac{\mathrm dn_N}{\mathrm dt}\right]_\mathrm{SBBN}
    + \left[\frac{\mathrm dn_N}{\mathrm dt}\right]_\mathrm{conv}
    + \left[\frac{\mathrm dn_N}{\mathrm dt}\right]_\mathrm{hadron}
    + \left[\frac{\mathrm dn_N}{\mathrm dt}\right]_\mathrm{\gamma}
\quad
(N = p,n)\,, \\
\frac{\mathrm dn_{A_i}}{\mathrm dt} + 3\,H\,n_{A_i}
& = \left[\frac{\mathrm dn_{A_i}}{\mathrm dt}\right]_\mathrm{SBBN}
    + \left[\frac{\mathrm dn_{A_i}}{\mathrm dt}\right]_\mathrm{hadron}
    + \left[\frac{\mathrm dn_{A_i}}{\mathrm dt}\right]_\gamma
\quad
(A_i = \mathrm D,\mathrm T,{}^3\mathrm{He},{}^4\mathrm{He},{}^6\mathrm{Li},{}^7\mathrm{Li})\,,
\end{aligned}
\end{equation}
where, as argued above, the relevant species emitted by PBH evaporation are $ \pi^+ $\,, $ \pi^- $\,, $ p $\,, $ \bar p $\,, $ n $\,, and $ \bar n $\,, as well as photons and neutrinos.
In these equations the subscript ``SBBN'' denotes the standard big bang nucleosynthesis contribution, which is followed using the \textsc{KAWANO} code 4.1 \cite{1992STIN...9225163K} with updated nuclear cross sections;
the subscript ``$ \mathrm{conv} $'' means interconversion between background neutrons and protons due to strong interactions;
``hadron'' represents both production and destruction through hadronic showers;
and ``$ \gamma $'' denotes production or destruction through photodissociation.

\subsubsection{
Interconversion of protons and neutrons due to mesons
}

Long-lived charged mesons are thermalized and scatter off the ambient nucleons with a thermally averaged cross section before they decay.
The cross section of the exothermic process is inversely proportional to the velocity $ v $ when it is small and so we can use the threshold value.
For an emitted (anti)nucleon, the efficiency of thermalization is greater.
Those (anti)nucleons are efficiently stopped by electron scattering until $ T \simeq 0.09\,\mathrm{MeV} $ for neutrons and $ T \simeq 0.03\,\mathrm{MeV} $ for protons, after which we may treat a nucleon-antinucleon pair as a kind of thermal meson, which induces interconversions via $ n \bar n + p \to n \cdots $ and $ p \bar p + n \to p + \cdots $\,.
For details, see Sec.~III of Ref.~\cite{Kohri:1999ex}.

\subsubsection{
Hadrodissociation
}

For $ T \lesssim 0.09\,\mathrm{MeV} $ for neutrons and antineutrons or $ T \lesssim 0.02\,\mathrm{MeV} $ for protons and antiprotons, high-energy nucleons may scatter off the background nuclei produced by BBN to induce hadrodissociation before losing kinetic energy through electromagnetic processes.
More precisely, we must specify the frequency of the hadronic reaction $ H_i + A_j \to A_k $ which occurs while the high-energy hadron is reducing its energy through the electromagnetic interactions.
Here we consider only the background protons, $ p_\mathrm{bg} $\,, and background helium nuclei, $ \alpha_\mathrm{bg} $\,, as target nuclei $ A_j $ because they are the dominant nuclear content at the relevant epoch.

There are three main effects in this regime:
(a) interconversion of background protons and neutrons through hadronic collisions (i.e.\ by energetic inelastic $ pp $ and $ pn $ scattering rather than by $ p \bar p $ and $ p \bar n $ discussed above);
(b) dissociation of background $ \alpha_\mathrm{bg} $\,, created in BBN, into energetic debris such as $ n $\,, $ p $\,, $ \mathrm D $, $ \mathrm T $, and $ {}^3\mathrm{He} $;
and (c) production of heavier nuclei, such as $ {}^6\mathrm{Li} $, $ {}^7\mathrm{Li} $, and $ {}^7\mathrm{Be} $, through collisions between these debris and background $ \alpha_\mathrm{bg} $\,.
For neutrons, we take into account the fact that they may decay before scattering off the background nuclei.
We consider 10 reactions relevant to (a) and (b) and these are summarized in Table~\ref{tab:reaction}.
As regards (c), we incorporate the nonthermal production of $ {}^{6}\mathrm{Li} $ by energetic $ \mathrm T $ or $ {}^3\mathrm{He} $ through $ \mathrm T + \alpha_\mathrm{bg} \to {}^6\mathrm{Li} + n $ and $ {}^3\mathrm{He} + \alpha_\mathrm{bg} \to {}^6\mathrm{Li} + p $\,.
Collisions of $ \alpha_\mathrm{bg} $ with high-energy $ {}^4\mathrm{He} $ may result in the production of $ {}^6\mathrm{Li} $, $ {}^7\mathrm{Li} $, or $ {}^7\mathrm{Be} $, although they do not affect the final constraint on the mass spectrum of PBHs.
\begin{table}[ht]
\caption{
Hadronic reactions with $ p_\mathrm{bg} $ and $ \alpha_\mathrm{bg} $ included in our calculation.
\label{tab:reaction}
}
\begin{ruledtabular}
\begin{tabular}{ccc}
Process
& $ i=n $
& $ i=p $ \\
\hline
($ i $,$ p_\mathrm{bg} $,1)
& $ n+p_\mathrm{bg} \to n+p $
& $ p+p_\mathrm{bg} \to p+p $ \\
($ i $,$ p_\mathrm{bg} $,2)
& $ n+p_\mathrm{bg} \to n+p+\pi^0 $
& $ p+p_\mathrm{bg} \to p+p+\pi^0 $ \\
($ i $,$ p_\mathrm{bg} $,3)
& $ n+p_\mathrm{bg} \to n+n+\pi^+ $
& $ p+p_\mathrm{bg} \to p+n+\pi^+ $ \\
($ i $,$ \alpha_\mathrm{bg} $,4)
& $ n+\alpha_\mathrm{bg} \to n+\alpha $
& $ p+\alpha_\mathrm{bg} \to p+\alpha $ \\
($ i $,$ \alpha_\mathrm{bg} $,5)
& $ n+\alpha_\mathrm{bg} \to \mathrm D+\mathrm T $
& $ p+\alpha_\mathrm{bg} \to \mathrm D+{}^3\mathrm{He} $ \\
($ i $,$ \alpha_\mathrm{bg} $,6)
& $ n+\alpha_\mathrm{bg} \to p+n+\mathrm T $
& $ p+\alpha_\mathrm{bg} \to 2p+\mathrm T $ \\
($ i $,$ \alpha_\mathrm{bg} $,7)
& $ n+\alpha_\mathrm{bg} \to n+2\mathrm D $
& $ p+\alpha_\mathrm{bg} \to p+2\mathrm D $ \\
($ i $,$ \alpha_\mathrm{bg} $,8)
& $ n+\alpha_\mathrm{bg} \to p+2n+\mathrm D $
& $ p+\alpha_\mathrm{bg} \to 2p+n+\mathrm D $ \\
($ i $,$ \alpha_\mathrm{bg} $,9)
& $ n+\alpha_\mathrm{bg} \to 2p+3n $
& $ p+\alpha_\mathrm{bg} \to 3p+2n $ \\
($ i $,$ \alpha_\mathrm{bg} $,10)
& $ n+\alpha_\mathrm{bg} \to n+\alpha+\pi^0 $
& $ p+\alpha_\mathrm{bg} \to p+\alpha+\pi^0 $
\end{tabular}
\end{ruledtabular}
\end{table}

The necessary procedure is then to evaluate the number of each stable nuclei species produced per single PBH evaporation and to compare this with observations.
Since the reaction rate depends on energy, it is important to calculate the energy spectrum of injected hadron species, as well as the number density.
In our analysis the energy spectra of primary protons and neutrons are calculated by \textsc{PYTHIA}.
For more details, see Secs.~VII and VIII of Ref.~\cite{Kawasaki:2004qu}.

\subsubsection{
Photodissociation
}

Even if high-energy quarks and gluons mainly produce hadronic particles initially, their kinetic energy is eventually converted into high-energy radiation through scattering.
Hence it is important to incorporate photodissociation.
Following Ref.~\cite{Kawasaki:1994af}, we consider the following processes in addition to the injection of particles from PBHs:
(i) double-photon pair creation, $ \gamma + \gamma_\mathrm{bg} \to e^+ + e^- $\,;
(ii) photon-photon scattering, $ \gamma + \gamma_\mathrm{bg} \to \gamma + \gamma$\,;
(iii) Compton scattering off background thermal electrons, $ \gamma + e^-_\mathrm{bg} \to \gamma + e^- $\,;
(iv) inverse Compton scattering off background photons, $ e^\pm + \gamma_\mathrm{bg} \to e^\pm + \gamma $\,;
(v) pair-creation off background protons, $ \gamma + p_\mathrm{bg} \to e^+ + e^- + p $\,.
The relevant terms in the Boltzmann equations are
\begin{equation}
\left[\frac{\mathrm dn_{A_i}}{\mathrm dt}\right]_\gamma
= -n_{A_i}\,\sum_j \int_{E_\gamma^{(\mathrm{th})}}\!\mathrm dE_\gamma\,
  \sigma_{A_i \to A_j}(E_\gamma)\,f_\gamma(E_\gamma)
  + \sum_j n_{A_j}\,
    \int_{E_\gamma^{(\mathrm{th})}}\!\mathrm dE_\gamma\,
    \sigma_{A_j \to A_i}(E_\gamma)\,f_\gamma(E_\gamma)\,,
\end{equation}
where $ \sigma_{A_i \to A_j} $ is the cross section for the reaction $ A_i + \gamma \to A_j + \cdots $ with threshold energy $ E_\gamma^{(\mathrm{th})} $ and $ f_\gamma(E_\gamma) $ is the photon energy distribution function, which is calculated taking above processes (i) through (v) into account.
The resulting spectrum has a cutoff at $ E_\gamma^\mathrm{max} \simeq m_e^2/(22\,T) $ \cite{Kawasaki:1994sc}, above which photons lose energy through the pair creation (i).
If $ E_\gamma^\mathrm{max} $ exceeds the threshold energies of light elements, high-energy photons can dissociate them and change their abundances.
Since the binding energies of $ \mathrm D $ and $ {}^4\mathrm{He} $ are $ B_\mathrm D = 2.2\,\mathrm{MeV} $ and $ B_\alpha = 20\,\mathrm{MeV} $, they are photodissociated only after $ 10^4\,\mathrm s $ and $ 10^6\,\mathrm s $, respectively.
Specifically, we incorporate the following photodissociation reactions:
$ \mathrm D + \gamma \to n + p $\,;
$ \mathrm T + \gamma \to n + \mathrm D,\ p + 2\,n $\,;
$ {}^3\mathrm{He} + \gamma \to p + \mathrm D,\ 2\,p + n $\,;
$ {}^4\mathrm{He} + \gamma \to p + \mathrm T,\ n + {}^3\mathrm{He},\ p + n + \mathrm D $.
We also include photodissociation of $ {}^6\mathrm{Li} $, $ {}^7\mathrm{Li} $, and $ {}^7\mathrm{Be} $.
For more details, see Sec.~IV of Ref.~\cite{Kawasaki:2004qu} or Refs.~\cite{Kawasaki:1994af,Holtmann:1998gd,Kawasaki:2000qr}.

\subsection{
Current status of observations of light elements
}

The observed light-element abundances are listed in Table~\ref{tab:elem} and we now summarize the status of the observations.
\begin{table}[ht]
\caption{
Observational constraints on abundances of light elements at $ 2\sigma $ to be used in the present paper.
\label{tab:elem}
}
\begin{ruledtabular}
\begin{tabular}{ccc}
Element
& Constraints
& Refs. \\
\hline
$ {}^4\mathrm{He} $
& $ Y_p = 0.2516 \pm 0.0080 $
& \cite{Izotov:1998mj,Izotov:2003xn,Olive:2004kq,Fukugita:2006xy,Peimbert:2007vm,Izotov:2007ed} \\
$ \mathrm D $
& $ \mathrm D/\mathrm H < 5.16 \times 10^{-5} $
& \cite{Burles:1997fa,O'Meara:2000dh} \\
$ {}^3\mathrm{He} $
& $ {}^3\mathrm{He}/\mathrm D < 1.37 $
& \cite{2003SSRv..106....3G} \\
$ {}^6\mathrm{Li} $
& $ {}^6\mathrm{Li}/{}^7\mathrm{Li} < 0.302 $
& \cite{Asplund:2005yt,Hisano:2009rc,*Hisano:2009rcE}
\end{tabular}
\end{ruledtabular}
\end{table}

\textit{Helium.}
Synthesis of $ {}^4\mathrm{He} $ is the most important consequence of BBN and its abundance should in principle be used as the most fundamental test of nonstandard BBN.
The primordial abundance of $ {}^4\mathrm{He} $ is inferred by using recombination lines from low-metallicity extragalactic $ \mathrm{HII} $ regions which yield $ \mathrm{HeII}/\mathrm{HII} $ ratios.
The primordial mass fraction, $ Y_p $\,, is obtained by extrapolating to the $ \mathrm O/\mathrm H \to 0 $ or $ \mathrm N/\mathrm H \to 0 $ limits.
The recent trend is that $ Y_p $ has a larger value than thought before.
Olive and Skillman \cite{Olive:2004kq} reanalyzed the seven $ \mathrm{HII} $ regions of Ref.~\cite{Izotov:1998mj}, using $ \mathrm{HeI} $ emission lines to determine the temperature and including underlying stellar absorption, to obtain $ Y_p = 0.2491 \pm 0.0091 $.
Fukugita and Kawasaki \cite{Fukugita:2006xy}, on the other hand, used the $ \mathrm{OIII} $ emission line in 33 $ \mathrm{HII} $ regions in Ref.~\cite{Izotov:2003xn} to determine the temperature.
Incorporating stellar absorption, they found $ Y_p = 0.250 \pm 0.004 $ but $ Y_p = 0.234 \pm 0.004 $ without stellar absorption.
Peimbert \textit{et al.} \cite{Peimbert:2007vm} used new computational results of $ \mathrm{HeI} $ emission lines to determine the temperature.
From five $ \mathrm{HII} $ regions of Ref.~\cite{Izotov:1998mj}, they found $ Y_p = 0.249 \pm 0.009 $.
Most recently, Izotov \textit{et al.} \cite{Izotov:2007ed} found $ Y_p = 0.2516 \pm 0.0011 $ from 93 $ \mathrm{HII} $ regions in $ \mathrm{HeBCD} $ and 271 $ \mathrm{HII} $ regions in SDSS DR5.
In this paper, we adopt their mean value and the $ 2\sigma $ error of Fukugita and Kawasaki \cite{Fukugita:2006xy}:
\begin{equation}
Y_p
= 0.2516 \pm 0.0080.
\end{equation}

\textit{Deuterium and helium-3.}
The primordial deuterium-to-hydrogen abundance ratio $ \mathrm D/\mathrm H $ is measured in high redshift QSO absorption systems.
Because such a Lyman-limit system is not expected to be contaminated by any Galactic or stellar chemical evolution, we can regard it as having primordial abundances.
There have been several recent measurements with improved modelling of the continuum level, the Lyman-$ \alpha $ forest, and the velocity structure of the absorption systems.
The weighted mean of these measurements, $ \mathrm D/\mathrm H = (2.82 \pm 0.26) \times 10^{-5} $ \cite{Burles:1997fa,O'Meara:2000dh}, agrees well with the prediction of standard BBN with the baryon-to-photon ratio inferred by observations of the CMB anisotropy \cite{Hinshaw:2008kr,Dunkley:2008ie,Komatsu:2008hk}.
Recently a new observation towards Q0913+07, giving the same weighted value but a smaller error $ \mathrm D/\mathrm H = (2.82 \pm 0.20) \times 10^{-5} $ was reported \cite{Pettini:2008mq}.
Among the data, the highest value is $ \mathrm D/\mathrm H = (3.98^{+0.59}_{-0.67}) \times 10^{-5} $.
In this paper we adopt this highest value to get a conservative $ 2\sigma $ upper bound
\begin{equation}
\frac{\mathrm D}{\mathrm H}
< 5.16 \times 10^{-5}.
\end{equation}
As for $ {}^3\mathrm{He} $, we use the upper limit on $ {}^3\mathrm{He}/\mathrm D $ because this ratio is an increasing function of cosmic time.
A recent measurement in protosolar clouds gives a $ 2\sigma $ upper bound \cite{2003SSRv..106....3G}
\begin{equation}
\frac{{}^3\mathrm{He}}{\mathrm D}
< 1.37.
\end{equation}

\textit{Lithium.}
The primordial value of $ {}^7\mathrm{Li}/\mathrm H $ is measured in old Population II halo stars.
However, there is a discrepancy between the observational values and theoretical predictions in BBN \cite{Cyburt:2008up,Cyburt:2008kw}.
Therefore we do not adopt a constraint from $ {}^7\mathrm{Li}/\mathrm H $ in this work.
$ {}^6\mathrm{Li} $ is much rarer than $ {}^7\mathrm{Li} $ and its abundance is believed to increase with cosmic time through cosmic-ray spallation, so one can obtain an upper bound on its primordial abundance by observing low-metallicity regions.
Asplund \textit{et al.} \cite{Asplund:2005yt} detected $ {}^6\mathrm{Li} $ in $ 9 $ out of $ 24 $ metal-poor halo dwarfs at more than $ 2\sigma $ significance.
In particular, they detected $ {}^6\mathrm{Li} $ in one very metal-poor star LP815-43 with $ [\mathrm{Fe}/\mathrm H] = -2.74 $ and inferred $ {}^6\mathrm{Li}/{}^7\mathrm{Li} = 0.046 \pm 0.022 $.
Based on this result and a possible error ($ +0.106 $) discussed in Ref.~\cite{Hisano:2009rc,*Hisano:2009rcE} for stellar destruction processes, we take a conservative $ 2\sigma $ upper bound
\begin{equation}
\frac{{}^6\mathrm{Li}}{{}^7\mathrm{Li}}
< 0.302.
\end{equation}

\subsection{
Constraints on $ \beta'(M) $ imposed by BBN
}

We now report our results on the constraints on the PBH mass spectrum imposed by BBN.
We have three parameters: the initial baryon-to-photon ratio $ \eta_\mathrm i $\,, the PBH initial mass $ M $ or (equivalently) its lifetime $ \tau $\,, and the initial PBH number density normalized to the entropy density, $ Y_\mathrm{PBH} \equiv n_\mathrm{PBH}/s $\,.
We start the BBN calculation at a cosmic temperature $ T = 100\,\mathrm{MeV} $.
The initial baryon-to-photon ratio needs to be set to an appropriate value, so we use the present one $ \eta = (6.225\pm 0.170) \times 10^{-10} $ \cite{Dunkley:2008ie}, after allowing for possible entropy production from PBH evaporations and photon heating due to $ e^+ e^- $ annihilations.
From Eq.~\eqref{eq:beta} the value of $ Y_\mathrm{PBH} $ is related to the (normalized) initial mass fraction $ \beta' $ by
\begin{equation}
\beta'
= 5.4 \times 10^{21}\,
  \left(\frac{\tau}{1\,\mathrm s}\right)^{1/2}\,
  Y_\mathrm{PBH}\,.
\end{equation}
In general $ \beta' $\,, $ \tau $\,, and $ Y_\mathrm{PBH} $ in this relation would all depend on the PBH mass $ M $\,, although we presuppose a monochromatic mass function in what follows.

\begin{figure}[ht]
\begin{center}
\includegraphics[scale=0.55]{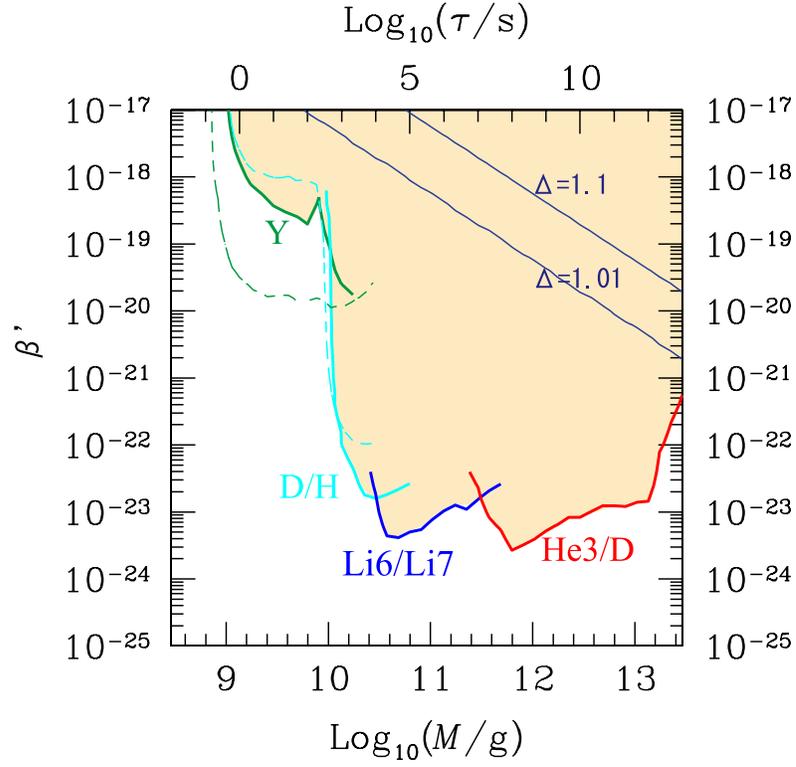}
\end{center}
\caption{
Upper bounds on $ \beta'(M) $ from BBN, with the broken lines giving the earlier limits in Ref.~\cite{Kohri:1999ex}.
\label{fig:bbn}
}
\end{figure}
Figure~\ref{fig:bbn} summarizes the results of our calculations.
First, we note that PBHs with lifetime smaller than $ 10^{-2}\,\mathrm s $ are free from BBN constraints because they evaporate well before weak freeze-out and leave no trace.
PBHs with $ M = 10^9\text{--}10^{10}\,\mathrm g $ and lifetime $ \tau = 10^{-2}\text{--}10^2\,\mathrm s $ are constrained by the extra interconversion between protons and neutrons due to emitted mesons and antinucleons, which increases the $ n/p $ freeze-out ratio as well as the final $ {}^4\mathrm{He} $ abundance.
For $ \tau = 10^2\text{--}10^7\,\mathrm s $, corresponding to $ M = 10^{10}\text{--}10^{12}\,\mathrm g $, hadrodissociation processes become important and the debris deuterons and nonthermally produced $ {}^6\mathrm{Li} $ put strong constraints on $ \beta(M) $\,.
Finally, for $ \tau = 10^7\text{--}10^{12}\,\mathrm s $, corresponding to $ M = 10^{12}\text{--}10^{13} \,\mathrm g $, energetic neutrons decay before inducing hadrodissociation.
Instead, photodissociation processes are operative and the most stringent constraint comes from overproduction of $ {}^3\mathrm{He} $ or $ \mathrm D $.
However, even these effects become insignificant after $ 10^{12}\,\mathrm s $.

For comparison, we have also shown the much weaker constraint imposed by the entropy production from evaporating PBHs, as first analyzed by Miyama and Sato \cite{Miyama:1978mp}.
The factor $ \Delta $ in Fig.~\ref{fig:bbn} is the ratio of the entropy density after and before PBH evaporations, which is calculated numerically.
We also show as dotted lines in Fig.~\ref{fig:bbn} the limits obtained about ten years ago by Kohri and Yokoyama \cite{Kohri:1999ex}.
It is noted that their helium limit on $ \beta(M) $ was stronger than ours, while their deuterium limit was weaker.
This is because the assumed helium abundance is now smaller, whereas hadrodissociation of helium produces more deuterium.

It is important to note that the limit on $ \beta(M) $ does not have a sudden cutoff above the mass $ M_1 = 10^{13}\,\mathrm g $ associated with the PBHs which evaporate at $ t_1 = 6.2 \times 10^{11}\,\mathrm s $.
This is because even PBHs with mass $ M_2 $ larger than $ M_1 $ will generate some emission before $ t_1 $\,.
However, the fraction of their energy involved is only
\begin{equation}
1 - \left[1 - \left(\frac{\tau_1}{\tau_2}\right)\right]^{1/3}
\approx
  \frac{1}{3}\,\left(\frac{\tau_1}{\tau_2}\right)
\approx
  \frac{1}{3}\,\left(\frac{M_1}{M_2}\right)^3
\end{equation}
for $ M_1 \ll M_2 $\,, so Eq.~\eqref{eq:beta} implies that the limit on $ \beta(M_2) $ is weaker than that on $ \beta(M_1) $ by a factor $ (M_2/M_1)^{7/2} $\,.
This explains the form of the upper cutoff above $ 10^{13}\,\mathrm g $ in Fig.~\ref{fig:bbn}.
Indeed, this same argument implies that one never has a sudden upper cutoff in $ \beta(M) $ for any limit which involves the total energy emitted by the PBH.

Finally we note that the neutron diffusion length at $ t = \tau $ is given by \cite{Applegate:1987hm}
\begin{equation}
d_n(\tau)
= 5.8 \times 10^2\,
  \left(\frac{T}{1\,\mathrm{MeV}}\right)^{-5/2}\,
  \left(\frac{\sigma_\mathrm t}{3 \times 10^{-30}\,\mathrm{cm^{2}}}\right)^{-1/2}\,
  \mathrm{cm},
\end{equation}
where $ \sigma_\mathrm t $ is a transport cross section.
From Eq.~\eqref{eq:beta}, the number of PBHs within the neutron diffusion volume is therefore given by \cite{Kohri:1999ex}
\begin{equation}
N_\mathrm d(\tau)
= 7.7 \times 10^{46}\,M_{10}^3\,\beta'\,d_n(\tau)^3\,H(\tau)^3
=
\begin{cases}
1.2 \times 10^{21}\,\beta'
& (\tau = 1\,\mathrm s)\,, \\
2.3 \times 10^{26}\,\beta'
& (\tau = 10^3\,\mathrm s)\,,
\end{cases}
\end{equation}
this scaling as $ M^{-6}\,T^{-15/2} \propto M^{-6}\,\tau^{15/4} \propto \tau^{7/4} $\,.
Comparison with Fig.~\ref{fig:bbn} shows that this number always exceeds $ 1 $ throughout BBN, so there are sufficiently many PBHs within the neutron diffusion length for the relevant parameter range for them to be observationally constrained.
Therefore we are justified in assuming that the emitted hadrons form a homogeneous background.

\section{
Constraints on PBHs from extragalactic photon background
\label{sec:photon}
}

\subsection{
Historical overview
}

One of the earliest works that applied the theory of black hole evaporation to astrophysics was carried out by Page and Hawking \cite{Page:1976wx}, who used the diffuse extragalactic $\gamma$-ray background observations to constrain the mean cosmological number density of PBHs which are completing their evaporation at the present epoch to be less than $ 10^4\,\mathrm{pc}^{-3} $, although their \emph{local} density could be much larger than this if they are clustered inside the Galactic halo.
This also corresponds to an upper limit on their density parameter $ \Omega_\mathrm{PBH} $ of around $ 10^{-8} $.
In fact, this limit was first noted by Chapline \cite{1975Natur.253..251C} and it was also studied by Carr \cite{Carr:1976jy}, though in less detail.
The limit has been subsequently refined but without entailing large quantitative changes.
For example, MacGibbon and Carr \cite{MacGibbon:1991vc} considered how it is modified by including quark and gluon emission effects and inferred $ \Omega_\mathrm{PBH} \le (7.6 \pm 2.6) \times 10^{-9}\,h^{-2} $\,.
(They gave the exponent of $ h $ as $ -1.95 \pm 0.15 $ rather than $ -2 $, because of the $ M $ dependence of $ f $, but the difference is negligible.)
Later they used EGRET observations between $ 30\,\mathrm{MeV} $ and $ 120\,\mathrm{GeV} $ \cite{Sreekumar:1997un} to derive a slightly stronger limit $ \Omega_\mathrm{PBH} \le (5.1 \pm 1.3) \times 10^{-9}\,h^{-2} $ \cite{Carr:1998fw}.
Using the modern value of $ h $ gives $ \Omega_\mathrm{PBH} \le (9.8 \pm 2.5) \times 10^{-9} $ and this corresponds to the ($ h $ independent) constraint $ \beta'(M_*) < 6 \times 10^{-26} $ from Eq.~\eqref{eq:omega}.
They also inferred from the form of the $ \gamma $-ray spectrum that PBHs do not provide the \emph{dominant} contribution to the background.
As discussed below, more recent EGRET and Fermi Large Area Telescope (LAT) data and more precise calculations of the PBH emission allow the limit on $ \beta'(M_*) $ to be improved but the second conclusion remains true.
This raises the point that one should in principle remove any known astrophysical sources of an extragalactic $\gamma$-ray background before calculating the PBH constraints.
For example, the contribution of blazars has recently been discussed by Inoue \textit{et al.} \cite{Inoue:2008pk,Inoue:2010tj}.
This is the strategy adopted by Barrau \textit{et al.} \cite{Barrau:2003nj}, who thereby obtain a limit $ \Omega_\mathrm{PBH} \le 3.3 \times 10^{-9} $ which is three times stronger than that in Ref.~\cite{Carr:1998fw}.
We do not attempt to make such a subtraction here but merely emphasize that our constraints on $ \beta'(M) $ are therefore more conservative than necessary.

\subsection{
Photon spectra
}

The photon emission has a primary and secondary component, as indicated by Eq.~\eqref{eq:prisec}.
The two spectra are calculated according to the prescription of Sec.~\ref{sec:spectra}.
One of our purposes here will be to determine the relative magnitude of these two components.
This is sensitive to the PBH mass and affects the associated $ \beta(M) $ limit.
In order to determine the present background spectrum of particles generated by PBH evaporations, we must integrate over both the lifetime of the black holes and their mass spectrum, allowing for the fact that particles generated in earlier cosmological epochs will be redshifted in energy by now.
This is quite a complicated calculation because the PBH mass is itself a function of time due to evaporation.
However, for the purpose of obtaining limits on $ \beta'(M) $\,, we are assuming a monochromatic mass spectrum, which considerably simplifies the analysis.

\begin{figure}[ht]
\begin{center}
\includegraphics{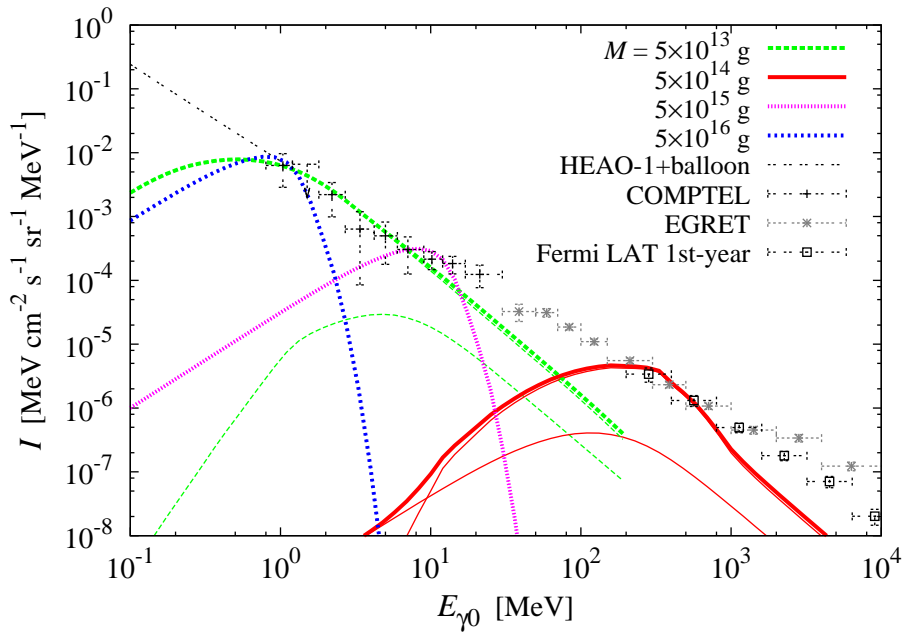}
\end{center}
\caption{
Fluxes corresponding to the upper limit on the PBH abundance for various values of $ M $\,.
All PBHs produce primary photons but $ M \lesssim M_* $ ones also produce secondary photons and this gives a stronger constraint on $ \beta $\,.
\label{fig:flux}
}
\end{figure}

If the PBHs all have the same initial mass $ M $\,, then the present-day photon flux is a superposition of the instantaneous emissions from all previous epochs.
If we approximate the number of emitted photons in the logarithmic energy bin $ \Delta E_\gamma \simeq E_\gamma $ by $ \dot N_\gamma(E_\gamma) \simeq E_\gamma\,(\mathrm d\dot N_\gamma/\mathrm dE_\gamma $), then the emission rate per volume ($ \mathrm{cm}^{-3}\,\mathrm s^{-1} $) at cosmological time $ t $ is
\begin{equation}
\frac{\mathrm dn_\gamma}{\mathrm dt}(E_\gamma,t)
\simeq
  n_\mathrm{PBH}(t)\,E_\gamma\,
  \frac{\mathrm d\dot N_\gamma}{\mathrm dE_\gamma}(M(t),E_\gamma)\,,
\end{equation}
where the $ t $ dependence of $ M $ just reflects the shrinkage due to evaporation.
Since the photon energy and density are redshifted by factors $ (1+z)^{-1} $ and $ (1+z)^{-3} $\,, respectively, the present number density of photons with energy $ E_{\gamma 0} $ is
\begin{equation}
\begin{aligned}
n_{\gamma 0}(E_{\gamma 0})
& = \int_{t_\mathrm{min}}^{\min(t_0,\tau)}\!\mathrm dt\,
    (1+z)^{-3}\,\frac{\mathrm dn_\gamma}{\mathrm dt}((1+z)\,E_{\gamma 0},t) \\
& = n_\mathrm{PBH}(t_0)\,E_{\gamma 0}\,
    \int_{t_\mathrm{min}}^{\min(t_0,\tau)}\!\mathrm dt\,(1+z)\,
    \frac{\mathrm d\dot N_\gamma}{\mathrm dE_\gamma}(M(t),(1+z)\,E_{\gamma 0})\,,
\end{aligned}
\end{equation}
where $ t_\mathrm{min} $ corresponds to the earliest time at which the photons freely propagate (see later) and $ n_\mathrm{PBH}(t_0) $ must be interpreted as the current PBH number density for $ M > M_* $ or the number density PBHs would have had now if they had not evaporated for $ M < M_* $\,.
The photon flux is then
\begin{equation}
I
\equiv
  \frac{c}{4\pi}\,n_{\gamma 0}
\end{equation}
in units of $ \mathrm s^{-1}\,\mathrm{cm}^{-2}\,\mathrm{sr}^{-1} $.
The calculated present-day fluxes of primary and secondary photons are shown in Fig.~\ref{fig:flux}, where the density $ n_\mathrm{PBH} $ for each $ M $ has the maximum value consistent with the observations.

Note that the highest-energy photons are associated with PBHs of mass $ M_* $\,.
Photons from PBHs with $ M > M_* $ are at lower energies because they are cooler, while photons from PBHs with $ M < M_* $ are at lower energies because (although initially hotter) they are redshifted.
The spectral shape differs depending on the mass $ M $ and can be analytically explained as follows.
Holes with $ M > M_* $ have a rather sharp peak which is well approximated by the instantaneous blackbody emission of the primary photons.
On the other hand, as indicated by Eq.~\eqref{eq:energytail1} and discussed in more detail in the Appendix, holes with $ M \le M_* $ have a high-energy $ n_{\gamma 0} \sim E_{\gamma 0}^{-2} $ tail for $ E_{\gamma 0} \gg T_\mathrm{BH}/(1+z(\tau)) $ due to their final phases of evaporation.
As $ M $ increases towards $ M_* $\,, the primary photons start to dominate and generate a redshifted \emph{low-energy} tail which scales as $ E_{\gamma 0}^{3/2} $\cite{MacGibbon:1991tj}.

Although most of the energy from both $ M > M_* $ and $ M < M_* $ PBHs currently appears below $ 100\,\mathrm{MeV} $, there is also a higher-energy component from any PBHs with $ M $ slightly larger than $ M_* $ which today have the mass $ m $ given by Eq.~\eqref{eq:mu}.
The relationship between the \emph{current} mass function ($ \mathrm dn/\mathrm dm $) and the \emph{formation} mass function ($ \mathrm dN/\mathrm dM $) can be approximated by
\begin{equation}
\frac{\mathrm dn}{\mathrm dm}
= \mathrm{min}
  \biggl[
   \left(\frac{m}{M_*}\right)^2\,
   \left(\frac{\mathrm dN}{\mathrm dM}\right)_*,
   \left(\frac{\mathrm dN}{\mathrm dM}\right)
  \biggr]\,,
\label{eq:masstail}
\end{equation}
where these are interpreted as comoving number densities.
The spectra are the same well above $ M_* $ but one has a low-mass tail with $ \mathrm dn/\mathrm dm \propto m^2 $ for $ m \ll M_* $\,.
In the more precise calculation presented in the Appendix, the slope flattens as $ m $ goes towards $ M_* $ and decreases to around $ 1 $ at $ M \approx 1.5\,M_* $\,.
It should be stressed that the low-mass tail would not be present if the formation mass function were \emph{precisely} monochromatic, so it is not accounted for by the $ 5 \times 10^{14}\,\mathrm g $ curve in Fig.~\ref{fig:flux}.
For if all the PBHs had exactly the mass $ M_* $\,, they would all evaporate at exactly the same time.
Indeed, the width of the mass function determines how much of the mass tail is present, as discussed in the Appendix.

So far we have implicitly assumed the PBHs are uniformly distributed throughout the Universe.
However, if the PBHs evaporating at the present epoch are clustered inside galactic halos (as is most likely), then there will also be a Galactic background generated by PBHs with $ M \ge M_* $\,.
As discussed in Sec.~\ref{sec:gal}, the Galactic spectrum is essentially dominated by the solid (red) curve in Fig.~\ref{fig:flux} but with the height increased by the local density enhancement.
While the mass tail makes a negligible contribution to the time-integrated extragalactic background, we will see that it is important for the Galactic background.

\subsection{
Limits on $ \beta'(M) $ imposed by observations of the extragalactic photon background
\label{constraints}
}

The origin of the diffuse x-ray and $ \gamma $-ray backgrounds is thought to be distant astrophysical sources, such as blazars, but so far no firm consensus has been established.
The relevant observations come from HEAO 1 and other balloon observations in the $ 3 \text{--}500\,\mathrm{keV} $ range \cite{1999ApJ...520..124G}, COMPTEL in the $ 0.8\text{--}30\,\mathrm{MeV} $ range \cite{2000AIPC..510..467W}, EGRET in the $ 30 \text{--}200\,\mathrm{MeV} $ range \cite{Strong:2004ry}, and Fermi LAT in the $ 200\,\mathrm{MeV}\text{--}102\,\mathrm{GeV} $ range \cite{collaboration:2010nz}.
In the intermediate region between HEAO 1 and COMPTEL, we adopt the fitting formula given in Ref.~\cite{1999ApJ...520..124G}.
Note that the EGRET results given in Ref.~\cite{Strong:2004ry} are somewhat different from the earlier ones reported in Ref.~\cite{Sreekumar:1997un}.
The measurements from the Fermi satellite seem more compatible with the earlier EGRET results.
All the observations are shown in Fig.~\ref{fig:flux}.
\begin{figure}[ht]
\begin{center}
\includegraphics{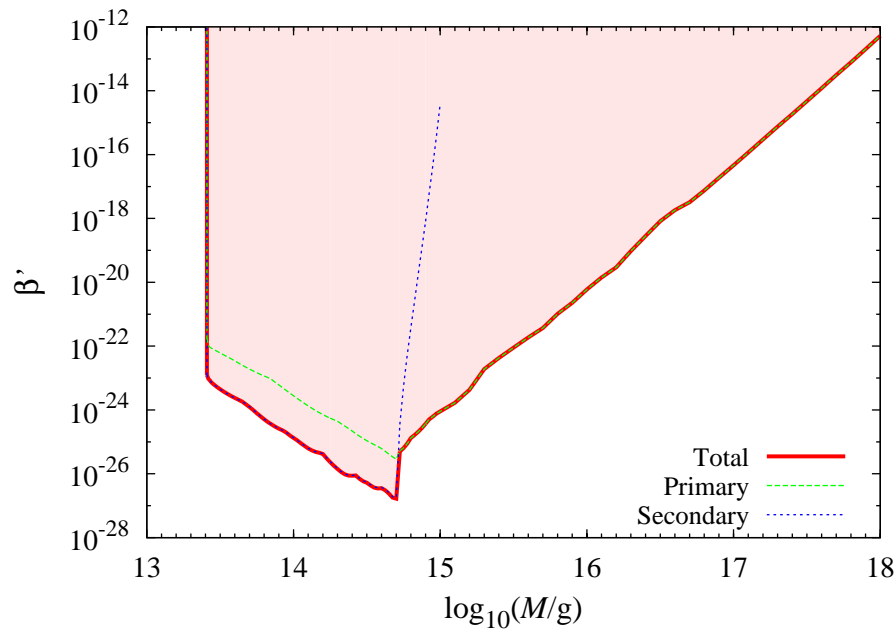}
\end{center}
\caption{
Upper bounds on $ \beta'(M) $ from the extragalactic photon background; these are conservative in the sense that no other contributors to the background have been subtracted.
\label{fig:photon}
}
\end{figure}

The upper limit on the PBH abundance is summarized in Fig.~\ref{fig:photon} and is set by the condition that the photon flux touches one of the upper tips of the $ 1\sigma $ error bars of each observation.
In what follows, we explain the form of this constraint analytically.
We should note that the peak of the flux for each $ M $ plays an essential role in our analysis.
We denote the energy and flux at the peak as $ \hat E_\gamma $ and $ \hat I $\,, respectively, with a subscript ``$ 0 $'' indicating current values.

In order to analyze the spectra of photons emitted from PBHs, different treatments are needed for PBHs with initial masses below and above $ M_* $\,.
We saw in Sec.~\ref{sec:spectra} that PBHs with $ M > M_* $ can never emit secondary photons at the present epoch, whereas those with $ M \le M_* $ will do so once the peak energy of their emission becomes comparable to the QCD scale, which applies once $ M $ falls below $ M_\mathrm q \approx 0.4 M_* \approx 2 \times 10^{14}\,\mathrm g $.
The ratio of the peak energies of primary and secondary photons
is inversely proportional to $ M $ because
\begin{equation}
(1+z(\tau))\,\hat E_{\gamma 0}^\mathrm{pri}
\approx
  6\,T_\mathrm{BH}
\propto
  M^{-1}\,,
\quad
(1+z(\tau))\,\hat E_{\gamma 0}^\mathrm{sec}
\simeq
  \frac{m_\pi}{2}
\approx
  68\,\mathrm{MeV}.
\end{equation}
This is consistent with the solid (red) curve in Fig.~\ref{fig:ratios}, which shows the ratio of $ \hat E_{\gamma 0}^\mathrm{sec}/\hat E_{\gamma 0}^\mathrm{pri} $ derived from our numerical calculations.

We now use simple analytical arguments to derive the form of the primary and secondary peak fluxes.
First, let us consider PBHs with $ M < M_* $\,.
Since the lifetime $ \tau $ for such PBHs is much shorter than $ t_0 $\,, the total number of primary photons emitted from a single hole is roughly $ M/T_\mathrm{BH} \propto M^2 $\,.
Since the comoving number density of evaporating PBHs is $ n_\mathrm{PBH}(\tau)/(1+z(\tau))^3 \propto M^{-3/2}\,\beta $ from Eq.~\eqref{eq:beta}, this being a good approximation even in the $ \Lambda $CDM cosmology, the present peak flux of primary photons scales as
\begin{equation}
\hat I^\mathrm{pri}
\propto
  \frac{n_\mathrm{PBH}(\tau)}{(1+z(\tau))^3}\,\frac{M}{T_\mathrm{BH}}
\propto
  M^{1/2}\,\beta
\quad
(M < M_*)\,.
\end{equation}
For $ M \ll M_* $\,, the temperature is always high enough to emit pions, which from Eq.~\eqref{eq:peakfluxaprx} will produce secondary photons with $ \mathrm d\dot N_\gamma^\mathrm{sec}/\mathrm dE_\gamma \propto T_\mathrm{BH} \propto M^{-1} $ at the peak energy $ \hat E_\gamma^\mathrm{sec} $ (see Fig.~\ref{fig:rate}).
Since the lifetime scales as $ M^3 $\,, this implies that the total number of secondary photons is proportional to $ M^2 $\,, so their peak flux scales as
\begin{equation}
\hat I^\mathrm{sec}
\propto
  \frac{n_\mathrm{PBH}(\tau)}{(1+z(\tau))^3}\,T_\mathrm{BH}\,\tau
\propto
  M^{1/2}\,\beta
\quad
(M \ll M_*)\,.
\end{equation}
This means that $ \hat I^\mathrm{sec}/\hat I^\mathrm{pri} $ is constant for $ M \ll M_* $\,, although the energy of the photons is different.
However, as the PBH mass increases towards $ M_* $\,, the temperature drops and the production of energetic (pion-producing) quarks and gluons becomes suppressed compared to less massive particles, so $ \hat I^\mathrm{sec}/\hat I^\mathrm{pri} $ decreases.
For masses exceeding $ M_* $\,, pion production is exponentially suppressed and the dependence of the primary flux on $ M $ is given by
\begin{equation}
\hat I^\mathrm{pri}
\propto
  n_\mathrm{PBH}(t_0)\,
  \left(\frac{M}{T_\mathrm{BH}}\right)\,
  \left(\frac{t_0}{\tau}\right)
\propto
  M^{-5/2}\,\beta
\quad
(M > M_*)\,.
\end{equation}
The dashed (green) line in Fig.~\ref{fig:ratios} shows the ratio of the primary and secondary peak fluxes found by our numerical calculations and is well explained by the above analytic considerations.

We now compare the predicted PBH flux with the data.
The observed x-ray and $ \gamma $-ray spectra roughly correspond to $ I^\mathrm{obs} \propto E_{\gamma 0}^{-(1+\epsilon)} $\,, where $ \epsilon $ is small and probably lies between $ 0.1 $ (the value favored in Ref.~\cite{Sreekumar:1997un}) and $ 0.4 $ (the value favored in Ref.~\cite{Strong:2004ry}).
For $ M < M_* $\,, the comparison of the primary and secondary peaks in Fig.~\ref{fig:ratios} shows that the limit is determined by the secondary flux.
The secondary peak energy is redshifted to $ \hat E_{\gamma 0}^\mathrm{sec} \simeq m_{\pi^0}/[2\,(1+z(\tau))] \propto M^2 $\,.
Putting $ \hat I^\mathrm{sec}(\beta,M) \le I^\mathrm{obs}(\hat E_{\gamma 0}^\mathrm{sec}) \propto (\hat E_{\gamma 0}^\mathrm{sec})^{-(1+\epsilon)} \propto M^{-2\,(1+\epsilon)} $\,, we can write the upper bound on $ \beta' $ as
\begin{equation}
\beta'(M)
\lesssim
  3 \times 10^{-27}\,
  \left(\frac{M}{M_*}\right)^{-5/2-2\,\epsilon}
\quad
(M < M_*)\,.
\label{photon1}
\end{equation}
For $ M > M_* $\,, secondary photons are not emitted.
The peak energy of the primary photons is $ \hat E_{\gamma 0}^\mathrm{pri} \sim T_\mathrm{BH} \propto M^{-1} $\,, so the observations imply a limit $ \hat I^\mathrm{pri}(\beta,M) \le I^\mathrm{obs} (\hat E_{\gamma 0}^\mathrm{pri}) \propto (\hat E_{\gamma 0}^{\mathrm{pri}})^{-(1+\epsilon)} \propto M^{1+\epsilon} $, which gives 
\begin{equation}
\beta'(M)
\lesssim
  4 \times 10^{-26}\,
  \left(\frac{M}{M_*}\right)^{7/2+\epsilon}
\quad
(M > M_*)\,.
\label{photon2}
\end{equation}
The precise numerical factors in the above limits have been calculated with our code but the $ M $ dependences qualitatively explain the slopes in Fig.~\ref{fig:photon} for $ \epsilon \approx 0.2\text{--}0.3 $.
The limit bottoms out at $ 3 \times 10^{-27} $ and, from Eq.~\eqref{eq:omega}, 
the associated limit on the density parameter is $ \Omega_\mathrm{PBH}(M_*) \le 5 \times 10^{-10} $, which might be compared with the limit $ \Omega_\mathrm{PBH}(M_*) \le 3 \times 10^{-9} $ obtained in Ref.~\cite{Barrau:2003nj}.

The limits given by Eqs.~\eqref{photon1} and \eqref{photon2} are both associated with observations below $ 100\,\mathrm{MeV} $.
Although the PBH flux includes a high-energy $ E_\gamma^{-2} $ tail for PBHs with $ M < M_* $\,, this does not affect the limit on $ \beta $ because the tail flux falls off faster than the observed background intensity.
However, we need to consider more carefully the constraint from observations above $ 100\,\mathrm{MeV} $ on PBHs with $ M $ slightly larger than $ M_* $\,, since these will generate the tail of low-mass holes described by Eq.~\eqref{eq:masstail} unless the initial mass function is precisely monochromatic.
A related point is that the sudden weakening of the limit shown in Fig.~\ref{fig:photon} as $ M $ increases through $ M_* $ is entirely a consequence of the assumption that the PBHs have a monochromatic mass function.
One would expect this discontinuity to be smoothed out in any realistic PBH formation scenario in which the mass function, while peaking above $ M_* $\,, has a spread of masses such that it incorporates some $ M < M_* $ holes.

These nonmonochromaticity effects are discussed further in the Appendix.
In particular, one must distinguish between a ``nearly monochromatic'' function centered at $ M_* $ with a finite width $ \mu\,M_* $\,, and a nearly monochromatic mass function centered at $ (1+\mu)\,M_* $ with a finite width $ \nu\,M_* $\,.
In either case, one needs to decide whether the definition of $ \beta(M) $\,, specified only for a monochromatic mass function in Eq.~\eqref{eq:beta}, includes an integral over the mass width or finely resolves the contribution from each part of the mass band.
While the second definition is more natural for $ \mu \gg 1 $, the first one is more natural for $ \mu \ll 1 $ and this is what we assume here.
For a mass function centered at $ M_* $ with $ \mu \ll 1 $, the total number density of tail photons generated by holes of current mass $ m $ at the peak energy is
\begin{equation}
\hat I^\mathrm{tail}
\propto
  n_\mathrm{PBH}(m,t_0)\,\frac{m}{T_\mathrm{BH}(m)}
\propto
  m^5\,\beta(M_*)
\quad
(0 < m \ll M_*)\,,
\label{photon3}
\end{equation}
so the condition $ \hat I^\mathrm{tail} \le I^\mathrm{obs}(\hat E_{\gamma 0}^\mathrm{tail}) \propto (\hat E_{\gamma 0}^{\mathrm{tail}})^{-(1+\epsilon)} \propto m^{1+\epsilon} $ gives an upper limit on $ \beta(M_*) $ larger than the limit on $ \beta(M_*) $ given by Eq.~\eqref{photon1} by a factor $ (m/M_*)^{\epsilon-4} $\,.
However, the limits may be comparable for $ \mu \sim 1 $ because the mass function becomes shallower there.
For a mass function centered at $ M > M_* $\,,
the same conclusion applies; the limit is weaker than the constraint already shown in Fig.~\ref{fig:photon} (associated with the earlier emission of these holes below $ 100\,\mathrm{MeV} $) because of the $ m^3 $ effect for $ \mu \ll 1 $ but comparable to it for $ \mu \sim 1 $.
Only for $ \nu > \mathrm{max}(1,\mu) $\,, which allows some PBHs to generate secondary emission, could the limit become stronger than the direct one.
In this case, the discontinuity at $ M_* $ in Fig.~\ref{fig:photon} would be smoothed out.

Finally, we determine the mass range over which the $ \gamma $-ray constraint applies.
Since photons emitted at sufficiently early times cannot propagate freely, there is a minimum mass $ M_\mathrm{min} $ below which the above constraint is inapplicable.
The dominant interactions between $ \gamma $ rays and the background Universe in the relevant energy range are pair-production off hydrogen and helium nuclei and Compton scattering off electrons \cite{Page:1976wx,MacGibbon:1991vc,2004PhRvD..70d3502C}.
The first dominates for $ E_\gamma > 100\,\mathrm{MeV} $ and is important below a redshift which is independent of $ E_\gamma $\,;
for the opacity of $ 0.0112\,\mathrm{cm}^2\,\mathrm g^{-1} $ appropriate for a $ 75\,\% $ hydrogen and $ 25\,\% $ helium mix \cite{Page:1976wx}, this redshift is given by \cite{MacGibbon:1991vc}
\begin{equation}
1 + z_\mathrm{max}
\approx
  1100\,
  \left(\frac{h}{0.72}\right)^{-2/3}\,
  \left(\frac{\Omega_\mathrm b}{0.05}\right)^{-2/3}\,,
\end{equation}
with the nucleon density parameter $ \Omega_\mathrm b $ and $ h $ both being normalized to modern values in the last expression.
It seems to be entirely coincidental that this is so close to the time of recombination.
The second process dominates below $ 100\,\mathrm{MeV} $ and, in this case, $ z_\mathrm{max} $ increases with $ E_\gamma $ and may be somewhat smaller.
The mass $ M_\mathrm{min} $ is then determined by the condition $ \tau(M_\mathrm{min}) = t(z_\mathrm{max}) $\,.
Using $ t(z_\mathrm{max}) = 400\,\mathrm{kyr} $ and allowing for the $ M $ dependence of $ f(M) $\,, we obtain
\begin{equation}
M_\mathrm{min}
= \left(\frac{t(z_\mathrm{max})}{t_0}\right)^{1/3}\,
  \left(\frac{f(M_\mathrm{min})}{f_*}\right)^{1/3}\,M_*
\approx
  3 \times 10^{13}\,\mathrm g.
\end{equation}
We therefore extend the limit down to this mass in Fig.~\ref{fig:photon}.
The maximum PBH mass for which the limit is useful is the value at which the photon background limit crosses the density constraint $ \Omega_\mathrm{PBH}(M) < 0.25 $.
Using Eq.~\eqref{eq:omega}, this mass is $ M_\mathrm{max} \approx 7 \times 10^{16}\,\mathrm g $.

\section{
Other constraints for evaporating black holes
\label{sec:other}
}

\begin{figure}[ht]
\begin{center}
\includegraphics{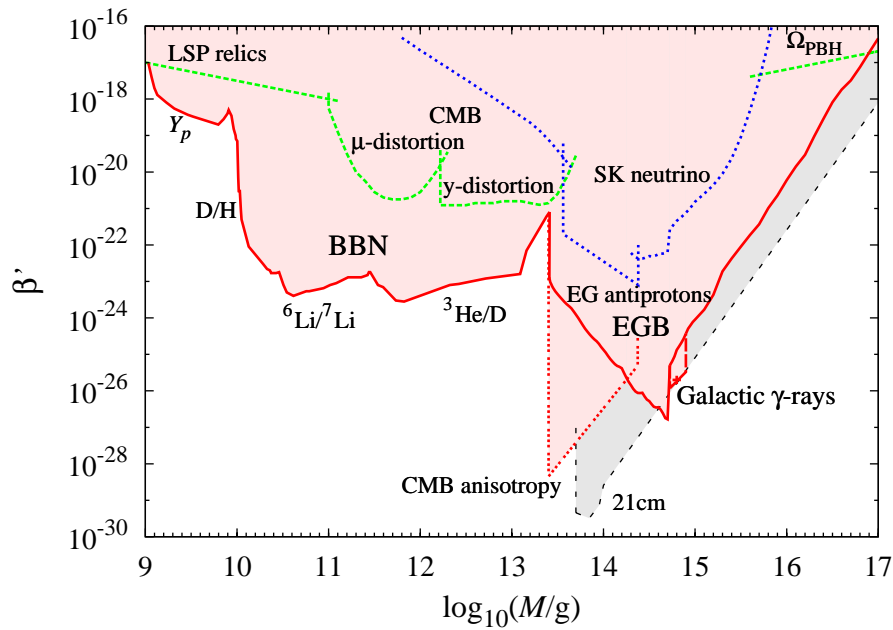}
\end{center}
\caption{
Combined BBN and EGB limits (solid red), compared to other constraints on evaporating PBHs from LSP relics and CMB distortions (short-dashed green), extragalactic antiprotons and neutrinos (dotted blue), the Galactic $ \gamma $-ray background, and CMB anisotropies (long-dashed and broken red), and the potential $ 21\,\mathrm{cm} $ limit (broken grey).
The density limit from the smallest unevaporated black holes is also shown (short-dashed green) to show where it crosses the background photon limit.
The solid red and dotted blue limits are original to this paper.
\label{fig:combined}
}
\end{figure}

The combined BBN and photon background constraints on $ \beta'(M) $ are shown by the solid (red) lines in Fig.~\ref{fig:combined}.
In this section we will summarize various other constraints in the same mass range, these being indicated by the short-dashed (green), dotted (blue), and broken and long-dashed (red) lines in Fig.~\ref{fig:combined}.
Most of the green constraints have been studied in detail elsewhere, so they are discussed only briefly, while the blue constraints---associated with extragalactic antiprotons and neutrinos---are original.
The nonsolid red constraints are also partly original: the long-dashed one is an adaptation of the Galactic $ \gamma $-ray background limit found in Ref.~\cite{Lehoucq:2009ge}, while the broken one is associated with the damping of CMB anisotropies and is an adaptation of a previously known limit associated with decaying particles.
The important point is that the BBN and photon background limits are the most stringent ones over almost the entire mass range $ 10^9\text{--}10^{17}\,\mathrm g $.
There is just a small band in the range $ 10^{13}\text{--}10^{14}\,\mathrm g $ where the CMB anisotropy damping limit dominates.
$ 21\,\mathrm{cm} $ observations \cite{Mack:2008nv} could potentially provide a stronger constraint in some mass range around $ 10^{14}\,\mathrm g $, as indicated by the broken (grey) line, but such limits do not exist at present.

\subsection{
Galactic gamma rays
\label{sec:gal}
}

If PBHs of mass $ M_* $ are clustered inside our own Galactic halo, as expected, then there should also be a Galactic $ \gamma $-ray background and, since this would be anisotropic, it should be separable from the extragalactic background.
The ratio of the anisotropic to isotropic intensity depends on the Galactic longitude and latitude, the Galactic core radius and the halo flattening.
Some time ago Wright \cite{1996ApJ...459..487W} claimed that such a halo background had been detected in EGRET observations between $ 30\,\mathrm{MeV} $ and $ 120\,\mathrm{MeV} $ \cite{Sreekumar:1997un} and attributed this to PBHs.
His detailed fit to the data, subtracting various other known components, required the PBH clustering factor to be $ (2\text{--}12) \times 10^5\,h^{-1} $\,, comparable to that expected, and the local PBH explosion rate to be $ \mathcal R = 0.07\text{--}0.42 \,\mathrm{pc}^{-3}\,\mathrm{yr}^{-1} $.
A more recent analysis of EGRET data between $ 70\,\mathrm{MeV} $ and $ 150\,\mathrm{MeV} $, assuming a variety of distributions for the PBHs, has been given by Lehoucq \textit{et al.} \cite{Lehoucq:2009ge}.
In the isothermal model, which gives the most conservative limit, the Galactic $ \gamma $-ray background requires $ \mathcal R \leq 0.06 \,\mathrm{pc}^{-3}\,\mathrm{yr}^{-1} $.
They claim this corresponds to $ \Omega_\mathrm{PBH}(M_*) \le 2.6 \times 10^{-9} $, which from Eq.~\eqref{eq:omega} implies $ \beta'(M_*) < 1.4 \times 10^{-26} $, a factor of 5 above the extragalactic background constraint obtained in Sec.~\ref{sec:photon}.
Lehoucq \textit{et al.} themselves claim that it corresponds to $ \beta'(M_*) < 1.9 \times 10^{-27} $ but this is because they appear to use an old and rather inaccurate formula relating $ \Omega_\mathrm{PBH} $ and $ \beta $\,.

It should be stressed that the Lehoucq \textit{et al.} analysis does not constrain PBHs of \emph{initial} mass $ M_* $ because these no longer exist.
Rather it constrains PBHs of \emph{current} mass $ M_* $ and, from Eq.~\eqref{eq:mu} with $ m = M_* $\,, this corresponds to an initial mass of $ 1.26\,M_* $\,.
However, as shown below, this value of $ M $ does not correspond to the \emph{strongest} limit on $ \beta(M) $\,.
This contrasts to the situation with the extragalactic background, where the strongest constraint on $ \beta(M) $ comes from the time-integrated contribution of the $ M_* $ black holes.
There would also be a Galactic contribution from PBHs which were slightly \emph{smaller} than $ M_* $ but sufficiently distant for their emitted particles to have only just reached us; since the light-travel time across the Galaxy is $ t_\mathrm {gal} \sim 10^5\,\mathrm{yr} $, this corresponds to PBHs initially smaller than $ M_* $ by $ (t_\mathrm{gal}/3\,t_0)\,M_* \sim 10^{-5}\,M_* $\,, so this extra contribution is negligible.

To examine this issue more carefully, we note that Eq.~\eqref{eq:mu} implies that the emission from PBHs with initial mass $ (1+\mu)\,M_* $ currently peaks at an energy $ E \approx 100\,(3\,\mu)^{-1/3}\,\mathrm{MeV} $ for $ \mu < 1 $, which is in the range $ 70\text{--}150\,\mathrm{MeV} $ for $ 0.7 > \mu > 0.08 $.
(Note that secondary emission can be neglected in this regime, this only being important for $ \mu < 0.02 $.)
So these black holes correspond to the ``tail'' population discussed in Sec.~\ref{constraints} and necessarily have the mass function indicated by Eq.~\eqref{eq:masstail}.
This means that the interpretation of the Galactic $ \gamma $-ray limit is sensitive to the nonmonochromaticity in the PBH mass function, so we must distinguish between the various situations described in the Appendix.
In particular, we need to distinguish between mass functions which center at $ M_* $ and some higher values of $ M $\,.
In the present context, we assume that the mass function is narrow, so that $ \beta(M) $ can be defined as the integral over the entire mass width.

The peak energy is above $ 150\,\mathrm{MeV} $ for $ \mu < 0.08 $, so the $ \gamma $-ray band is in the Rayleigh--Jeans region.
From Eq.~\eqref{eq:grey} with $ \sigma \propto M^4 $\,, the flux of an individual hole scales as $ m^3 \propto \mu $ in this regime.
As clarified in the Appendix, although this flux must be weighted by the number density of the holes, $ n(m) \propto m^3 \propto \mu $ for $ \mu \ll 1 $, this factor is necessarily balanced by the ratio $ (\Delta M/\Delta m) $ of the mass widths at formation and now, so the limit on $ \beta(M) $ scales as $ \mu^{-1} $\,.
For $ 0.7 > \mu > 0.08 $, the current number flux of photons from each PBH scales as $ m^{-1} \propto \mu^{-1/3} $, so the effective limit on $ \beta(M) $ scales as $ \mu^{1/3} $\,.
The observed $ \gamma $-ray band enters the Wien part of the spectrum for $ \mu > 0.7 $, so the limit on $ \beta(M) $ weakens exponentially for $ M > 1.7\,M_* $\,.
(Although not mentioned in Ref.~\cite{Lehoucq:2009ge}, one could in principle get a stronger limit in this mass regime from observations at energies lower than $ 70\,\mathrm{MeV} $, but we do not attempt this here.)
Hence the largest contribution to the Galactic background and the strongest constraint on $ \beta(M) $ comes from PBHs with $ M \approx 1.08\,M_* $ and has the form indicated in Fig.~\ref{fig:combined}.
The height of the limit depends on the halo concentration.
Although the tail holes are less numerous than the $ M_* $ holes by a factor $ (m/M_*)^3 \approx 3\,\mu $\,, this is more than compensated by the halo concentration factor of $ 10^6 $ for $ \mu > 3 \times 10^{-5} $.
This is quite a complicated problem, so we have essentially adopted the model of Lehoucq \textit{et al.} in determining the height.

\subsection{
Galactic antiprotons
}

Since the ratio of antiprotons to protons in cosmic rays is less than $ 10^{-4} $ over the energy range $ 100\,\mathrm{MeV}\text{--}10\,\mathrm{GeV} $, whereas PBHs should produce them in equal numbers, PBHs could only contribute appreciably to the antiprotons \cite{Carr:1976jy,Turner:1981ez,Kiraly:1981ci}.
It is usually assumed that the observed antiprotons are secondary particles, produced by spallation of the interstellar medium by primary cosmic rays.
However, the spectrum of secondary antiprotons should show a steep cutoff at kinetic energies below $ 2\,\mathrm{GeV} $, whereas the spectrum of PBH antiprotons should continue down to $ 0.2\,\mathrm{GeV} $.
Also any primary antiproton fraction should tend to $ 0.5 $ at low energies.
Both these features provide a distinctive signature of any PBH contribution.

The black hole temperature must be much larger than $ T_\mathrm{BH}(M_*) $ to generate antiprotons, so the local cosmic-ray flux from PBHs should be dominated by the ones just entering their explosive phase at the present epoch.
Such PBHs should be clustered inside our halo, so any charged particles emitted will have their flux enhanced relative to the extragalactic spectra by a factor $ \zeta $ which depends upon the halo concentration factor and the time for which particles are trapped inside the halo by the Galactic magnetic field.
This time is rather uncertain and also energy dependent.
At $ 100\,\mathrm{MeV} $ one expects roughly $ \zeta \sim 10^3 $ for electrons or positrons and $ \zeta \sim 10^4 $ for protons or antiprotons \cite{Carr:1998fw}.

MacGibbon and Carr \cite{MacGibbon:1991vc} originally calculated the PBH density required to explain the interstellar antiproton flux at $ 1\,\mathrm{GeV} $ and found a value somewhat larger than the density associated with the $ \gamma $-ray limit.
After the BESS balloon experiment measured the antiproton flux below $ 0.5\,\mathrm{GeV} $ \cite{Yoshimura:1995sa}, Maki \textit{et al.} \cite{Maki:1995pa} tried to fit this in the PBH scenario by using Monte Carlo simulations of cosmic-ray propagation.
They found that the local PBH-produced antiproton flux was mainly due to PBHs exploding within a few kpc and used the observational data to infer a limit on the local PBH explosion rate of $ \mathcal R < 0.017\,\mathrm{pc}^{-3}\,\mathrm{yr}^{-1} $.
A more recent attempt to fit the antiproton data came from Barrau \textit{et al.} \cite{Barrau:2002ru}, who compared observations by BESS95 \cite{Yoshimura:1995sa}, BESS98 \cite{Orito:1999re}, CAPRICE \cite{Boezio:2001ac}, and AMS \cite{Jacholkowska:2007fb} with the spectrum from evaporating PBHs.
According to their analysis, PBHs with $ \beta'(M_*) \approx 5 \times 10^{-28} $ would be numerous enough to explain the observations.
This is well below the photon background limit of $ 3 \times 10^{-27} $ obtained in this paper, which suggests that the PBHs required to explain the extragalactic photon background would overproduce antiprotons.
However, these results are based on the assumption that the PBHs have a spherically symmetric isothermal profile with a core radius of $ 3.5\,\mathrm{kpc} $.
A different clustering assumption would lead to a different constraint on $ \beta'(M_*) $\,.

PBHs might also be detected by their antideuteron flux.
Barrau \textit{et al.} \cite{Barrau:2002mc} argue that AMS and GAPS would be able to detect the antideuterons from PBH explosions if their local density were as large as $ 2.6 \times 10^{-34}\,\mathrm g\,\mathrm{cm}^{-3} $ and $ 1.4 \times 10^{-35}\,\mathrm g\,\mathrm{cm}^{-3} $, respectively.
If a null result were maintained up to these levels, it would imply $ \beta'(M_*) < 1.5 \times 10^{-26}\,(\zeta/10^4)^{-1} $ and $ 8.2 \times 10^{-28}\,(\zeta/10^4)^{-1} $\,, respectively.
However, these are only potential and not actual limits.

The most recent data come from the BESS Polar-I experiment, which flew over Antarctica in December 2004 \cite{Abe:2008sh}.
The reported antiproton flux lies between that of BESS(95+97) and BESS(2000), and the $ \bar p/p $ ratio ($ r \approx 10^{-5} $) is similar to that reported by BESS(95+97).
Although the latter had indicated an excess flux below $ 400\,\mathrm{MeV} $, this was not found in the BESS Polar-I data.
However, given the magnitude of the error bars, we might still expect a constraint on the local PBH number density similar to that discussed above, in which case the constraint on $ \beta(M_*) $ becomes
\begin{equation}
\beta'(M_*)
< 3.9 \times 10^{-25}\,
  \left(\frac{\zeta}{10^4}\right)^{-1}\,,
\end{equation}
where we have used Eq.~\eqref{eq:density}.
For reasonable values of $ \zeta $\,, this is much weaker than the $ \gamma $-ray background limit, but the value of $ \zeta $ is anyway very uncertain and so this limit is not shown explicitly in Fig.~\ref{fig:combined}.

\subsection{
Extragalactic antiprotons
}

Galactic antiprotons can constrain PBHs only in a very narrow range around $ M \approx M_* $ since they diffusively propagate to the Earth on a time much shorter than the cosmic age.
However, there will also be a spectrum of cosmic-ray antiprotons from PBHs sufficiently light that they evaporated well before the epoch of galaxy formation ($ \sim 1\,\mathrm{Gyr} $).
Although such pregalactic PBHs would not be clustered, the antiprotons they emitted could still be around today and may occasionally enter the Galaxy and be detectable at Earth, so we require that this does not happen frequently enough to exceed the observational limits.
One might assume that the maximum mass relevant for this constraint corresponds to the PBHs evaporating at galaxy formation:
\begin{equation}
M
\lesssim
  \left(\frac{1\,\mathrm{Gyr}}{t_0}\right)^{1/3}\,
  \left(\frac{f(M)}{f_*}\right)^{1/3}\,M_*
\approx
  2 \times 10^{14}\,\mathrm g.
\end{equation}
This mass scale is not necessarily relevant because even the antiprotons generated \emph{inside} galaxies may escape on a cosmological time scale and become part of the extragalactic background.
However, the leakage time depends on the size of the cosmic-ray region, which is rather uncertain, and could well be comparable to $ t_0 $\,.
We therefore adopt the above upper mass limit.
The minimum mass relevant for this constraint is determined as follows.
The mean rate of $ \bar p p $ annihilations in the cosmological background, where the target $ p $'s are in hydrogen and helium nuclei, is
\begin{equation}
\Gamma_{\bar pp}(t)
= \langle\sigma_{\bar pp}\,v_{\bar p}\rangle\,n_p(t)
\approx
  2 \times 10^{-22}\,\left(\frac{t}{t_0}\right)^{-2}\,\mathrm s^{-1},
\end{equation}
where we have used $ \langle\sigma_{\bar pp}\,v_{\bar p}\rangle/c \approx 4 \times 10^{-26}\,\mathrm{cm}^2 $ and $ n_p(t_0) \approx 2 \times10^{-7}\,\mathrm{cm}^{-3} $.
The condition that an antiproton emitted by a PBH with lifetime $ \tau $ survives until now is then
\begin{equation}
\int_\tau^{t_0}\!\Gamma_{\bar pp}(t')\,\mathrm dt'
\approx
 \left(\frac{t_0}{\tau}\right)\,
 \left(\frac{t_0}{5 \times 10^{21}\,\mathrm s}\right)
\approx
  \frac{1.3\,\mathrm{Myr}}{\tau}
< 1.
\end{equation}
Using Eq.~\eqref{eq:lifetime}, we infer that the extragalactic $ \bar p $ limit applies for
\begin{equation}
M
\gtrsim
  \left(\frac{1.3\,\mathrm{Myr}}{t_0}\right)^{1/3}\,
  \left(\frac{f(M)}{f_*}\right)^{1/3}\,
  M_*
\approx
  4 \times 10^{13}\,\mathrm g.
\end{equation}
We have seen there is no primary emission of antiprotons and the analog of Eq.~\eqref{eq:peakfluxaprx} implies that the current number density from secondary emission of PBHs with mass $ M $ is
\begin{equation}
\begin{aligned}
n_{\bar p}(M, t_0)
& = \frac{n(\tau)\,\tau}{(1+z(\tau))^3}\,
    \frac{\mathrm d\dot N_{\bar p}}{\mathrm d\ln E_{\bar p}}
    (E_{\bar p}=100\,\mathrm{MeV}) \\
& \approx
    2 \times 10^{22}\,
    \frac{n(\tau)}{(1+z(\tau))^3}\,
    \left(\frac{\tau}{1\,\mathrm s}\right)\,
    \left(\frac{T_\mathrm{BH}}{1\,\mathrm{GeV}}\right) \\
& \approx
    6 \times 10^{11}\,
    \beta'(M)\,
    \left(\frac{f(M)}{f_*}\right)^{-1}\,
    \left(\frac{M}{M_*}\right)^{1/2}\,
    \mathrm{cm}^{-3}.
\end{aligned}
\end{equation}
Since the antiproton-to-proton ratio $ r $ is required to be less than $ 10^{-5} $, one finds the upper bound
\begin{equation}
\beta'(M)
\lesssim
  3 \times 10^{-24}\,
  \left(\frac{r}{10^{-5}}\right)\,
  \left(\frac{f(M)}{f_*}\right)\,
  \left(\frac{M}{M_*}\right)^{-1/2}
\quad
(4 \times 10^{13}\,\mathrm g \lesssim M \lesssim 2 \times 10^{14}\,\mathrm g)\,.
\end{equation}
This constraint is shown by the dotted (blue) line in Fig.~\ref{fig:combined}.
Although it is much weaker than the Galactic antiproton constraint, it is independent of the rather uncertain parameter $ \zeta $\,.

\subsection{
PBH explosions
}

The extragalactic $\gamma$-ray background limit implies that the PBH explosion rate $ \mathcal R $ could be at most $ 10^{-6}\,\mathrm{pc}^{-3}\,\mathrm{yr}^{-1} $ if the PBHs are uniformly distributed or $ 10\,\mathrm{pc}^{-3}\,\mathrm{yr}^{-1} $ if they are clustered inside galactic halos \cite{Page:1976wx,1979Natur.277..199P}.
The latter figure might be compared to the Lehoucq \textit{et al.} Galactic $ \gamma $-ray limit of $ 0.06\,\mathrm{pc}^{-3}\,\mathrm{yr}^{-1} $ \cite{Lehoucq:2009ge} and the Maki \textit{et al.} antiproton limit of $ 0.02\,\mathrm{pc}^{-3}\,\mathrm{yr}^{-1} $ \cite{Maki:1995pa}.
We now compare these limits to the direct observational constraints on the explosion rate.

In the standard model of particle physics, where the number of elementary particle species never exceeds around $ 100 $, it has been appreciated for a long time that the likelihood of detecting the final explosive phase of PBH evaporations is very low \cite{Semikoz:1994uz}.
However, the physics of the QCD phase transition is still uncertain and the prospects of detecting explosions would be improved in less conventional particle physics models.
For example, in a Hagedorn-type picture \cite{Carter:1976di}, where the number of particle species exponentiates at the quark-hadron temperature, the upper limit on $ \mathcal R $ is reduced to $ 0.05\,\mathrm{pc}^{-3}\,\mathrm{yr}^{-1} $ \cite{Fichtel:1994sf}, which is comparable to the antiproton limit.

Even without the Hagedorn effect, something dramatic is likely to occur at the QCD temperature since the number of species being emitted increases dramatically \cite{Halzen:1991uw}.
For this reason, Cline and colleagues have long argued that the formation of a fireball at the QCD temperature could explain some of the short-period $ \gamma $-ray bursts (i.e.\ those with duration less than $ 100\,\mathrm{ms} $) \cite{Cline:1992ps,Cline:1995sz,Cline:1996zg}.
In Ref.~\cite{Cline:2001tq} they claim to find $ 42 $ BATSE candidates of this kind and the fact that their distribution matches the spiral arms suggests that they are Galactic.
In Ref.~\cite{Cline:2005xb} they claim that there is a class of short-period KONUS bursts which has a much harder spectrum than usual and identify these with exploding PBHs.
More recently they have found a further $ 8 $ candidates in the Swift data \cite{Cline:2006nv}.
Overall they claim that the BATSE, KONUS, and Swift data correspond to a $ 4.5 \sigma $ effect and that several events exhibit the time structure expected of PBH evaporations \cite{Cline:2009ni}.
One distinctive feature of $ \gamma $-ray bursts generated by PBH explosions would be a temporal delay between the high and low-energy pulses, an effect which might be detectable by Fermi LAT \cite{Ukwatta:2009xk}.

It is clearly important to understand how likely PBHs are to resemble $ \gamma $-ray bursts from a theoretical perspective.
We have seen that evaporating black holes form QCD photospheres according to Heckler \cite{Heckler:1995qq,Heckler:1997jv}.
A rather different way of producing a $ \gamma $-ray burst is to assume that the outgoing charged particles form a plasma due to turbulent magnetic field effects at sufficiently high temperatures \cite{1996MNRAS.283..626B}.
MacGibbon \textit{et al.} \cite{MacGibbon:2007yq} have questioned both of these claims and argued that the observational signatures of a cosmic or Galactic halo background of PBHs or an individual high-temperature black hole remain essentially those of the standard Hawking model, with little change to the detection probability.
However, they accept that a photosphere might form for a short period around the QCD temperature, so perhaps the best strategy is to accept that our understanding of such effects is incomplete and focus on the empirical aspects of the $ \gamma $-ray burst observations.

At much higher energies, several groups have looked for $ 1\text{--}100\,\mathrm{TeV} $ photons from PBH explosions using cosmic-ray detectors.
However, in this case, the constraints are also strongly dependent on the theoretical model \cite{Bugaev:2007py}.
In the standard model the upper limits on the explosion rate are $ 5 \times 10^8\,\mathrm{pc}^{-3}\,\mathrm{yr}^{-1} $ from the CYGNUS array \cite{Alexandreas:1993zx}, $ 8 \times 10^6\,\mathrm{pc}^{-3}\,\mathrm{yr}^{-1} $ from the Tibet array \cite{1996A&A...311..919T}, $ 1 \times 10^6\,\mathrm{pc}^{-3}\,\mathrm{yr}^{-1} $ from the Whipple Cerenkov telescope \cite{Linton:2006yu}, and $ 8 \times 10^8\,\mathrm{pc}^{-3}\,\mathrm{yr}^{-1} $ from the Andyrchy array \cite{2008PAZ....34..P563,*Petkov:2008rz}.
These limits correct for the effects of burst duration and array ``dead time.''
Such limits are far weaker than the ones associated with observations at $ 100\,\mathrm{MeV} $.
They would be even weaker in the $ 45\,\mathrm{GeV} $ QED photosphere model advocated by Heckler, since there are then far fewer TeV particles \cite{2008PZETF....87..3P,*Petkov:2008rv}.
For example, Bugaev \textit{et al.} \cite{Bugaev:2009ad} have used Andyrchy data to obtain constraints of $ 1 \times 10^9\,\mathrm{pc}^{-3}\,\mathrm{yr}^{-1} $ in the Daghigh--Kapusta model \cite{Kapusta:2000xt,Daghigh:2001gy,Daghigh:2002fn,Daghigh:2006dt} and $ 5 \times 10^9\,\mathrm{pc}^{-3}\,\mathrm{yr}^{-1} $ in the Heckler model \cite{Heckler:1995qq}.
Because of the uncertainties, we do not show any of these limits in Fig.~\ref{fig:combined}.

\subsection{
CMB distortion and anisotropy
}

The effects of PBH evaporations on the CMB were first analyzed by Zel'dovich \textit{et al.} \cite{1977PAZh....3..208Z,*1977SvAL....3..110Z}.
They pointed out that photons emitted sufficiently early will be completely thermalized and merely contribute to the photon-to-baryon ratio.
The requirement that this does not exceed the observed ratio of around $ 10^9 $ leads to a limit
\begin{equation}
\beta'(M)
< 10^9\,\left(\frac{M}{M_\mathrm{Pl}}\right)^{-1}
\approx
  10^{-5}\,\left(\frac{M}{10^9\,\mathrm g}\right)^{-1}
\quad
(M < 10^9\,\mathrm g)\,,
\label{eq:entropy}
\end{equation}
so only PBHs below $ 10^4\,\mathrm g $ could generate \emph{all} of the CMB.
They also noted that photons from PBHs in the range $ 10^{11}\,\mathrm g < M < 10^{13}\,\mathrm g $, although partially thermalized, will produce noticeable distortions in the CMB spectrum unless
\begin{equation}
\beta'(M)
< \left(\frac{M}{M_\mathrm{Pl}}\right)^{-1}
\approx
  10^{-16}\,\left(\frac{M}{10^{11}\,\mathrm g}\right)^{-1}
\quad
(10^{11}\,\mathrm g < M < 10^{13}\,\mathrm g)\,,
\label{eq:distort}
\end{equation}
this corresponding to the condition $ \alpha(M) < 1 $.
In the intermediate mass range, there is a transition from limit \eqref{eq:entropy} to the much stronger limit \eqref{eq:distort}.
Subsequently the form of these distortions has been analyzed in greater detail.
If an appreciable number of photons are emitted after the freeze-out of double-Compton scattering ($ t \gtrsim 7 \times 10^6\,\mathrm s $), corresponding to $ M > 10^{11} \mathrm g $, the distribution of the CMB photons develops a nonzero chemical potential, leading to a $ \mu $ distortion.
On the other hand, if the photons are emitted after the freeze-out of the single-Compton scattering ($ t \gtrsim 3 \times 10^9\,\mathrm s $), corresponding to $M > 10^{12} \mathrm g $, the distribution is modified by a $ y $ distortion.
These constraints were first calculated in the context of decaying particle models \cite{Hu:1993gc}.
In the PBH context, recent calculations of Tashiro and Sugiyama \cite{Tashiro:2008sf} show that the CMB distortion constraints are of order $ \beta'(M) < 10^{-21} $ for some range of $ M $\,.
The precise form of the constraints is shown in Fig.~\ref{fig:combined}.
We note that they are weaker than the BBN constraint but stronger than the constraint given by Eq.~\eqref{eq:distort}.
There are also CMB distortion constraints associated with the accretion of larger nonevaporating black holes and these are discussed in Sec.~\ref{sec:large}.

Another constraint on PBHs evaporating after the time of recombination is associated with the damping of small-scale CMB anisotropies.
The limit can be obtained by modifying an equivalent calculation for decaying particles, as described by Zhang \textit{et al.} \cite{Zhang:2007zzh}.
Their constraint can be written in the form
\begin{equation}
\log_{10}\zeta
< -10.8 - 0.50\,x + 0.085\,x^2 + 0.0045\,x^3\,,
\quad
x
\equiv
  \log_{10}\left(\frac{\Gamma}{10^{-13}\,\mathrm s^{-1}}\right)\,,
\label{eq:zeta}
\end{equation}
where $ \Gamma $ is the decay rate, which corresponds to $ \tau(M)^{-1} $ in our case, and $ \zeta $ is equivalent to the fraction of the CDM in PBHs, which is simply related to $ \beta(M) $\,, times the fraction of the emitted energy which goes into heating the matter.
The last factor, which includes the effects of the electrons and positrons as well as the photons, will be denoted by $ f_\mathrm H $ and depends on the redshift \cite{MacGibbon:1991vc,2004PhRvD..70d3502C}.
Most of the heating will be associated with the electrons and positrons; they are initially degraded by inverse Compton scattering off the CMB photons but after scattering have an energy 
\begin{equation} 
\gamma^2\,E_\mathrm{CMB}
\approx
  300\,
  \left(\frac{\gamma}{10^3}\right)^2\,(1+z)\,\mathrm{eV}
\approx
  5 \times 10^5\,
  \left(\frac{M}{10^{13}\,\mathrm g}\right)^{-2}\,(1+z)\,\mathrm{eV}
\approx
  20\,(1+z)^2\,\mathrm{eV},
\end{equation} 
where $ \gamma $ is the Lorentz factor and in the last expression we assume that the mass of a PBH evaporating at redshift $ z $ in the matter-dominated era is $ M \approx M_*\,(1+z)^{-1/2} $\,.
Since this energy is always above the ionization threshold for hydrogen ($ 13.6\,\mathrm{eV} $), we can assume that the heating of the electrons and positrons is efficient before reionization.
Using Eq.~\eqref{eq:omega} for $ \beta(M) $ and Eq.~\eqref{eq:tau} for $ \tau(M) $\,, one can now express Eq.~\eqref{eq:zeta} as a limit on $ \beta(M) $\,.
In the mass range of interest, the rather complicated cubic expression in $ M $ can be fitted by the approximation
\begin{equation}
\beta'(M)
< 3 \times 10^{-30}\,
  \left(\frac{f_\mathrm H}{0.1}\right)^{-1}\,
  \left(\frac{M}{10^{13}\,\mathrm g}\right)^{3.1}
\quad
(2.5 \times 10^{13}\,\mathrm g \lesssim M \lesssim 2.4 \times 10^{14}\,\mathrm g)\,,
\label{eq:cmb}
\end{equation}
where $ f_\mathrm H \approx 0.1 $ is the fraction of emission which comes out in electrons and positrons.
Here the lower mass limit corresponds to black holes evaporating at recombination and the upper one to those evaporating at a redshift $ 6 $ \cite{Fan:2006dp}, after which the ionization ensures the opacity is too low for the emitted electrons and positrons to heat the matter much.
Equation~\eqref{eq:cmb} is stronger than all the other available limits in this mass range but does not seem to have been pointed out before.

Finally we note that PBH evaporations could also be potentially constrained by their effect on the form of the recombination lines in the CMB spectrum \cite{Sunyaev:2009qn}, just as in the annihilating dark matter scenario \cite{Chluba:2009uv,Slatyer:2009yq}, although we do not discuss this further here.

\subsection{
Relic neutrinos
\label{sec:neutrino}
}

\begin{figure}[ht]
\begin{center}
\includegraphics{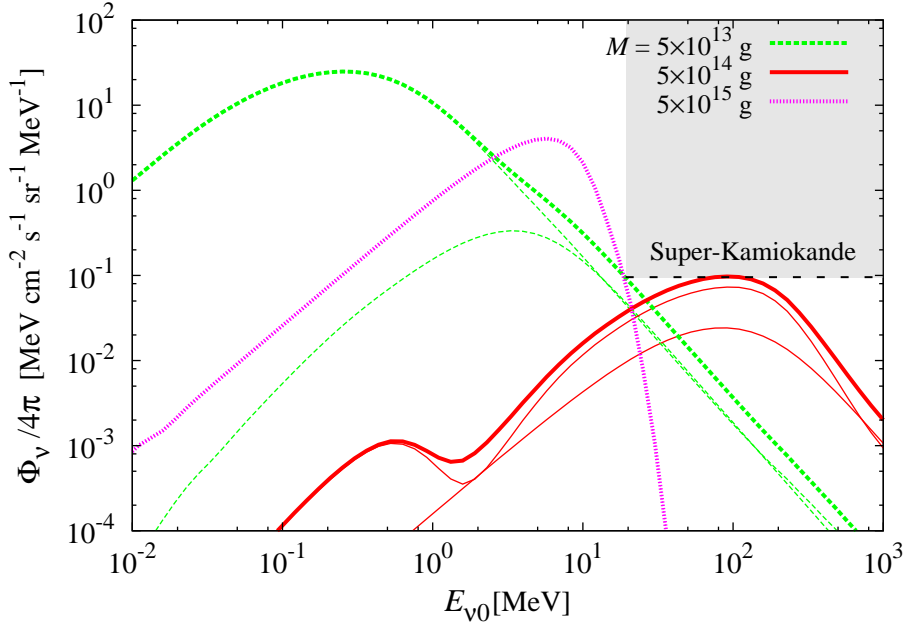}
\end{center}
\caption{
An illustration of the maximum $ \bar\nu_e $ flux allowed by the SK limit for three different masses.
For $ M \lesssim M_* $\,, the lower (upper) curves represent primary (secondary) components and the thick curves denote their sum.
$ M \gtrsim M_* $ holes emit primary neutrinos only.
\label{fig:nu}
}
\end{figure}

Neutrinos may either be emitted directly as blackbody radiation (primaries) or they may result from the decay of emitted pions, leptons, neutrons, and antineutrons (secondaries).
As a result, their emission spectra are similar to those for photons up to a normalization factor.
The neutrino background can in principle constrain PBHs whose lifetime is more than the time of neutrino decoupling ($ \tau \gtrsim 1\,\mathrm s $).
This corresponds to a minimum mass of $ M_{\mathrm{min},\nu} \approx 10^9\,\mathrm g $.
However, the low-energy neutrinos which we have to use are poorly limited by observations.
In Super-Kamiokande (SK), the null detection of relic $ \bar\nu_e $'s implies $ \Phi_{\bar\nu_e} \le 1.2\,\mathrm{cm}^{-2}\,\mathrm s^{-1} $ above the threshold $ E_{\bar\nu_e 0} = 19.3\,\mathrm{MeV} $ \cite{Malek:2002ns}.
As seen in Fig.~\ref{fig:nu}, this energy corresponds to the high-energy tail for light PBHs.
The constraint associated with the relic neutrinos is shown by the dotted (blue) line in Fig.~\ref{fig:combined} and is much weaker than the BBN and photon background limits.
Although this is an original limit, similar calculations have been made by Bugaev and collaborators \cite{Bugaev:2002yt,Bugaev:2008gw}.
However, they assume that the PBHs have a continuous mass function and they link their model with a particular inflationary scenario.
They also consider the possibility that positrons from the same PBHs could explain the $ 511\,\mathrm{keV} $ annihilation line from center of the Galaxy, as first suggested by Okele and Rees \cite{1980A&A....81..263O}.
We do not show the associated constraint in Fig.~\ref{fig:combined} because it is very model dependent.

\subsection{
Other relics
\label{sec:LSPs}
}

Evaporating PBH should produce any other particles predicted in theories beyond the standard model.
The number of PBHs is therefore limited by both the abundance of stable massive particles \cite{Green:1999yh} or the decay of long-lived ones \cite{Khlopov:2004tn}.
In supersymmetry (SUSY) or supergravity (SUGRA), the lightest supersymmetric particle (LSP) is stable and becomes a candidate for the dark matter.
If LSPs are produced by the evaporation of the PBHs, in order not to exceed the observed CDM density at present, one obtains the upper bound \cite{Lemoine:2000sq}:
\begin{equation}
\beta'(M)
\lesssim
  10^{-18}\,\left(\frac{M}{10^{11}\,\mathrm g}\right)^{-1/2}\,
  \left(\frac{m_\mathrm{LSP}}{100\,\mathrm{GeV}}\right)^{-1}
\quad
(M < 10^{11}\,(m_\mathrm{LSP}/100\,\mathrm{GeV})^{-1}\,\mathrm g)\,.
\label{eq:stable}
\end{equation}
This constraint is shown in Fig.~\ref{fig:combined} but depends on the mass of the LSP and is therefore subject to considerable uncertainty \cite{Kohri:2005wn,Kawasaki:2008qe}.
In addition, unstable particles such as the gravitino or neutralino might be produced by evaporating PBHs.
The decay of these particles into lighter ones also affects BBN \cite{Kawasaki:2004yh,Kawasaki:2004qu} and this gives another constraint \cite{Khlopov:2004tn}:
\begin{equation}
\beta'(M)
\lesssim
  5 \times 10^{-19}\,
  \left(\frac{M}{10^{9}\,\mathrm g}\right)^{-1/2}\,
  \left(\frac{Y}{10^{-14}}\right)\,
  \left(\frac{x_\phi}{0.006}\right)^{-1}
\quad
(M < 10^{9}\,\mathrm g)\,,
\end{equation}
where $ Y $ is the limit on the number density to entropy density ratio and $ x_\phi $ is the fraction of the luminosity going into quasistable massive particles, both being normalized to reasonable values.
This limit is not shown explicitly in Fig.~\ref{fig:combined} but it has a similar form to Eq.~\eqref{eq:stable}.

\subsection{
Reionization and $ 21\,\mathrm{cm} $ signature
}

The 5-year WMAP results give the optical depth as $ \tau \sim 0.1 $ for CMB photons emitted from the last scattering surface.
On the other hand, recent observations of the Gunn--Peterson troughs and a $ \gamma $-ray burst around $ z \sim 6 $ imply that reionization of the Universe occurred at $ z \sim 6 $ \cite{Fan:2006dp}.
Thus PBHs cannot be so numerous that they lead to reionization earlier than $ z \sim 6 $.
In principle, this leads to a constraint on PBHs with 
\begin{equation}
M
\ge
  M_*\,
  \left(\frac{t_\mathrm{dec}}{t_0}\right)^{1/3}\,
  \left(\frac{f(M)}{f_*}\right)^{1/3}
\approx
  2 \times 10^{13}\,\mathrm g.
\end{equation}
An increase in the ionization of the intergalactic medium would also produce a $ 21\,\mathrm{cm} $ signature.
Mack and Wesley \cite{Mack:2008nv} have shown that future observations of $ 21\,\mathrm{cm} $ radiation from high redshift neutral hydrogen could place important constraints on PBHs in the mass range $ 5 \times 10^{13}\,\mathrm g < M < 10^{17}\,\mathrm g $.
This is essentially due to the coincidence that photons emitted from PBHs during $ 30 < z < 300 $ peak in the energy range in which the intergalactic medium has low optical depth.
Any process which heats the intergalactic medium in this period will produce a signal but the ionizing flux of photons and electrons and positrons from PBHs would generate a distinctive feature in the $ 21\,\mathrm{cm} $ brightness temperature.
PBHs with $ 5 \times 10^{13}\,\mathrm g < M < 10^{14}\,\mathrm g $ evaporate in $ 30 < z < 90 $ and would raise the $ 21\,\mathrm{cm} $ brightness temperature, thereby reducing the absorption seen against the CMB.
PBHs with $ M \sim 10^{14}\,\mathrm g $ would raise the spin temperature above the CMB, so that the $ 21\,\mathrm{cm} $ appears in emission rather than absorption.
PBHs with $ 10^{14}\,\mathrm g < M < 10^{17}\,\mathrm g $ would have a less pronounced effect.
The latter limit is shown in Fig.~8 of their paper and can be expressed in the form
\begin{equation}
\beta'(M)
< 3 \times 10^{-29}\,
  \left(\frac{M}{10^{14}\,\mathrm g}\right)^{7/2}
\quad
(M > 10^{14}\,\mathrm g)\,.
\end{equation}
It bottoms out at a mass of around $ 10^{14}\,\mathrm g $ and is well below the photon background limit.
The associated limits are shown in Fig.~\ref{fig:combined} but only as a broken curve since they are potential rather than actual.

\section{
Constraints on nonevaporating PBHs
\label{sec:large}
}

\begin{figure}[ht]
\begin{center}
\includegraphics{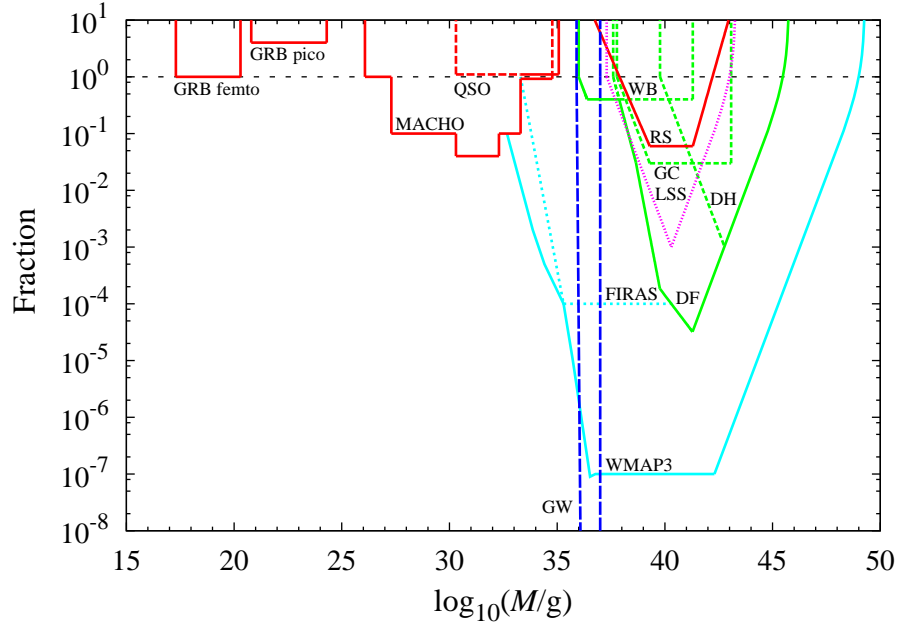}
\end{center}
\caption{
Constraints on $ f(M) $ for a variety of dynamical (green), lensing (red), and astrophysical (blue) effects associated with large PBHs, the dominant limit for each type of effect being shown as a solid line.
The effects are femtolensing and picolensing of gamma-ray bursts (GRB), microlensing of stars (MACHO) and quasars (QSO), millilensing of compact radio sources (RS), wide binary disruption (WB), globular cluster disruption (GC), dynamical friction (DF), disk heating (DH), generation of large-scale structure through Poisson fluctuations (LSS), accretion effects on the CMB (FIRAS, WMAP3), and gravitational waves (GW).
The lines on the right correspond to having one PBH in the relevant volume (Galactic halo, Lyman-$ \alpha $ forest, CMB anisotropy scale), the limits not applying below these lines.
\label{fig:large}
}
\end{figure}

For completeness, we now review the various constraints associated with PBHs which are too large to have evaporated by now.
We also include a discussion of Planck-mass relics, although these are not large---indeed they are the smallest conceivable objects in nature.
Many of the limits assume that PBHs cluster in the Galactic halo in the same way as other CDM particles.
In this case, Eq.~\eqref{eq:omega} implies that the fraction of the halo in PBHs is related to $ \beta'(M) $ by
\begin{equation}
f
\equiv
  \frac{\Omega_\mathrm{PBH}}{\Omega_\mathrm{CDM}}
\approx
  4.8\,\Omega_\mathrm{PBH}
= 4.11 \times 10^8\,\beta'(M)\,\left(\frac{M}{M_\odot}\right)^{-1/2}\,,
\label{eq:f}
\end{equation}
where we assume $ \Omega_\mathrm{CDM} = 0.21 $ and this $ f $ is to be distinguished from the one appearing in the previous sections.
Our limits on $ f(M) $ are summarized in Fig.~\ref{fig:large}.
Many of them have been described elsewhere, so we only discuss these briefly, although we update them where appropriate.
Further details can be found in the papers by Josan \textit{et al.} \cite{Josan:2009qn} (see their Table~1), Mack \textit{et al.} \cite{Mack:2006gz} (see their Fig.~4), and Ricotti \textit{et al.} \cite{Ricotti:2007au} (see their Fig.~9).
Note that some of the limits are extended into the $ f > 1 $ domain, although this is obviously excluded by the independent density constraint.
However, it is still useful to see where the limits are located since they could become stronger in the future as the observational data improve.

\subsection{
Lensing constraints
}

Microlensing observations of stars in the large and small magellanic clouds probe the fraction of the Galactic halo in MACHOs of a certain mass range \cite{Paczynski:1985jf}.
We assume that PBHs cluster in the same way as other CDM particles, so that Eq.~\eqref{eq:f} applies.
The optical depth of the halo, $ \tau_\mathrm L $\,, is defined as the probability that any given star is amplified by at least $ 1.34 $ at a given time.
Although the relation between $ \tau_\mathrm L $ and $ f $ depends on the halo model, the so-called S model \cite{Alcock:2000ph}, which is often adopted in the analysis, gives $ \tau^{(\mathrm{LMC})}_\mathrm L = 4.7 \times 10^{-7}\,f $ and $ \tau^{(\mathrm{SMC})}_\mathrm L = 1.4\,\tau^{(\mathrm{LMC})}_\mathrm L $\,.
Although the initial motivation for microlensing surveys was to search for brown dwarfs with $ 0.02\,M_\odot < M < 0.08\,M_\odot $\,, the possibility that the halo is dominated by these objects was soon ruled out by both the MACHO \cite{Alcock:2000ke} and EROS \cite{Tisserand:2006zx} surveys.
Instead MACHO observed $ 17 $ events apparently induced by compact objects with $ M \sim 0.5\,M_\odot $ and contributing about $ 20\,\% $ of the halo mass \cite{Alcock:2000ph}.
Recently similar claims have been made by the POINT-AGAPE collaboration, which detected $ 6 $ microlensing events in a survey of the Andromeda galaxy \cite{CalchiNovati:2005cd}.

This raises the possibility that some of the halo dark matter could be PBHs formed at the QCD phase transition \cite{Jedamzik:1996mr,Widerin:1998my,Jedamzik:1999am}.
However, recent results suggest that even a $ 20\,\% $ halo contribution of $ M \sim 0.5\,M_\odot $ PBHs is excluded \cite{Hamadache:2006fw}.
For example, the EROS-2 collaboration monitored brighter stars in a wider solid angle and obtained fewer events than expected.
They thus obtained more stringent constraints on $ f $ and argued that some of the MACHO events were due to self-lensing or halo clumpiness \cite{Tisserand:2006zx}.
Specifically, they excluded $ 6 \times 10^{-8}\,M_\odot < M < 15\,M_\odot $ MACHOs from dominating the halo;
combining the MACHO and EROS results \cite{Allsman:2000kg} extends the upper bound to $ 30\,M_\odot $\,.
Even stronger constraints are obtained for an intermediate mass range.
The constraints are summarized as follows:
\begin{equation}
f(M)
<
\begin{cases}
1
& (6 \times 10^{-8}\,M_\odot < M < 30\,M_\odot)\,, \\
0.1
& (10^{-6}\,M_\odot < M < 1\,M_\odot)\,, \\
0.04
& (10^{-3}\,M_\odot < M < 0.1\,M_\odot)\,,
\end{cases}
\end{equation}
The latest OGLE-II data \cite{Wyrzykowski:2009ep,Novati:2009kq} yield a constraint which is consistent with this but somewhat weaker.

Other lensing constraints come from femtolensing and picolensing of gamma-ray bursts \cite{Marani:1998sh}, assuming the bursts are at a redshift $ z \sim 1 $, which gives $ f < 1 $ for $ 10^{-16}\,M_\odot < M < 10^{-13}\,M_\odot $ and $ f < 4 $ for $ 10^{-12}\,M_\odot < M < 10^{-9}\,M_\odot $\,, respectively, microlensing of quasars, which gives $ f < 1 $ for $ 10^{-3}\,M_\odot < M < 60\,M_\odot $ \cite{1994ApJ...424..550D}, and millilensing of compact radio sources, which gives \cite{Wilkinson:2001vv}
\begin{equation}
f(M)
<
\begin{cases}
(M/2 \times 10^5\,M_\odot)^{-2}
& (M \lesssim 10^6\,M_\odot)\,, \\
0.06
& (10^6\,M_\odot \lesssim M \lesssim 10^8\,M_\odot)\,, \\
(M/4 \times 10^8\,M_\odot)^2
& (M \gtrsim 10^8\,M_\odot)\,.
\end{cases}
\end{equation}
These constraints have been summarized by Josan \textit{et al.} \cite{Josan:2009qn} and further relevant references can be found in their paper.
Note that Hawkins \cite{1993Natur.366..242H} once claimed evidence for a critical density of jupiter-mass objects from observations of quasar microlensing and associated these with PBHs formed at the quark-hadron transition.
However, the status of his observations is no longer clear, so this is not included in Fig.~\ref{fig:large}.

\subsection{
Dynamical constraints
}

Binary star systems with wide separation are vulnerable to disruption from encounters with PBHs or any other type of MACHO \cite{1985ApJ...290...15B,1987ApJ...312..367W}.
Observations of wide binaries in the Galaxy therefore constrain the abundance of halo PBHs.
By comparing the result of simulations with observations, Yoo \textit{et al.} \cite{Yoo:2003fr} essentially ruled out MACHOs with $ M > 43\,M_\odot $ from providing most of the halo mass.
However, Quinn \textit{et al.} \cite{Quinn:2009zg} have recently performed a more careful analysis of the radial velocities of these binaries and found that one of the widest-separation ones is spurious.
The resulting constraint now reads
\begin{equation}
f(M)
<
\begin{cases}
(M/400\,M_\odot)^{-1}
& (400\,M_\odot < M \lesssim 10^3\,M_\odot )\,, \\
0.4
& (10^3\,M_\odot \lesssim M < 10^8\,M_\odot)\,.
\end{cases}
\end{equation}
Other dynamical constraints have been studied by Carr and Sakellariadou \cite{Carr:1997cn}.
An argument similar to the binary disruption one shows that the survival of globular clusters against tidal disruption by passing PBHs gives a limit
\begin{equation}
f(M)
<
\begin{cases}
(M/3 \times 10^4\,M_\odot)^{-1}
& (3 \times 10^4\,M_\odot < M < 10^6\,M_\odot )\,, \\
0.03
& (10^6\,M_\odot < M < 6 \times 10^9\,M_\odot)\,,
\end{cases}
\end{equation}
although this depends sensitively on the mass and radius of the cluster.
The limit cuts off above $ 6 \times 10^9\,M_\odot $ because the encounter becomes nonimpulsive.
The upper limit of $ 3 \times 10^4\,M_\odot $ on the mass of objects dominating the halo is consistent with the numerical calculations of Moore \cite{1993ApJ...413L..93M}.

Other dynamical limits come into play at higher mass scales.
Halo objects will overheat the stars in the Galactic disc unless one has
\begin{equation}
f(M)
< \left(\frac{M}{3 \times 10^6\,M_\odot}\right)^{-1}\,.
\end{equation}
The upper limit of $ 3 \times 10^6\,M_\odot $ agrees with earlier calculations by Lacey and Ostriker \cite{1985ApJ...299..633L}, although they argued that black holes with $ 2 \times 10^6\,M_\odot $ could \emph{explain} some features of disc heating.
Positive evidence from a qualitatively similar effect may come from the recent claim of Totani \cite{Totani:2009af} that elliptical galaxies are puffed up by dark halo objects of $ 10^5\,M_\odot $\,.
Another limit arises because halo objects will be dragged into the nucleus of the Galaxy by the dynamical friction of the spheroid stars and halo objects themselves (if they have an extended mass function), this leading to excessive nuclear mass unless \cite{Carr:1997cn}
\begin{equation}
f(M)
<
\begin{cases}
(M/2 \times 10^4\,M_\odot)^{-10/7}\,(r_\mathrm c/2\,\mathrm{kpc})^2
& (M < 5 \times 10^5\,M_\odot)\,, \\
(M/4 \times 10^4\,M_\odot)^{-2}\,(r_\mathrm c/2\,\mathrm{kpc})^2
& (5 \times 10^5\,M_\odot \ll M < 2 \times 10^6\,(r_\mathrm c/2\,\mathrm{kpc})\,M_\odot)\,, \\
(M/0.1\,M_\odot)^{-1/2}
& (M \gg 2 \times 10^6\,(r_\mathrm c/2\,\mathrm{kpc})\,M_\odot)\,.
\end{cases}
\end{equation}
The different mass regimes correspond to the drag being dominated by spheroid stars (low $ M $), by halo objects (high $ M $) and by some combination of the two (intermediate $ M $).
This limit is sensitive to the halo core radius $ r_\mathrm c $ but the dominant constituent of the halo must be smaller than $ 2 \times 10^4\,(r_\mathrm c/2\,\mathrm{kpc})^{1.4}\,M_\odot $\,.
However, there is a caveat here in that holes drifting into the nucleus might be ejected by the slingshot mechanism if there is already a binary black hole there \cite{Hut:1992iy}.
Note that all these dynamical limits are only meaningful if there is at least one PBH per Galactic halo, corresponding to 
\begin{equation}
f(M)
> \left(\frac{M}{M_\mathrm{halo}}\right),
\quad
M_\mathrm{halo}
\approx
  3 \times 10^{12}\,M_\odot\,,
\end{equation}
so none of them applies \emph{below} this line.
This is what Carr and Sakellariadou term the ``incredulity limit'' \cite{Carr:1997cn}.
The excluded region is therefore the upper right triangle in Fig.~\ref{fig:large}.

The effects of PBH collisions on astronomical objects---including the Earth \cite{1973Natur.245...88J}---have been a subject of long-standing interest \cite{Carr:1997cn}.
For example, Zhilyaev \cite{Zhilyaev:2007rx} has suggested that collisions with stars could produce $ \gamma $-ray bursts and Khriplovich \textit{et al.} \cite{Khriplovich:2008er} have examined whether terrestrial collisions could be detected acoustically.
Recently, Roncadelli \textit{et al.} \cite{Roncadelli:2009qj} have suggested that halo PBHs could be captured and swallowed by stars in the Galactic disk.
The stars would eventually be accreted by the holes, producing a lot of radiation and a population of subsolar black holes which could not be of more conventional (nonprimordial) origin.
They infer a constraint $ f < (M/3 \times 10^{26}\,\mathrm g) $\,, which would give a \emph{lower} limit on the mass of any objects dominating the halo.
However, a careful analysis of the collisions of PBHs with main-sequence stars, red giant cores, white dwarfs, and neutron stars by Abramowicz \textit{et al.} \cite{Abramowicz:2008df} suggests that these effects are never important.
Collisions are either too rare (for $ M > 10^{20}\,\mathrm g $) or produce too little power to be detected (for $ M < 10^{20}\,\mathrm g $).
Although captures of the kind envisaged by Roncadelli \textit{et al.} could occur for $ M > 10^{28}\,\mathrm g $, this range is already excluded by the microlensing constraints.
We therefore do not show any collisional constraints in Fig.~\ref{fig:large}.

Each of these dynamical constraints is subject to certain provisos but it is interesting that they all correspond to an upper limit on the mass of the objects which dominate the halo in the range $ 500\text{--}2 \times 10^4\,M_\odot $\,, the binary disruption limit being the strongest.
This is particularly relevant for constraining models, such as that proposed by Frampton \textit{et al.} \cite{Frampton:2010sw}, in which the dark matter is postulated to comprise PBHs in this mass range.
However, apart from the Galactic disc and elliptical heating arguments of Refs.~\cite{1985ApJ...299..633L,Totani:2009af}, it must be stressed that none of these dynamical effects gives \emph{positive} evidence for MACHOs.
Furthermore, none of them requires the MACHOs to be PBHs.
Indeed, they could equally well be clusters of smaller objects \cite{1987ApJ...316...23C};
this is particularly pertinent in light of the claim by Dokuchaev \textit{et al.} \cite{Dokuchaev:2004kr} and Chisholm \cite{Chisholm:2005vm} that PBHs could form in tight clusters, giving an overdensity well in excess of that provided by the halo concentration alone.

\subsection{
Large-scale structure constraints
}

Sufficiently large PBHs could have important consequences for large-scale structure formation because of the Poisson fluctuations in their number density.
This was first pointed out by M\'esz\'aros \cite{Meszaros:1975ef}, although he overestimated the magnitude of the effect \cite{1977A&A....56..377C}.
Subsequently, various authors have invoked this effect \cite{1983ApJ...275..405F,1983ApJ...268....1C,Afshordi:2003zb,Mack:2006gz}.
In particular, Afshordi \textit{et al.} \cite{Afshordi:2003zb} use observations of the Lyman-$ \alpha $ forest to obtain an upper limit of about $ 10^4\,M_\odot $ on the mass of any PBHs which provide the CDM.
Although this conclusion is based on detailed simulations, one can understand their result heuristically and extend it to the case where the PBHs only provide a fraction $ f(M) $ of the dark matter by noting that the initial density perturbation associated with the Poisson fluctuations in the number of PBHs on a mass scale $ M_{\mathrm{Ly}\alpha} $ is
\begin{equation}
\delta(M)
\sim
  f(M)\,\left(\frac{f(M)\,M_{\mathrm{Ly}\alpha}}{M}\right)^{-1/2}
\sim
  10^{-5}\,
  f(M)^{1/2}\,\left(\frac {M}{M_\odot}\right)^{1/2}\,
  \left(\frac{M_{\mathrm{Ly}\alpha}}{10^{10}\,M_\odot}\right)^{-1/2}\,.
\end{equation}
Since these fluctuations can grow between the redshift of CDM domination ($ z_\mathrm{eq} \sim 4000 $) and the redshift at which Lyman-$ \alpha $ clouds are observed ($ z_{\mathrm{Ly}\alpha} \sim 4 $) by a factor $ z_\mathrm{eq}/z_{\mathrm{Ly}\alpha} \sim 10^3 $, imposing the condition that $ \delta(M) $ has not grown to more than order unity (as required for Lyman-$ \alpha $ clouds) leads to the constraint shown in Fig.~\ref{fig:large}:
\begin{equation}
f(M)
< \left(\frac{M}{10^4\,M_\odot}\right)^{-1}\,.
\label{eq:cluster}
\end{equation}
This form of the statistical clustering limit was also derived in Ref.~\cite{1977A&A....56..377C} and is shown in Fig.~5 of Ref.~\cite{Mack:2006gz}.
Note that this limit requires at least one PBH per Lyman-$\alpha$ mass, which corresponds to
\begin{equation}
f(M)
> \left(\frac{M}{M_{\mathrm{Ly}\alpha}}\right)\,,
\quad
M_{\mathrm{Ly}\alpha}
\approx
  10^{10}\,M_\odot\,.
\end{equation}
The data from SDSS are now more extensive \cite{McDonald:2004eu}, thereby reducing the PBH contribution to the fluctuations (i.e.\ reducing the value of $ \delta $ at $ z = z_{\mathrm{Ly}\alpha} $).
Indeed, the limiting mass might be reduced to $ 10^3\,M_\odot $ \cite{Gonzales:2009}, which is comparable to the strongest dynamical limits discussed above, although we do not use this in Fig.~\ref{fig:large}.
A similar effect can allow clusters of large PBHs to evolve into the supermassive black holes thought to reside in galactic nuclei \cite{1984MNRAS.206..801C,Duechting:2004dk,Khlopov:2004sc};
if one replaces $ M_{\mathrm{Ly}\alpha} $ with $ 10^8\,M_\odot $ and $ z_{\mathrm{Ly}\alpha} $ with $ 10 $ in the above analysis, one reduces the limiting mass in Eq.~\eqref{eq:cluster} to $ 600\,M_\odot $\,.
However, this limit is not shown in Fig.~\ref{fig:combined}.

\subsection{
Accretion constraints
}

There are good reasons for believing that PBHs cannot grow very much during the radiation-dominated era.
Although a simple Bondi-type argument suggests that they could grow as fast as the horizon \cite{1966AZh....43..758Z,*1967SvA....10..602Z}, this does not account for the background cosmological expansion and a fully relativistic calculation shows that such self-similar growth is impossible \cite{Carr:1974nx,1978ApJ...219.1043B,1978ApJ...225..237B}.
Consequently there is very little growth at all.
The only exception might be if the Universe were dominated by a ``dark energy'' fluid with $ p < -(1/3)\,\rho\,c^2 $\,, as in the quintessence scenario, since self-similar black hole solutions do exist in this situation \cite{Harada:2007tj,Maeda:2007tk,Carr:2010wk}.
This may support the claim of Bean and Magueijo \cite{Bean:2002kx} that intermediate mass PBHs might accrete quintessence efficiently enough to evolve into the supermassive black holes in galactic nuclei.

Even if PBHs cannot accrete appreciably in the radiation-dominated era before decoupling, massive ones might still do so in the period after decoupling and the Bondi-type analysis \emph{should} then apply.
The associated accretion and emission of radiation could have a profound effect on the thermal history of the Universe, as first analyzed by Carr \cite{1981MNRAS.194..639C}.
Recently this possibility has been investigated in great detail by Ricotti \textit{et al.} \cite{Ricotti:2007au}, who study the effect of such PBHs on the ionization and temperature evolution of the Universe.
They point out that the x rays emitted by gas accretion onto the black holes could produce measurable effects in the spectrum and anisotropies of the CMB.
Using FIRAS data to constrain the first and WMAP data to constrain the second, they improve the constraints on $ f(M) $ by several orders of magnitude for $ M > 0.1\,M_\odot $\,.
These limits are shown in Fig.~9 of Ref.~\cite{Ricotti:2007au}, although their precise form clearly depends on details of the accretion model.
They should also be updated in principle to comply with the WMAP 7-year results.
Note that this limit requires at least one PBH on the scale associated with the relevant CMB anisotropies; for $ l=100 $ modes, this corresponds to 
\begin{equation}
f(M)
> \left(\frac{M}{M_{l=100}}\right)\,,
\quad
M_{l=100}
\approx
  10^{16}\,M_\odot\,.
\end{equation}
In related work, Mack \textit{et al.} \cite{Mack:2006gz} have considered the growth of large PBHs through the capture of dark matter halos and suggested that their accretion could give rise to ultraluminous x-ray sources.
The latter possibility has also been explored by Kawaguchi \textit{et al.} \cite{Kawaguchi:2007fz}.

\subsection{
Gravitational-wave constraints
}

A population of massive PBHs would be expected to generate a background of gravitational waves \cite{1980A&A....89....6C}.
This would be especially interesting if there were a population of binary halo PBHs coalescing at the present epoch due to gravitational radiation losses \cite{Nakamura:1997sm,Ioka:1998gf}.
If such waves were detected, it would provide a unique probe of the halo distribution \cite{Inoue:2003di}.
The LIGO data already place weak constraints on such scenarios \cite{Abbott:2006zx}.
However, a different type of gravitational-wave constraint on $ \beta(M) $ has recently been pointed out by Saito and Yokoyama \cite{Saito:2008jc,Saito:2009jt}.
This is because the second-order tensor perturbations generated by the scalar perturbations which produce the PBHs are surprisingly large.
The associated frequency is around $ 10^{-8}\,(M/10^3\,M_\odot)\,\mathrm{Hz} $, while the limit on $ \beta(M) $ just relates to the amplitude of the density fluctuations at the horizon epoch and is formally of order $ 10^{-52} $.
The limit from pulsar timing data is shown in Fig.~\ref{fig:large} and already excludes PBHs with $ 10^2\,M_\odot < M < 10^4\,M_\odot $ from providing an appreciable amount of dark matter.
The potential limits from LIGO, VIRGO, and BBO are not shown explicitly but cover the mass range down to $ 10^{20}\,\mathrm g $.
This effect has also been noted by several other authors \cite{Assadullahi:2009jc,Bugaev:2009zh}.
These limits apply if and only if the PBHs are generated by super-Hubble scale fluctuations, like the ones created during inflation.
We also note that the limiting value of $ \beta $ depends on the fluctuations
being Gaussian.
Although this is questionable in the context of the large-amplitude fluctuations relevant to PBH formation, the studies in Refs.~\cite{Hidalgo:2007vk,Saito:2008em,Hidalgo:2009fp}
show that non-Gaussian effects are not expected to be large.
Gravitational-wave detectors in space, like LISA, might also constrain PBHs \emph{indirectly}.
This is because LISA could detect isolated PBHs in the mass range $ 10^{14}\text{--}10^{20}\,\mathrm g $ by measuring the gravitational impulse induced by any nearby passing one \cite{Adams:2004pk,Seto:2004zu}.

\subsection{
Planck-mass relic constraints
}

If PBH evaporations leave stable Planck-mass relics, these might also contribute to the dark matter.
This was first pointed out by MacGibbon \cite{MacGibbon:1987my} and has subsequently been explored in the context of inflationary scenarios by several authors \cite{Barrow:1992hq,Carr:1994ar,Green:1997sz,Alexeyev:2002tg,Chen:2002tu,Barrau:2003xp,Chen:2004ft,Nozari:2005ah}.
If the relics have a mass $ \kappa\,M_\mathrm{Pl} $\,, where $ M_\mathrm{Pl} $ is the Planck mass, and reheating occurs at a temperature $ T_\mathrm R $\,, then the requirement that they have less than the critical density implies \cite{Carr:1994ar}
\begin{equation}
\beta'(M)
< 8 \times 10^{-28}\,\kappa^{-1}\,\left(\frac{M}{M_\mathrm{Pl}}\right)^{3/2}
\label{eq:betarelic}
\end{equation}
for the mass range
\begin{equation}
\left(\frac{T_\mathrm R}{T_\mathrm{Pl}}\right)^{-2}
< \frac{M}{M_\mathrm{Pl}}
< 10^{11}\,\kappa^{2/5}\,.
\label{eq:relic}
\end{equation}
Note that we would now require the density to be less than $ \Omega_\mathrm{CDM} \approx 0.25 $, which strengthens the limit by a factor of $ 4 $.
The lower mass limit arises because PBHs generated before reheating are diluted exponentially.
The upper mass limit arises because PBHs larger than this dominate the total density before they evaporate, in which case the final cosmological photon-to-baryon ratio is determined by the baryon asymmetry associated with their emission.
Recently Alexander and M\'esz\'aros \cite{Alexander:2007gj} have advocated an extended inflationary scenario in which evaporating PBHs naturally generate the dark matter, the entropy, and the baryon asymmetry of the Universe.
This triple coincidence applies providing inflation ends at $ t \sim 10^{-23}\,\mathrm s $, so that the PBHs have an initial mass $ M \sim 10^6\,\mathrm g $.
This just corresponds to the upper limit indicated in Eq.~\eqref{eq:relic}, which explains one of the coincidences.
The other coincidence involves the baryon asymmetry generated in the evaporations.
It should be stressed that the limit \eqref{eq:betarelic} still applies even if there is no inflationary period but then extends all the way down to the Planck mass.

\section{
Conclusions
\label{sec:conc}
}

\begin{figure}[ht]
\begin{center}
\includegraphics{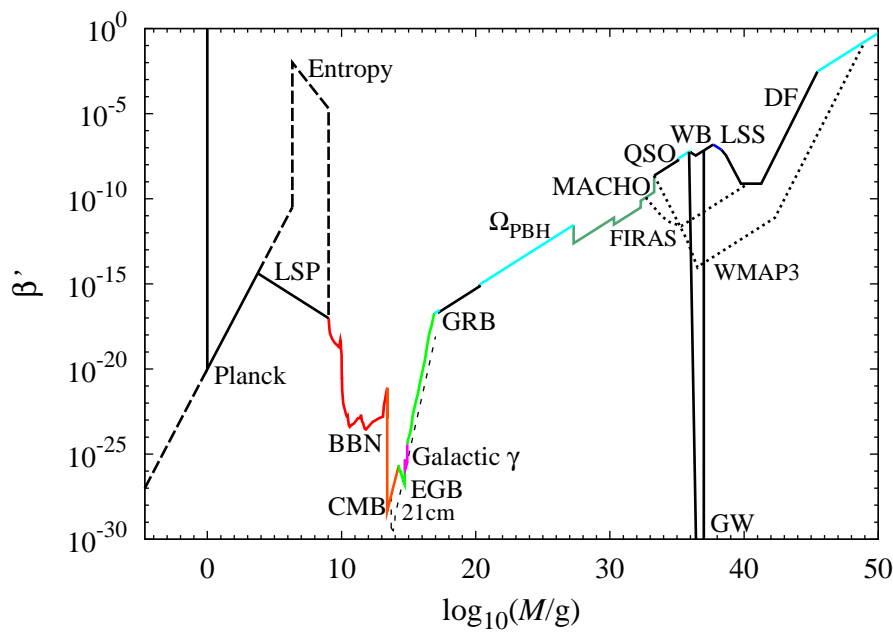}
\end{center}
\caption{
Master $ \beta'(M) $ constraints diagram for the mass range $ 1\text{--}10^{50}\,\mathrm g $, the acronyms being specified in the caption to Fig.~\ref{fig:large}.
\label{fig:master}
}
\end{figure}

All the limits considered in this paper are brought together in a master $ \beta'(M) $ diagram in Fig.~\ref{fig:master}.
In particular, the constraints on $ f(M) $ discussed in the previous section have been converted into limits on $ \beta'(M) $ using Eq.~\eqref{eq:f}.
We also include the relic limit associated with Eq.~\eqref{eq:betarelic}---with the broken line to the left applying if there is no inflation---and the entropy limit associated with Eq.~\eqref{eq:entropy}.
The latter is also shown broken since it is much weaker than the LSP constraint, albeit more secure.
Most of the limits are associated with various caveats, but where they are reasonably firm, only the dominant one is indicated for each value of $ M $\,.
Nevertheless, we include several overlapping ones at high masses.
Figure~\ref{fig:master} covers the entire mass range from $ 1\text{--}10^{50}\,\mathrm g $ and involves a wide variety of physical effects.
This reflects the fact that PBHs provide a unique probe of the early Universe, gravitational collapse, high-energy physics, and quantum gravity.
In particular, they can probe scales and epochs inaccessible by any other type of cosmological observation.

Although none of the effects discussed in this paper provides positive evidence for PBHs, Fig.~\ref{fig:master} illustrates that even the nondetection of PBHs allows one to infer important constraints on the early Universe.
In particular, the limits on $ \beta (M) $ can be used to constrain all the PBH formation mechanisms described in Sec.~\ref{sec:intro}.
Thus, for example, they constrain models involving inflation, a dustlike phase, and the collapse of cosmic strings or domain walls.
They also restrict the form of the primordial inhomogeneities (whatever their source) and their possible non-Gaussianity.
Finally, they constrain less conventional models, such as those involving a variable gravitational constant or extra dimensions.
However, it must be emphasized that the form of the $ \beta(M) $ limits may itself change in such models, so it is not just a matter of applying the form of the limits derived in this paper directly.
These issues are too broad to address here but they provide much scope for future work.

\begin{acknowledgments}
We are grateful to Jens Chluba, Masahiro Kawasaki, Jane MacGibbon, Pat McDonald, Takeo Moroi, Toshikazu Shigeyama, and Alexei Starobinsky for helpful input.
B.J.C.\ thanks the Research Center for the Early Universe (RESCEU), University of Tokyo, and the Canadian Institute for Theoretical Astrophysics (CITA), University of Toronto, for hospitality received during this work.
He also acknowledges support from an STFC award.
The work of K.K.\ is supported in part by an STFC grant and Grant-in-Aid for Scientific Research on Priority Areas No.~18071001.
Y.S.\ is supported in part by Grant-in-Aid for JSPS Fellows No.~19-7852.
J.Y.\ was supported in part by JSPS Grant-in-Aid for Scientific Research No.~19340054 and the Grant-in-Aid for Scientific Research on Innovative Areas No.~21111006.
This work has also benefited from exchange visits supported by a Royal Society and JSPS bilateral grant.
\end{acknowledgments}

\appendix*

\section{
Link between mass tail, energy tail, and nonmonochromaticity of PBH mass function
}

The high-energy tail generated by PBH evaporations has been studied in detail by MacGibbon \cite{MacGibbon:1991tj}, so we only summarize the argument briefly here.
We confine attention to the photon tail for simplicity.
The form of this tail can be derived by integrating Eq.~\eqref{eq:bb} over time (i.e.\ $ M $) for fixed $ E $\,:
\begin{equation}
\frac{\mathrm dN_\gamma}{\mathrm dE}
= \int_{M_\mathrm i}^0\!
  \frac{\mathrm d\dot N_\gamma}{\mathrm dE}\,
  \frac{\mathrm dt}{\mathrm dM}\,
  \mathrm dM 
\propto
  E^2\,
  \int_0^{M_\mathrm i}\!
  \frac{\sigma_s(E,M)\,M^2\,\mathrm dM}{e^{E\,M} - 1}\,,
\label{eq:energytail}
\end{equation}
where we have used the relation $ \mathrm dM/\mathrm dt \propto M^{-2} $\,, $ \sigma_s(E,M) $ is the absorption cross section appearing in Eq.~\eqref{eq:grey}, which scales as $ M^2 $ for $ E > M^{-1} $ and $ E^2\,M^4 $ for $ E < M^{-1} $ \cite{Page:1976df}, $ M_\mathrm i $ is the initial mass of the hole and (for the current discussion) $ M $ is the evolving mass.
We are here neglecting redshift effects, because most emission occurs at fixed redshift for a given value of $ M_\mathrm i $\,.
However, this could be accounted for by replacing $ E $ by the \emph{present} photon energy $ E_0 = E\,(1+z_{\mathrm{evap}})^{-1} $\,.
We also drop the suffix $ \gamma $ on $ E $\,.

Let us first assume $ M_\mathrm i > M_\mathrm q \approx 0.4 M_* $\,, so that secondary emission is initially unimportant.
For $ E < M_\mathrm i^{-1} $\,, the mass integral just involves the Rayleigh--Jeans part of the spectrum, so it is dominated by the upper limit and scales as $ M_\mathrm i^6\,E $\,, leading to $ \mathrm dN_\gamma/\mathrm dE \propto E^3 $\,.
For $ E > M_\mathrm i^{-1} $\,, the exponential term cuts the integral off above a mass $ M \sim E^{-1} $\,, so the integral scales as $ E^{-5} $\,, leading to $ \mathrm dN_\gamma/\mathrm dE \propto E^{-3} $\,.
The time-integrated spectrum of photons from a PBH with $ M_\mathrm i > M_\mathrm q $ can therefore be expressed as
\begin{equation}
\frac{\mathrm dN_\gamma}{\mathrm dE}
\propto
\begin{cases}
E^3 & (E < M_\mathrm i^{-1})\,, \\ 
E^{-3} & (M_\mathrm i^{-1} < E < \Lambda_\mathrm{QCD})\,.
\end{cases}
\label{eq:int}
\end{equation}
Once secondary emission becomes important, as is always the case for $ M_\mathrm i < M_\mathrm q $\,, the analysis of MacGibbon and Webber \cite{MacGibbon:1990zk} shows that 
\begin{equation}
\frac{\mathrm d\dot N_\gamma}{\mathrm dE}
\propto
\begin{cases}
E^{-1} & (m_\pi < E < M^{-1})\,, \\ 
E^2\,M^2\,e^{-E\,M} & (E > M^{-1})\,.
\end{cases}
\end{equation}
The form around the peak at $ m_\pi $ reflects the low-energy fragmentation function, which is roughly Gaussian, although we do not give this explicitly.
One now uses this expression in Eq.~\eqref{eq:energytail} and integrates over $ M $\,.
For $ E < M_\mathrm i^{-1} $\,, the integral is dominated by the emission when $ M \sim M_\mathrm i $\,, so $ \mathrm dN_\gamma/\mathrm dE $ scales as $ E^{-1} $\,.
For $ E > M_\mathrm i^{-1} $\,, the integral is dominated by holes with $ M \sim E^{-1} $\,, so $ \mathrm dN_\gamma/\mathrm dE $ scales as $ E^{-3} $\,.
Therefore the $ E^{-3} $ tail given by Eq.~\eqref{eq:int} simply extends into the $ E > \Lambda_\mathrm{QCD} $ regime for $ M_\mathrm i > M_\mathrm q $\,.
However, for $ M_\mathrm i < M_\mathrm q $\,, we obtain
\begin{equation}
\frac{\mathrm dN_\gamma}{\mathrm dE}
\propto
\begin{cases}
E^{-1} & (M_\mathrm i^{-1} > E > \Lambda_\mathrm{QCD})\,, \\ 
E^{-3} & (E > M_\mathrm i^{-1})\,,
\end{cases}
\end{equation}
where the form around $ m_\pi $ is not shown explicitly but again reflects the low-energy fragmentation function.
The important qualitative point is that one has the same high-energy $ E^{-3} $ tail as before, although there is now an intermediate $ E^{-1} $ regime.

\begin{figure}[ht]
\begin{center}
\includegraphics[scale=0.6]{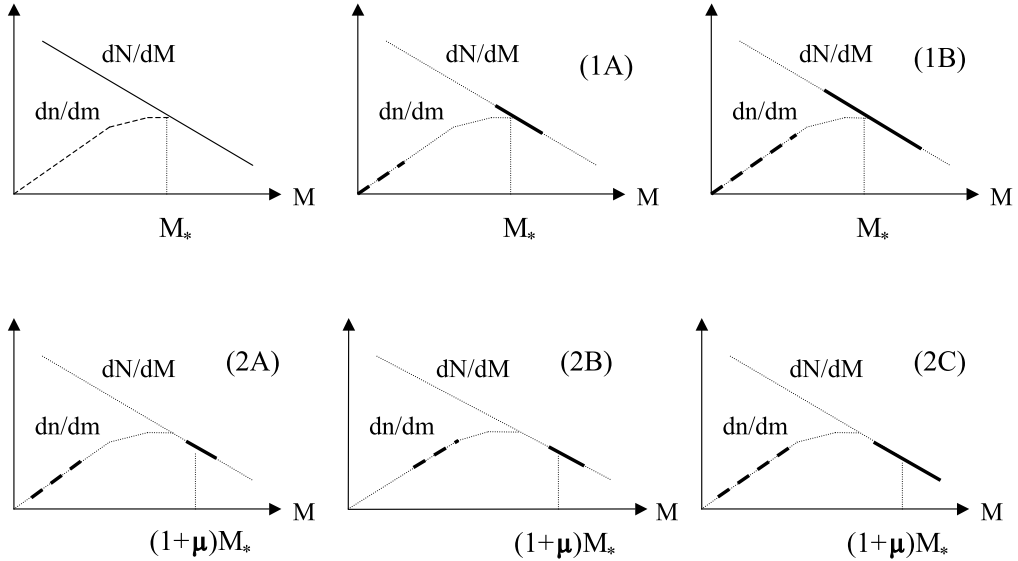}
\vspace{-3cm}
\end{center}
\caption{
This shows the relationship between the PBH mass functions at formation and currently.
For a nearly monochromatic initial mass function centered at $ M_* $\,, the extent of the current low-mass tail depends on the initial mass width.
It covers only low masses if $ \Delta M \ll M_* $ [case (1A)] but the entire range from $ 0 $ to $ M_* $ if $ \Delta M \sim M_* $ [case (1B)].
For a nearly monochromatic mass function centered above $ M_* $\,, we have either the low end of the tail [case (2A)], part of the upper end [case (2B)] or the entire tail [case (2C)].
\label{massfunction}
}
\end{figure}

There is also a contribution to the photon background above $ 100\,\mathrm{MeV} $ from the current remnants of PBHs which were initially slightly larger than $ M_* $\,.
This is distinct from the high-energy tail discussed above, although (as we will see) the two contributions are related.
This can be understood as a consequence of Eq.~\eqref{eq:mu}, since this implies that the \emph{current} mass function ($ \mathrm dn/\mathrm dm $) is related to the \emph{formation} mass function ($ \mathrm dN/\mathrm dM $) by
\begin{equation}
\frac{\mathrm dn}{\mathrm dm}
= \left(\frac{m}{M_*}\right)^2\,
  \left(\frac{1}{1+\mu(m)}\right)^2\,
  \left(\frac{\mathrm dN}{\mathrm dM}\right)
\approx
  \left(\frac{m}{M_*}\right)^2\,
  \left(\frac{\mathrm dN}{\mathrm dM}\right)_*
\quad
(m \ll M_*)\,,
\label{eq:spectrum}
\end{equation}
both mass functions being comoving.
The first expression is exact, with $ \mu(m) $ being implicitly determined by Eq.~\eqref{eq:mu}, while the second expression applies for $ \mu \ll 1 $.
In the latter case, $ m \approx (3\,\mu)^{1/3}\,M_* $ and the integrated comoving number density of holes with mass between $ 0 $ and $ m $ can be approximated by 
\begin{equation}
n(<m)
\approx
  \frac{1}{3}\,
  \left(\frac{m}{M_*}\right)^3\,
  M_*\,
  \left(\frac{\mathrm dN}{\mathrm dM}\right)_*
= \frac{1}{3}\,
  \left(\frac{m}{M_*}\right)^3\,
  n_\mathrm{PBH}(M_*)
\quad
(m \ll M_*)\,,
\label{current}
\end{equation}
where we have used Eq.~\eqref{eq:ndef}.
A correction would be required if the spectrum did not extend all the way down to zero mass (i.e.\ if the initial spectrum did not extend all the way down to $ M_* $).
We describe this low-mass part of the present spectrum as the ``mass tail.''
Note that $ \mathrm dn/\mathrm dm \approx \mathrm dN/\mathrm dM $ for $ \mu \gg 1 $ and this reflects the fact that $ m \approx M $ once $ \mu $ goes much above $ 1 $.
[For example, Eq.~\eqref{eq:mu} implies the ratio $ m/M $ is $ (26/27)^{1/3} $ for $ \mu = 2 $ and $ (63/64)^{1/3} $ for $ \mu = 3 $.]
There is no simple analytic expression for $ n(m) $ in the intermediate case ($ \mu \sim 1 $) but the overall current mass function can be approximated by 
\begin{equation}
\frac{\mathrm dn}{\mathrm dm}
= \mathrm{min}
  \biggl[
   \left(\frac{m}{M_*}\right)^2\,\left(\frac{\mathrm dN}{\mathrm dM}\right)_*,
   \left(\frac{\mathrm dN}{\mathrm dM}\right)
  \biggr]\,.
\end{equation}
In a more precise calculation, Eqs.~\eqref{eq:spectrum} and \eqref{eq:mu} imply that the slope of the mass function decreases as $ m $ approaches $ M_* $\,, with the exponent of $ m $ being $ 2/(1+\mu)^3 $\,.
The relationship between $ \mathrm dn/\mathrm dm $ and $ \mathrm dN/\mathrm dM $ is represented qualitatively in Fig.~\ref{massfunction}, although it should be noted that this assumes $ \mathrm dN/\mathrm dM $ is a decreasing function of $ M $\,.

It should be stressed that the low-mass tail would not be present if the formation mass function were \emph{precisely} monochromatic.
For if all the PBHs had exactly the mass $ M_* $\,, they would all evaporate at exactly the same time.
Indeed, the width of the mass function determines how much of the mass tail is present.
In order to understand these nonmonochromaticity effects, we must distinguish between two different situations: (1) a ``nearly monochromatic'' function centered at $ M_* $ but with a finite width $ \mu\,M_* $\,, so that it extends up to $ (1+\mu)\,M_* $\,; and (2) a ``nearly monochromatic'' function centered at $ (1+\mu)\,M_* $ but with a finite width $ \nu\,M_* $ (where $ \mu $ may be different from $ \nu $).
In the first situation, there are two subcases: for $ \mu \ll 1 $ [case (1A)], we only have the lower part of the $ m^2 $ tail (i.e.\ the spectrum does not extend all the way up to $ M_* $); for $ \mu \sim 1 $ [case (1B)], we have the entire tail.
In the second situation, there are three subcases: for $ \mu \ll 1 $ and $ \nu \ll 1 $ [case (2A)], we only have part of the lower mass tail; for $ \mu \sim 1 $ and $ \nu \ll 1 $ [case (2B)], we only have part of the tail close to $ M_* $; for $ \nu > \mathrm{max}(1,\mu) $ [case (2C)], we have the entire tail.
Clearly cases (2A) and (2B) are most consistent with the rationale of the present analysis, which is to assume a monochromatic mass function.
The different possible forms of the current mass function $ \mathrm dn/\mathrm dm $ in these subcases are illustrated in Fig.~\ref{massfunction}.

Since the photon production rate of an individual hole is $ \dot N_\gamma \propto m^{-1} $\,, the instantaneous flux from the tail population is $ I \propto n(m)\,m^{-1} \propto m^2 \propto E^{-2} $\,, which is reminiscent of the high-energy tail from the PBHs with $ M \le M_* $\,.
However, the connection between the mass tail for $ M \ge M_* $ and energy tail for $ M \le M_* $ is a subtle one, which requires some clarification.
All PBHs generate an $ E^{-3} $ energy tail \emph{eventually} but only those with $ M \le M_* $ produce one by the present epoch and PBHs with mass slightly above $ M_* $ do not produce the \emph{entire} energy tail because they have still not completed their evaporation (i.e.\ the highest-energy part is missing).
But it is precisely these unevaporated remnants which provide the mass tail, so the energy and mass tails are complementary in the sense that the latter becomes important just as the former becomes unimportant.

Finally, it should be stressed that the definition of $ \beta(M) $ requires some care for an extended mass function, the definition given by Eq.~\eqref{eq:beta} only applying in the monochromatic case.
One needs to decide whether to integrate $ \mathrm dN/\mathrm dM $ over the entire mass band or over smaller ranges within this band.
While the second definition is more natural for $ \mu \gg 1 $, the first one is more natural for $ \mu \ll 1 $ and this is what we assume in this paper.
In discussing the value of $ \beta(M) $ for $ M \sim M_* $\,, one must also distinguish between the width of the mass band at formation, $ \Delta M $\,, and its width at the current epoch, $ \Delta m $\,.
For a narrow mass function, the PBH collapse fraction at formation is
\begin{equation}
\beta(M_*)
\propto
  \left(\frac{\mathrm dN}{\mathrm dM}\right)_*\,
  \Delta M
= \left(\frac{\mathrm dn}{\mathrm dm}\right)\,
  \Delta m\,.
\end{equation}
Although $ \mathrm dn/\mathrm dm $ is less than $ \mathrm dN/\mathrm dM $ by a factor $ (m/M_*)^2 $ for $ m \ll M_* $\,, this is necessarily balanced by the factor $ (\Delta M/\Delta m) $\,, since the PBH number density itself is preserved prior to complete evaporation.
So there is no extra $ (m/M_*)^3 $ factor in any constraint on $ \beta(M) $\,.
However, the current \emph{integrated} number density of black holes between $ 0 $ and $ m $ is less the number density of the original $ M_* $ black holes by a factor $ (m/M_*)^3 $ because this corresponds to PBHs with initial mass in some small range above $ M_* $\,.

\bibliographystyle{apsrev4-1}
\bibliography{pbh}

\end{document}